\documentclass[%
 reprint,
superscriptaddress,
 amsmath,amssymb,
 aps,
showkeys
]{revtex4-1}

\usepackage{hyperref}
\usepackage{graphicx}
\usepackage{dcolumn}
\usepackage{bm}

\begin{document}


\title{Addressing the Grid-size Sensitivity Issue in Large-eddy Simulations of Stable Boundary Layers}

\author{Yi Dai}
\affiliation{Faculty of Civil Engineering and Geosciences, Delft University of Technology, Delft, the Netherlands}

\author{Sukanta Basu}%
\email{sukanta.basu@gmail.com}
\affiliation{Faculty of Civil Engineering and Geosciences, Delft University of Technology, Delft, the Netherlands}

\author{Bj\"{o}rn Maronga }
\affiliation{Geophysical Institute, University of Bergen, Bergen, Norway}%
\affiliation{Institute of Meteorology and Climatology, Leibniz University Hannover, Hannover, Germany}

\author{Stephan R de Roode}%
\affiliation{Faculty of Civil Engineering and Geosciences, Delft University of Technology, Delft, the Netherlands}

\date{\today}

\begin{abstract}
In this study, we have identified certain fundamental limitations of a mixing length parameterization used in a popular turbulent kinetic energy-based subgrid-scale model. Replacing this parameterization with a more physically realistic one significantly improves the overall quality of the large-eddy simulations (LESs) of stable boundary layers. For the range of grid sizes considered here (specifically, 1 m -- 12.5 m), the revision dramatically reduces the grid-size sensitivity of the simulations. Most importantly, the revised scheme allows us to reliably estimate the first- and second-order statistics of a well-known LES intercomparison case, even with a coarse grid-size of O(10 m). 
\end{abstract}

\maketitle


\section{Introduction} \label{intro}

The first large-eddy simulation (LES) intercomparison study \citep{beare06}, organized under the auspices of the Global Energy and Water Exchanges (GEWEX) Atmospheric Boundary Layer Study (GABLS), has had a lasting impact on the stable boundary layer (SBL) research field. In the past decade and a half, the key findings from this study (henceforth referred to as GABLS1-LES) were cited by numerous papers; a few examples are: 
 
\begin{quote}
``Adequate WSBL [weakly stable boundary layer] resolution is attained with 2-m grids \citep{beare06}, but higher resolution is required for moderate and very stable stratification (e.g., SBL depths of less than 50 m or so).'' \citep{fernando10}
\end{quote}

\begin{quote}
``Even for weak to moderately stable conditions, LES of the NBL [nocturnal boundary layer] requires a grid spacing of O(1 m) \citep{beare06}, which greatly increases the computational burden.'' \citep{vanstratum2015}
\end{quote}

\begin{quote}
``In particular, it is often observed that grid convergence for simulations of the stable boundary layer is lacking, see \cite{beare06} and \cite{sullivan16}. The latter used fine grid spacings down to [0.39] m (pseudo-spectral code) and still reported a sensitivity of their results to the grid spacing. Until now, a convincing explanation for this behaviour has been lacking, creating a limitation for the application of LES models for simulating the stable boundary layer.'' \citep{maronga19}
\end{quote}

Similar statements on the grid-size sensitivity can be found in other peer-reviewed publications and are often heard a few times in any contemporary workshop or conference session on SBLs. Such an overwhelming consensus among the SBL-LES community at large is somewhat disconcerting given the fact that a handful of papers established a while ago that certain dynamic (tuning-free) subgrid-scale (SGS) models perform rather well (in terms of first- and second-order statistics) with coarse resolutions. As a matter of fact, around the time of publication by Beare et al (2006)\cite{beare06}, Basu and Port{\'e}-Agel (2006)\cite{basu06} demonstrated that the GABLS1-LES case can be simulated reliably by a dynamic SGS parameterization called the locally-averaged scale-dependent dynamic (LASDD) model. They concluded:
\begin{quote}
    ``Moreover, the simulated statistics obtained with the LASDD model show relatively little resolution dependence for the range of grid sizes considered here. In essence, it is shown here that the new LASDD model is a robust subgrid-scale parameterization for reliable, tuning-free simulations of stable boundary layers, even with relatively coarse resolutions.'' \citep{basu06}
\end{quote}
Later on, Stoll and Port{\'e}-Agel\cite{stoll08} and Lu  and Port{\'e}-Agel\cite{lu13} compared different dynamic SGS schemes and also reported negligible sensitivity to grid-sizes.

In this study, we revisit the GABLS1-LES case study and probe into the inherent cause of the grid-size sensitivity of a static (non-dynamic) SGS parameterization. This parameterization was originally proposed by Deardorff (1980)\cite{deardorff80} and appears in many LES codes \citep[e.g.,][]{moeng07,DALES,PALM,gibbs16}. Henceforth, we refer to this parameterization as D80. Several past SBL-LES studies have reported grid-size sensitivity with the D80 SGS scheme \citep[e.g.,][]{jimenez05,beare06,deroode17,maronga19}. After extensive numerical experiments, we have identified the SGS mixing length ($\lambda$) parameterization in this scheme to be at the root of the grid-size sensitivity issue. We have found that a rather simple (yet physically realistic) modification of $\lambda$ alleviates the grid-size sensitivity substantially. In addition, the first- and second-order statistics from the LES runs utilizing this revised parameterization (named D80-R) agrees well with the ones produced by a pseudo-spectral LES code utilizing a dynamic SGS model. Most importantly, the D80-R scheme allows us to reliably simulate the GABLS1-LES case even with a coarse grid-size of O(10 m).

The organization of this paper is as follows. In the following section we describe the D80 parameterization and its fundamental limitations. Physical interpretation and analytical derivations pertaining to the D80-R parameterization are included in Sect.~3. Technical details of the simulations are provided in Sect.~4. Results based on the D80 and D80-R SGS schemes are documented in Sect.~5. A few concluding remarks are made in Sect.~6. In Appendix~1, we have documented a few SGS models which do not use grid size as a mixing length scale; instead, similar to the D80-R parameterization, they use flow-physics-dependent mixing length scale formulation. To further confirm the strength of the D80-R approach, simulated results from an independent LES code are included in Appendix~2. 

\section{SGS Parameterizations}

The SGS parameterization (D80) by Deardorff \cite{deardorff80} utilizes the following mixing length scale: 

\begin{equation}
    \lambda = \min \left(\Delta_g, L_b \right),
    \label{lambda}
\end{equation}
where, $\Delta_g = \left(\Delta x \Delta y \Delta z\right)^{1/3}$. $L_b$ is the so-called buoyancy length scale and is typically represented as follows: 
\begin{equation}
L_b = c_n \frac{\overline{e}^{1/2}}{N},    
\label{Lb}
\end{equation}
where, $\overline{e}$ denotes SGS turbulent kinetic energy (TKE). $N$ is the Brunt-V\"{a}is\"{a}la frequency. Deardorff \cite{deardorff80} assumed $c_n$ to be equal to 0.76. Many LES studies still assume this constant value \citep[e.g.,][]{jimenez05,DALES,PALM} though there are a handful of exceptions. For  example, Gibbs and Fedorovich \cite{gibbs16} assumed $c_n = 0.5$. 

The eddy viscosity ($K_m$), eddy diffusivity ($K_h$), and energy dissipation rate ($\overline{\varepsilon}$) are all assumed to be functions of $\lambda$ as shown below:

\begin{subequations}
\begin{equation}
    K_m = c_m \lambda \overline{e}^{1/2},
\end{equation}
\begin{equation}
    K_h = c_h \lambda \overline{e}^{1/2},
\end{equation}
\begin{equation}
    \overline{\varepsilon} = \frac{c_\varepsilon}{\lambda} \overline{e}^{3/2}.
    \label{EDR0}
\end{equation}
\end{subequations}
The unknown coefficients ($c_m$, $c_h$, and $c_\varepsilon$) are either prescribed or parameterized as follows: 

\begin{subequations}
\begin{equation}
c_m = 0.12,    
\end{equation}
\begin{equation}
c_h = \left(1 + 2\frac{\lambda}{\Delta_g} \right) c_m,
\label{ch}
\end{equation}
\begin{equation}
c_\varepsilon = 0.19 + 0.51 \frac{\lambda}{\Delta_g}.   
\label{cv}
\end{equation}
\label{coeff}
\end{subequations}
Please note that the values of the coefficients in Eq.~\ref{coeff} somewhat vary between different studies \citep[e.g.,][]{deardorff80,moeng88,saiki00}. 

For neutral condition, Eq.~\ref{lambda} reduces to: $\lambda = \Delta_g$. As a direct consequence, $K_m$, $K_h$, and $\overline{\varepsilon}$ become stability independent, as would be physically expected. Furthermore, SGS Prandtl number ($Pr_S = K_m/K_h = c_m/c_h$) equals to 0.33 for neutral condition. 

For stably stratified condition, one expects that $K_m$, $K_h$, and $\overline{\varepsilon}$ should have a clear dependence on $N$. Deardorff \cite{deardorff80} accounted for such dependence by introducing $L_b$ in Eq.~\ref{lambda}. Specifically, he stated (we have changed the variable notation for consistency): 
\begin{quote}
    ``In past work it has been assumed that $\lambda = \Delta_g$, which fails to take account of the possibility that in a stably stratified region $\lambda$ could become much smaller than the grid interval.'' \citep{deardorff80}
\end{quote}
For the very stable case, in the limit of $1/N \to 0$, the D80 SGS model predicts $\lambda = L_b \to 0$, $c_\varepsilon \to 0.19$, and $Pr_S \to 1$. Also, $K_m$ and $K_h$ are expected to approach negligibly small values for such conditions. 

The behavior of the D80 SGS model in the weakly and moderately stable boundary layer is rather problematic and is the focus of the present study. For such cases (including the GABLS1-LES case), $L_b$ is typically on the order of 10 m in the lower and middle parts of the SBLs. As mentioned earlier, it has become customary to perform SBL-LES runs with grid-sizes of 1--5 m or finer these days. In such simulations, by virtue of Eq.~\ref{lambda}, $\lambda$ becomes equal to $\Delta_g$ for lower and middle parts of the SBL. In other words, the `min' operation in Eq.~\ref{lambda} is only utilized in the upper part of the SBL, and in the free atmosphere. Most importantly, akin to the neutral condition, $K_m$, $K_h$, and $\overline{\varepsilon}$  become spuriously independent of stability in the lower and middle parts of the SBL. 

There is another fundamental problem with Deardorff's SGS mixing length parameterization. It is not influenced by the presence of a surface. Deardorff \cite{deardorff80} recognized this problem and proposed a solution of increasing $c_\varepsilon$ near the surface \citep[see also][]{moeng84}.
In this context, Gibbs and Fedorovich \cite{gibbs16} stated: 
\begin{quote}
    ``To this end, it is unclear, however, whether the parameter adjustments incorporated in the original D80 scheme were based on some clear physical reasoning or were intended to merely produce more plausible effects close to the surface.''
\end{quote}
To the best of our knowledge, such ad-hoc solutions are not implemented in recent LES codes \cite[e.g.,][]{DALES,PALM,deroode17}. In other static SGS models (e.g., the Smagorinsky-Lilly model and its variants), empirical wall functions are utilized to explicitly account for the near-surface shear effects \cite[e.g.,][]{mason90,brown94}. Interestingly, it is not a common practice to use wall functions with the D80 SGS scheme. In the case of dynamic SGS models, wall functions are not needed as the estimated SGS coefficient steadily decreases as one approaches any surface, and thus, reduces the SGS mixing length in a self-consistent manner \cite[e.g.,][]{basu06,stoll08}. 

In the present study, we replace the mixing length parameterization in D80 (i.e., Eq.~\ref{lambda}) with the following formulation: 
\begin{equation}
    \frac{1}{\lambda} = \frac{1}{\kappa z} + \frac{1}{L_b},
    \label{brost}
\end{equation}
where, $\kappa$ is the von K\'{a}rm\'{a}n constant. Both the effects of stability and near-surface are nicely captured by this equation. In Sect.~3, we will show that the $\kappa z$ term of this equation can be derived from a well-known spectral scaling. 

The origin of Eq.~\ref{brost} can be traced back to the work of Blackadar \cite{blackadar62} and Brost and Wyngaard \cite{brost78}. Blackadar \cite{blackadar62} introduced the following length scale: 

\begin{equation}
    \frac{1}{\lambda} = \frac{1}{\kappa z} + \frac{1}{\lambda_0},
    \label{blackadar}
\end{equation}
where, $\lambda_0$ is an asymptotic length scale. Brost and Wyngaard \cite{brost78} and following studies \citep[e.g.,][]{baas08} assumed: 

\begin{equation}
    \lambda_0 = L_b^* \approx \frac{\overline{E}^{1/2}}{N},
    \label{brost2}
\end{equation}
where, $\overline{E}$ is the total TKE. 

Please note that despite of apparent similarity, the length scales $L_b$ and $L_b^*$ are quite different. $L_b$ is proportional to SGS TKE ($\overline{e}$), and thus, implicitly depend on the filter-size ($\Delta_f$) (see Sect.~3). In contrast, $L_b^*$ is used in Reynolds Averaged Navier-Stokes (RANS) models, and by definition, it does not depend $\Delta_f$.

Hereafter, we refer to Deardorff's SGS parameterization in conjunction with Eq.~\ref{brost} as the D80-R parameterization. In this parameterization, with increasing resolution, both $\overline{e}$ and $L_b$ decrease. As a result, $\lambda$, $K_m$, and $K_h$ also decrease. Additional effect of grid-size is also felt via $c_h$, and $c_\varepsilon$ coefficients. In terms of SGS Prandtl number ($Pr_S$), both the D80 and D80-R parameterizations suffer from unphysical prescription in different ways. This issue is discussed in detail in Sect.~5.

\section{Physical Interpretation and Analytical Derivation}

In this section, we make direct associations between the energy spectra of turbulence and several key elements of the D80-R parameterization. In addition, we introduce a generalized form Eq.~\ref{brost} which has the potential to further extend the regime of application of the proposed D80-R parameterization. 

\subsection{Dependence of $\overline{e}$ on Filter-size ($\Delta_f$)}
\label{Theory1}

A simple model of longitudinal velocity spectrum, spanning all the scales of turbulence, can be written as \cite[see page 232 of][]{pope00}:

\begin{equation}
    E_u\left(k\right) \sim \overline{\varepsilon}^{2/3} k^{-5/3} \Phi_\eta\left(k\eta \right) \Phi_{L_I}\left(k L_I \right),
    \label{SpectraU}
\end{equation}
where, $k$ is wavenumber. $E_u$ is the energy spectrum for longitudinal velocity component. $L_I$ and $\eta$ denote integral length scale and Kolmogorov's scale, respectively. The non-dimensional functions $\Phi_{L_I}$ and $\Phi_{\eta}$ represent the buoyancy range and dissipation ranges, respectively. In the inertial-range, both these functions are close to unity. For small values of $k L_I$ and large values of $k \eta$, they deviate from unity.  

If the filter size ($\Delta_f$) is within the inertial-range (as inherently assumed in LES), the subgrid-scale variance of longitudinal velocity component ($\sigma_{us}^2$) can be estimated from Eq.~\ref{SpectraU} as: 

\begin{equation}
    \sigma_{us}^2 \sim \overline{\varepsilon}^{2/3} \int_{k_\Delta}^{k_\eta} k^{-5/3} \Phi_\eta\left(k\eta \right) dk,
    \label{VarU}
\end{equation}
where, $k_\eta$ is the dissipation wavenumber. The wavenumber associated with the filter is: $k_\Delta = \frac{\pi}{\Delta_f}$. Within the range of $k_\Delta$ to $k_\eta$, $\Phi_{L_I}$ is unity. Since the contribution of dissipation range is typically small towards $\sigma_{us}^2$, one can assume $\Phi_{\eta} \approx 1$ in Eq.~\ref{VarU}. 

Owing to isotropy in the inertial-range, Eq.~\ref{VarU} can be integrated as follows: 

\begin{equation}
    \overline{e} = \frac{3}{2} \sigma_{us}^2 \sim \overline{\varepsilon}^{2/3} \left[ \frac{1}{k_\Delta^{2/3}} - \frac{1}{k_\eta^{2/3}}  \right].
    \label{TKESGS1}
\end{equation}
Since $k_\eta \gg k_\Delta$, we can simplify this equation as: 

\begin{equation}
    \overline{e} \sim \overline{\varepsilon}^{2/3} \Delta_f^{2/3}.
    \label{TKESGS2}
\end{equation}
Eq.~\ref{TKESGS2} can be re-written as: 
\begin{equation}
    \overline{\varepsilon} \sim \frac{\overline{e}^{3/2}}{\Delta_f}.
    \label{EDR1}
\end{equation}
This equation is identical to Eq.~\ref{EDR0}, if one replaces $\lambda$ with $\Delta_f$. 

\subsection{Estimation of $\overline{e}$ in Surface Layer}
\label{Theory2}

The derivations in Sect.~\ref{Theory1} assume $\Delta_f$ falls within the inertial-range. However, in the surface layer, the inertial-range is rather limited. For an extensive range of scales, the longitudinal velocity spectrum follows a $k^{-1}$ power law. Thus, in most LES studies, it is likely that $\Delta_f$ falls within the $k^{-1}$ range in the surface layer and not within the inertial range. 

Following Tchen (1953)\cite{tchen53}, numerous studies have reported the $k^{-1}$ scaling in the literature; please refer to a comprehensive list in Table 1 of Katul and Chu (1998)\cite{katul98}. By combining the $k^{-1}$ scaling with the inertial-range scaling (i.e., $k^{-5/3}$), the energy spectrum for the longitudinal velocity spectrum can be written as:  

\begin{subequations}
\begin{equation}
E_u \sim u_*^2 k^{-1} \mbox{\hspace{0.9in}for\hspace{0.2in}} k_\Delta \le k \le k_o    
\end{equation}

\begin{equation}
E_u \sim \overline{\varepsilon}^{2/3} k^{-5/3} \Phi_\eta\left(k\eta \right) \mbox{\hspace{0.2in}for\hspace{0.2in}} k_o \le k \le k_\eta    
\end{equation}
\label{SLScaling}
\end{subequations}

\noindent where, the friction velocity is denoted by $u_*$. The crossover wavenumber $k_o$ equals to $\frac{1}{\gamma z}$. For unstable condition, $\gamma$ was found to be equal to unity by Kader and Yaglom \cite{kader91} and others. For stable condition, the value of $\gamma$ decreases from unity with increasing stability \citep{banerjee16}.   

By integrating Eq.~\ref{SLScaling}, we can estimate the SGS variance of longitudinal velocity component as follows:

\begin{equation}
    \sigma_{us}^2 = c_1 u_*^2 \int_{k_\Delta}^{k_o} k^{-1} dk + c_2 \overline{\varepsilon}^{2/3} \int_{k_o}^{k_\eta} k^{-5/3} \Phi_\eta\left(k\eta \right) dk, 
    \label{SLVar1}
\end{equation}
where, $c_1$ and $c_2$ are unknown constants. Due to strong anisotropy in the surface layer, we cannot estimate SGS TKE ($\overline{e}$) from the $\sigma_{us}^2$. However, it is expected that $\overline{e}$ will be proportional to $\sigma_{us}^2$. 

The first integration term of Eq.~\ref{SLVar1} can be simplified as: 
\begin{align}
    \sigma_{us1}^2 \sim u_*^2 \int_{k_\Delta}^{k_o} k^{-1} dk & = u_*^2 \log\left( \frac{k_o}{k_\Delta}\right) \nonumber \\  & = u_*^2 \log\left( \frac{\Delta_f}{\pi\gamma z}\right).
    \label{SLVar2}
\end{align}
Whereas, the second term of Eq.~\ref{SLVar1} amounts to:

\begin{align}
    \sigma_{us2}^2 \sim \overline{\varepsilon}^{2/3} \int_{k_o}^{k_\eta} k^{-5/3} dk & \sim \overline{\varepsilon}^{2/3} \left[ \frac{1}{k_o^{2/3}} - \frac{1}{k_\eta^{2/3}}  \right] \nonumber \\ & \sim \overline{\varepsilon}^{2/3} \left(\gamma z \right)^{2/3}
    \label{SLVar3}
\end{align}

Please note that, according to Eq.~\ref{SLVar3}, $\sigma_{us2}^2$ is not dependent on $\Delta_f$, rather it is dependent on $z$. Due to logarithmic operation in Eq.~\ref{SLVar2}, $\sigma_{us1}^2$ weakly depends on $\Delta_f$. Thus, $\overline{e}$ is expected to be weakly dependent on $\Delta_f$ in the surface layer. More importantly, Eq.~\ref{EDR1} cannot be used to estimate $\overline{\varepsilon}$ in the surface layer. 

In general, both the terms $\sigma_{us1}^2$ and $\sigma_{us2}^2$ contribute to $\overline{e}$. It is not possible to neglect either of them for further simplification. However, if we assume that both these variances are proportional to $\overline{e}$, we get qualitatively similar results regarding the mixing length. 

Based on Eq.~\ref{SLVar3}, we can write: 
\begin{subequations}
\begin{equation}
    \overline{e} \sim \sigma_{us2}^2 \sim \overline{\varepsilon}^{2/3} \left(\gamma z \right)^{2/3},  
\end{equation}
and
\begin{equation}
    \overline{\varepsilon} = \frac{c_\gamma}{z} \overline{e}^{3/2},  
\end{equation}
\end{subequations}
where, $c_\gamma$ is an unknown constant. By comparing Eq.~\ref{EDR0} with this equation, we can assert that the mixing length scale ($\lambda$) in D80 and D80-R should be proportional to $z$ in the surface layer instead of $\Delta_f$. In the following section, we arrive at the same conclusion via a different route. 

\subsection{Shear-based Mixing Length Scale}

Let us assume that $S$ represents the magnitude of velocity shear. Thus, gradient Richardson number ($Ri_g$) equals to $N^2/S^2$. For shear-dominated flows, Hunt et al \cite{hunt88,hunt89} proposed a length scale: 

\begin{equation}
    L_H^* \sim \frac{\overline{E}^{1/2}}{S},
    \label{Hunt0}
\end{equation}
where, $\overline{E}$ is the total TKE. It is a common knowledge that the effects of shear are more prevalent than the buoyancy effects near the surface. For such a situation, $L_H^*$ is a more relevant length scale than $L_b^*$ (defined earlier in Eq.~\ref{brost2}). 

For SGS modeling, if $\overline{E}$ is replaced with $\overline{e}$, an analogous shear-based length scale can be defined: 
\begin{equation}
    L_H \sim \frac{\overline{e}^{1/2}}{S},
    \label{Hunt1}
\end{equation}
In contrast to $L_H^*$, the length scale $L_H$ implicitly depends on $\Delta_f$ due to its explicit relationship to $\overline{e}$.

Earlier we showed that, Eq.~\ref{SLVar2} holds near the surface. Now, if we assume $\overline{e}$ is proportional to $\sigma_{us1}^2$, we can combine Eq.~\ref{SLVar2} with Eq.~\ref{Hunt1} and get: 

\begin{equation}
    \lambda \sim L_H \sim \frac{u_*}{S} \left[ \log\left(\frac{\Delta_f}{\pi \gamma z} \right)\right]^{1/2} = \frac{\kappa z}{\phi_m} \left[ \log\left(\frac{\Delta_f}{\pi \gamma z} \right)\right]^{1/2},
\end{equation}
where, $\phi_m$ is the non-dimensional velocity gradient. 

For finite-difference-based LES codes, $\Delta_f$ is typically 4--6 times larger than $\Delta_g$. For isotropic grids, the height of vertical levels ($z$) equal to $m \Delta_g$; where, $m$ are half-integers. For vertically stretched grids, $z$ could be a small fraction of $\Delta_g$ for the first few levels. For stably stratified condition, $\phi_m \geq 1$. As mentioned earlier, for such situation, $\gamma \leq 1$. Based on these estimates, it is reasonable to state that $\lambda$ is proportional to $z$ in the surface layer; the exact value of the proportionality constant is unknown. In this study, based on other usage in the literature (specifically in RANS modeling), we assumed the proportionality constant to be equal to $\kappa$. 

\subsection{A Generalized Mixing Length Parameterization}

The proposed D80-R mixing length parameterization (i.e., Eq.~\ref{brost}) is valid for stable condition. However, for near-neutral condition, as $N$ approaches zero, the value of $\lambda$ could become very large. To account for such stability regimes, one can adopt a more generic parameterization for the mixing length. Essentially, one can combine $L_H$ and $L_b$ in the following manner: 
\begin{equation}
    \frac{1}{\lambda} = \frac{c_H}{L_H} + \frac{c_b}{L_b},
    \label{Generalized}
\end{equation}
where, the unknown coefficients $c_H$ and $c_b$ should be prescribed. For near-neutral condition, the term involving $L_H$ will dominate and will lead to realistic $\lambda$ values. As discussed in the previous section, this term will also perform well in the surface layer. 

Unfortunately, we do not know how to estimate optimal values of $c_H$ and $c_b$ in a meaningful way. Running numerous large-eddy simulations for various case studies with different combinations of $c_H$ and $c_b$ is not a computationally viable option. Hopefully, an efficient strategy will emerge in the near-future. For the time-being, we utilize an approximation of Eq.~\ref{Generalized}, i.e., Eq.~\ref{brost}, as a working substitute for stable boundary layer simulations. 

\section{Description of the Simulations}
In this work, we simulate the GABLS1-LES case study using the Dutch Atmospheric Large-Eddy Simulation \citep[DALES;][]{DALES} and the PALM model system \citep{PALM,maronga20}. Since the configurations of the GABLS1-LES case study are well-known in the literature, we mention them in a succinct manner. The boundary layer is driven by an imposed, uniform geostrophic wind of 8 m s$^{-1}$, with a surface cooling rate of 0.25 K~h$^{-1}$ and attains a quasi-steady state in about 8--9 h with a boundary layer depth of approximately 200 m. The initial mean potential temperature is 265 K up to 100 m with an overlying inversion of 0.01 K m$^{-1}$. The Coriolis parameter is set to $1.39 \times 10^{-4}$ s$^{-1}$, corresponding to latitude 73$^{\circ}$ N. Both the aerodynamic roughness length ($z_0$) and the scalar roughness length ($z_{0h}$) are assumed to be equal to 0.1 m. 

For all runs, the computational domain is fixed at 400 m $\times$ 400 m $\times$ 400 m. A wide range of isotropic $\Delta_g$ values are used to investigate the aforementioned grid-size sensitivity issue. For the DALES code, in order to avoid any temporal discretization error, the time step $\Delta t$ is kept at a constant value of 0.1~s for all the simulations. In contrast, an adaptive time-stepping approach is used by the PALM model system. In addition to the results from the finite-difference-based DALES and PALM codes, we also report results from a pseudo-spectral code (called MATLES) utilizing  the LASDD SGS model along with a grid-size of 3.5~m and a fixed time step of 0.075~s.  


\begin{table*}[ht]
\centering
\caption{SGS coefficients and mixing length values in the DALES and PALM codes}
\label{T1}
\begin{tabular}{@{\extracolsep{4pt}}lcccccccc@{}}
\hline\noalign{\smallskip}
& \multicolumn{4}{c}{D80}                                                   & \multicolumn{4}{c}{D80-R}                                              \\
\hline\noalign{\smallskip}
& \multicolumn{2}{c}{$N^2 > 0$} & \multicolumn{2}{c}{$N^2 \le 0$} &  \multicolumn{2}{c}{$N^2 > 0$} & \multicolumn{2}{c}{$N^2 \le 0$} \\ \cline{2-3} \cline{4-5} \cline{6-7} \cline{8-9}
 & $c_h$ & $\lambda$ & $c_h$ & $\lambda$ & $c_h$ & $\lambda$ & $c_h$ & $\lambda$ \\ 
\hline\noalign{\smallskip}
DALES & Eq.~\ref{ch} & Eq.~\ref{lambda}~~ & Eq.~\ref{ch} & $\Delta_g$ & Eq.~\ref{ch} & Eq.~\ref{brost} & Eq.~\ref{ch} & $\Delta_g$\\
PALM & Eq.~\ref{ch} & Eq.~\ref{lambda}* & Eq.~\ref{ch} & $\Delta_g$ & $c_h = c_m$ & Eq.~\ref{brost} & Eq.~\ref{ch} & $\Delta_g$\\
\hline\noalign{\smallskip}
\end{tabular}
\newline * PALM uses $\min(1.8z,\Delta_g,L_b)$.
\end{table*}

In terms of the numerical schemes in the DALES and PALM codes, a third-order Runge-Kutta scheme is used for time integration in conjunction with a fifth-order advection scheme in the horizontal direction \citep{wicker02}. In the vertical direction, a second-order and a fifth-order scheme (which reduces to a second-order scheme near the surface) are used by the DALES and PALM codes, respectively. The MATLES code uses a second-order Adams-Bashforth scheme for time advancement. 

In all the simulations, the lower boundary condition is based on the Monin-Obukhov similarity theory (MOST). As discussed by \cite{basu17}, in order to apply MOST, the first model level in an LES model should not be at a height lower than $\sim 50 z_0$. In this study, for simplicity, this requirement has been violated for all the runs involving high-resolution. At this point, the impact of this violation on the simulated statistics is not noticeable and a thorough investigation on surface layer physics will be conducted in a follow-up study.  

Before discussing the results, we point out that some of the prescribed SGS constants differ between the DALES and the PALM codes. For example, $c_m$ is assumed to be equal to 0.12 in DALES; whereas, PALM sets it at 0.1. The other specifications related to $c_h$ and $\lambda$ depend on local $N^2$ values and are listed in Table~\ref{T1}. In addition, the coefficients in Eq.~\ref{cv} are also slightly different in DALES and PALM. In spite of these differences, the results from DALES and PALM codes follow the same trends as depicted in the following section and in Appendix~1.

\section{Results}

The left and right panels of all the following figures represent the statistics from the D80- and D80-R-based runs, respectively. As prescribed in the GABLS1-LES study, all the statistics are averaged over the last one hour (i.e., 8--9 h) of the simulations. 

\begin{figure*}[ht!]
\centering
  \includegraphics[width=0.49\textwidth]{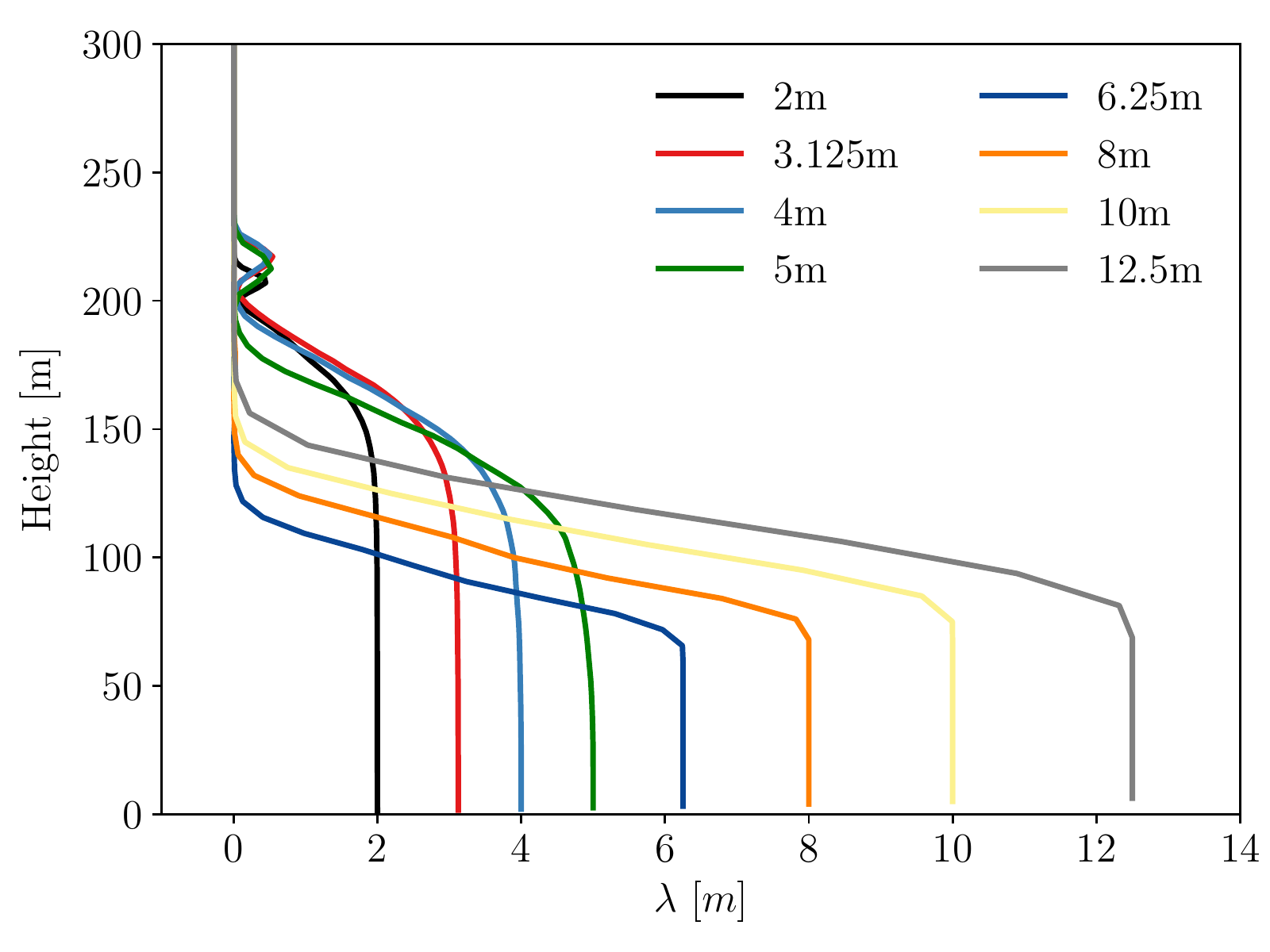}
  \includegraphics[width=0.49\textwidth]{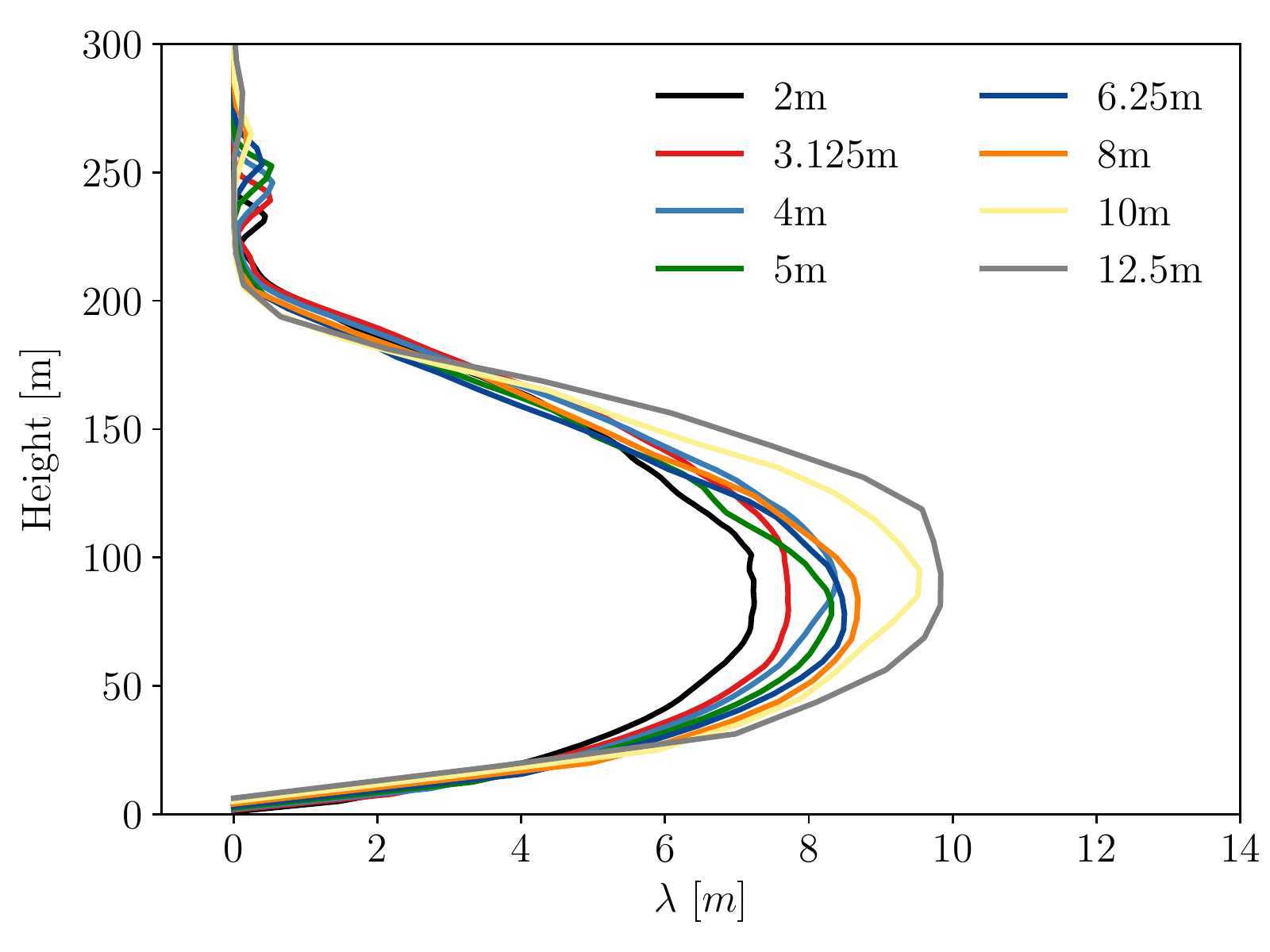}
  \includegraphics[width=0.49\textwidth]{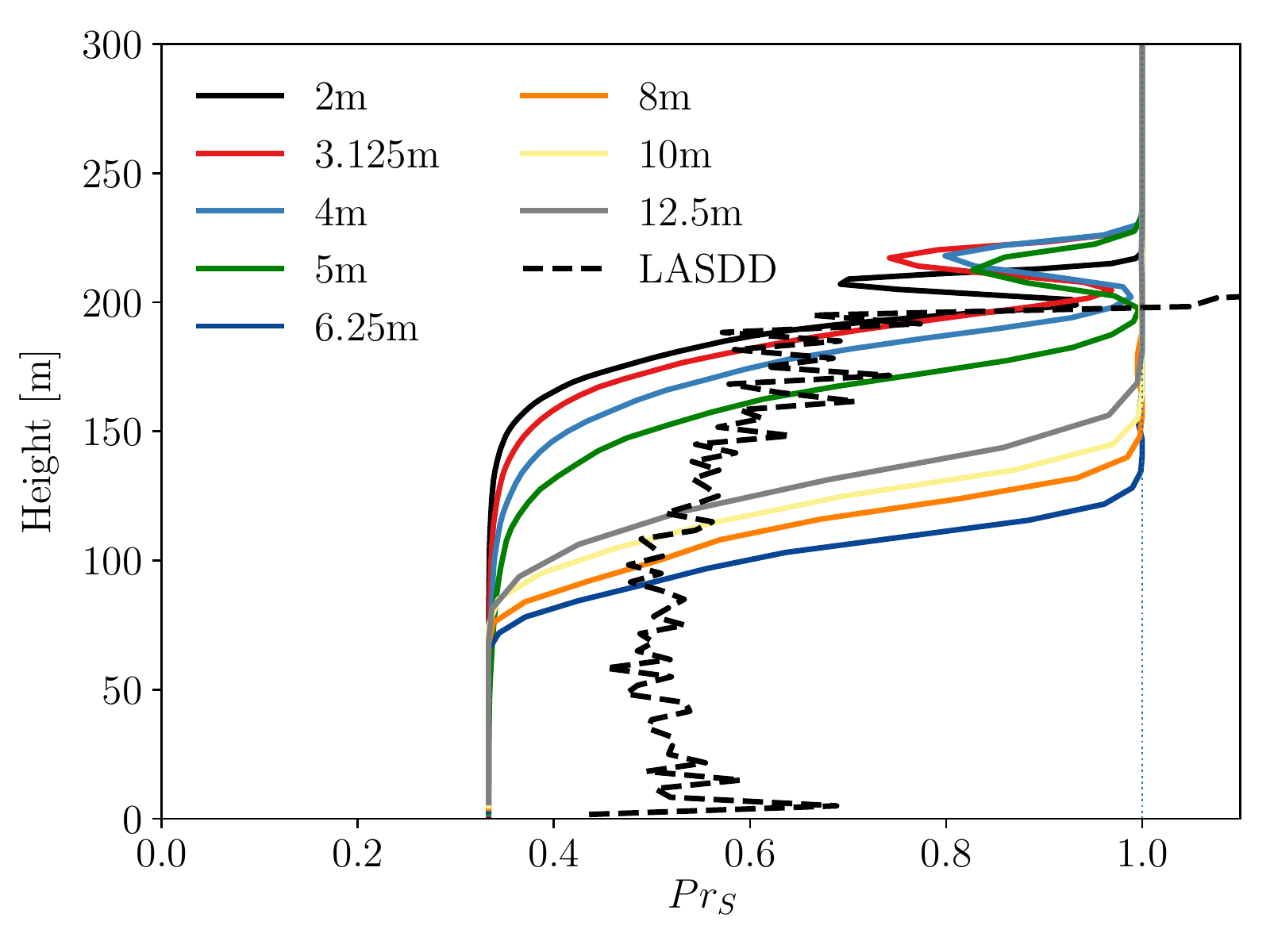}
  \includegraphics[width=0.49\textwidth]{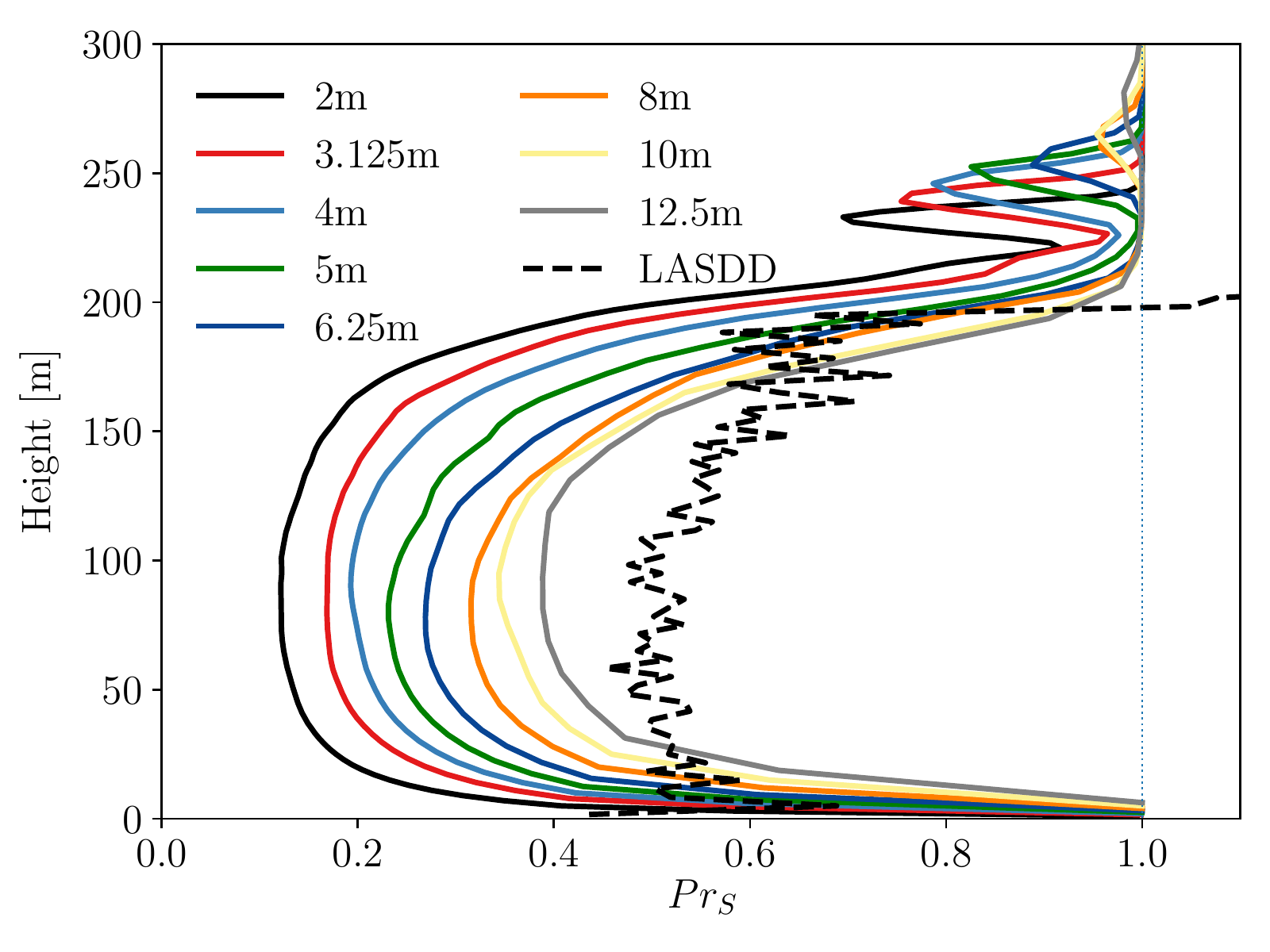}
\caption{Vertical profiles of mixing length ($\lambda$) (top panel) and turbulent Prandtl number (bottom panel) from the D80 (left panel) and D80-R (right panel) based simulations using the DALES code. Different colored lines correspond to different grid sizes ($\Delta_g$). Results from the MATLES code are overlaid (dashed black lines) for comparison.}
\label{fig:lambda}      
\end{figure*}

Prior to discussing the first- and second-order statistics, it is important to highlight the differences between the original and the revised runs in terms of the SGS mixing length ($\lambda$) profiles. From Fig.~\ref{fig:lambda} (top panel), it is evident that in the D80-based runs (left panel), $\lambda$ equals to $\Delta_g$ for the lower and middle parts of the SBL. In contrast, the D80-R-based runs show clear dependence on the distance from the surface. In both types of runs, $\lambda$ values monotonically decrease to zero in the upper part of the SBL and in the free atmosphere due to increasing stratification (as quantified by $N$). The $\lambda$ profiles from the D80-R-based simulations also show clear dependence on grid size in the core region of the SBL. This is due to the fact that SGS TKE decreases with increasing resolution, and in turn, decreases $\lambda$.  

In the D80-based runs, due to the underlying condition of $\lambda = \Delta_g$, $Pr_S$ becomes exactly equal to 0.33 for the lower part of the SBL (bottom-left panel of Fig.~\ref{fig:lambda}). Whereas, in the case of D80-R,  $\lambda$ is typically smaller than $\Delta_g$ near the surface. Thus, $Pr_S$ is much larger than 0.33. However, in the middle part of the SBL, depending on stratification, $\lambda$ may become larger than $\Delta_g$. For such cases, in D80-R-based runs, $Pr_S$ can be even smaller than 0.33  (bottom-right panel of Fig.~\ref{fig:lambda}). In the upper part of the SBL, due to stronger stratification, $\lambda$ becomes much smaller than $\Delta_g$, and as a consequence, $Pr_S$ monotonically approaches to unity for both the D80 and D80-R cases. In the case of the LASDD SGS model, the dynamically estimated $Pr_S$ values remain more or less constant (around 0.5--0.6) inside the SBL and becomes greater than 1 in the inversion layer. 

Gibbs and Fedorovich \cite{gibbs16} recognized the problem with the $Pr_S$ in D80 parameterization. They proposed to use $Pr_S = 1$ when $N^2$ is locally positive. For locally negative $N^2$ values they proposed an empirical formulation for $Pr_S$. In the present study, thus, the PALM model is employed with $Pr_S = 1$ for $N^2 > 0$ in all the D80-R-based runs. These results are shown in Appendix~1. Please note that for $N^2 \le 0$, both the DALES and the PALM models always use Eq.~\ref{ch} (i.e., they effectively assume $Pr_S$ = 0.33); we have not incorporated the empirical formulation by Gibbs and Dedorovich \cite{gibbs16}.

\begin{figure*}[ht!]
\centering
  \includegraphics[width=0.49\textwidth]{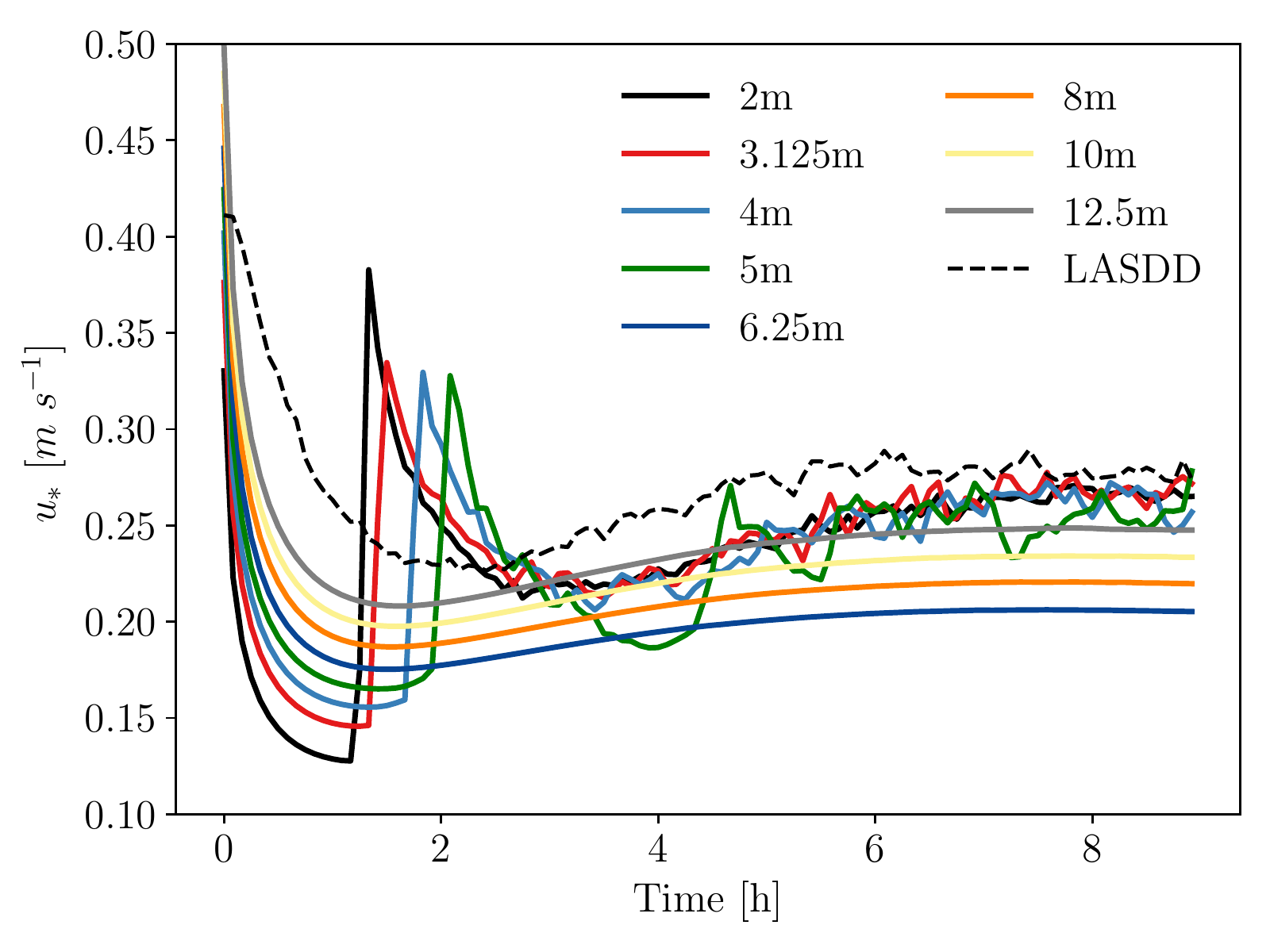}
  \includegraphics[width=0.49\textwidth]{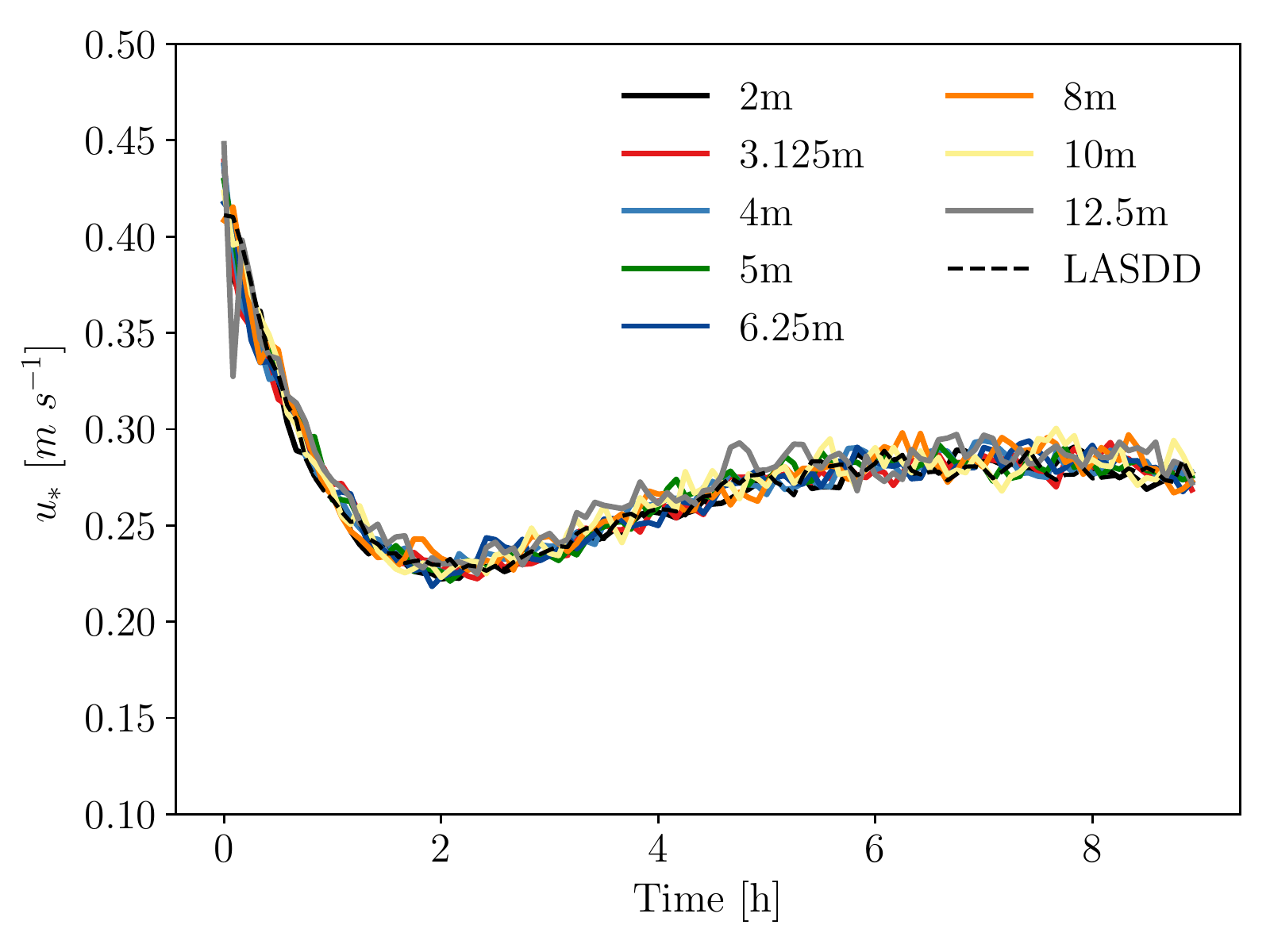}
  \includegraphics[width=0.49\textwidth]{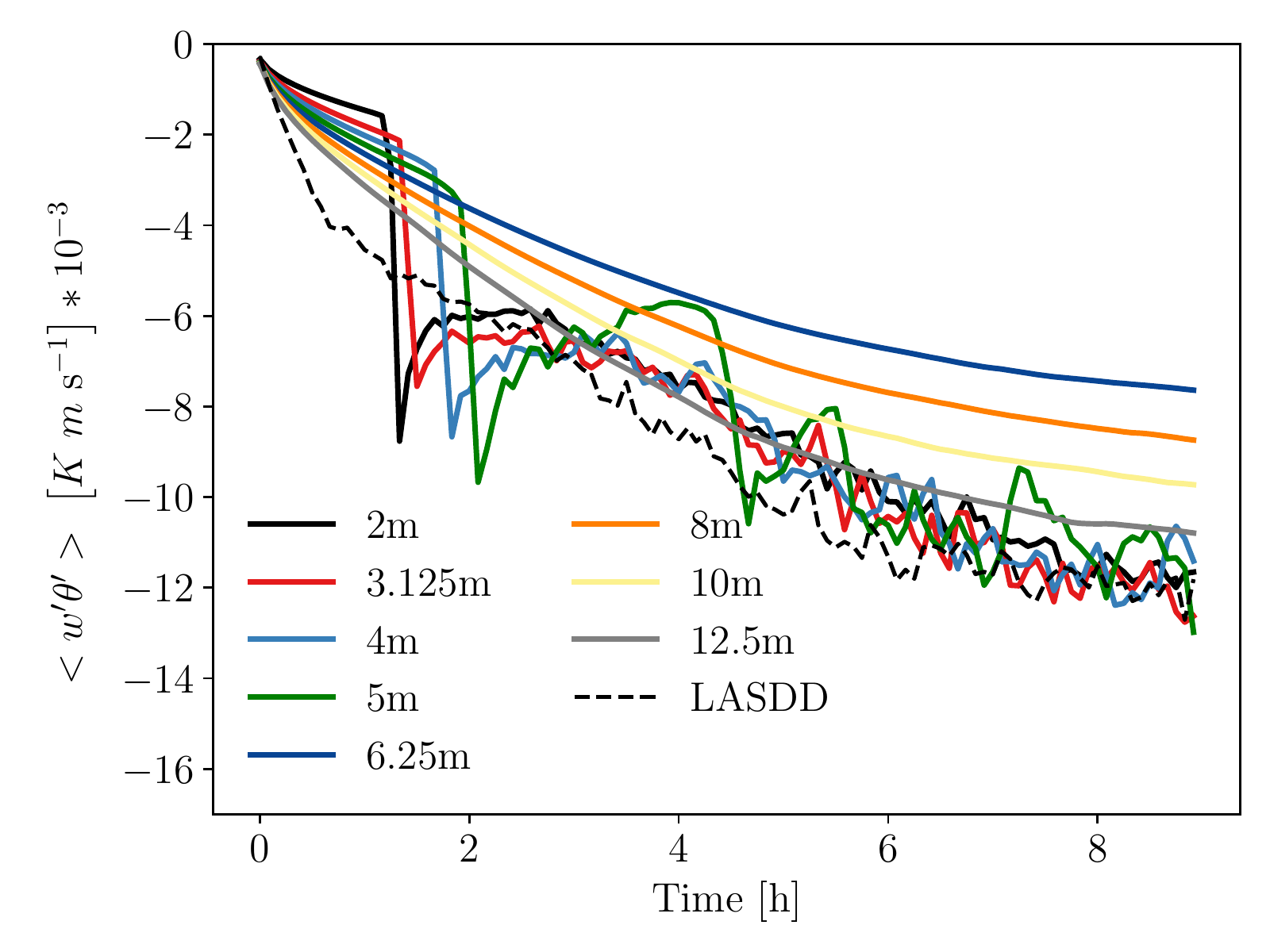}
  \includegraphics[width=0.49\textwidth]{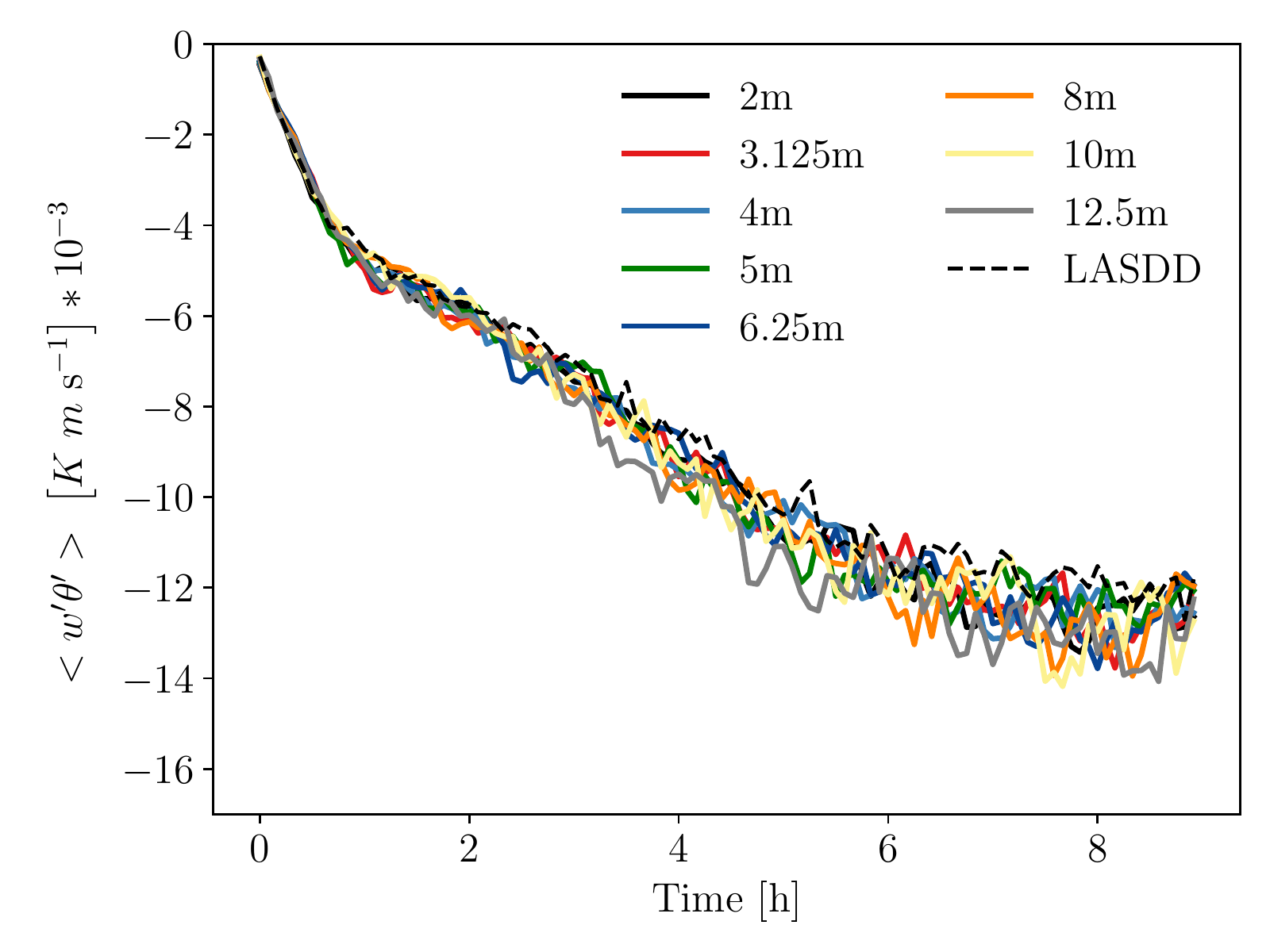}
\caption{Time series of surface friction velocity (top panel) and sensible heat flux (bottom panel) from the D80 (left panel) and D80-R (right panel) based simulations using the DALES code. Different colored lines correspond to different grid sizes ($\Delta_g$). Results from the MATLES code are overlaid (dashed black lines) for comparison.}
\label{fig:surface}      
\end{figure*}

Simulated time series of surface friction velocity and sensible heat flux are documented in Fig.~\ref{fig:surface}. In the D80-based runs, no temporal fluctuations of surface fluxes are visible for $\Delta_g \ge $ 6.25~m. When $\Delta_g$ is finer than 6.25~m, temporal fluctuations do appear approximately 1--2 hours into the simulation. Increasing grid-resolution helps in systematically reducing this turbulence `kick-off' time. In the D80-R-based runs, similar artifacts are not visible in the dynamical evolution of surface fluxes. In the D80-based runs, for $\Delta_g \ge $ 6.25~m, $\lambda$ (= $\Delta_g$) values are excessively large near the surface and simply do not allow for the sustenance of turbulence. The inclusion of the surface dependence term in Eq.~\ref{brost} reduces the mixing length in a meaningful way and promotes the production and transport of turbulence.  

In Fig.~\ref{fig:surface}, the simulated time series from the MATLES code are also overlaid. In the D80-R-based runs, the agreement between the DALES- and MATLES-based results is remarkable.  All the time series of surface friction velocity reach quasi-equilibrium stage after approximately 5~h of simulation. The surface sensible flux time series reach quasi-equilibrium stage at a later time ($\sim$ 6 h). 

\begin{figure*}[ht!]
\centering
  \includegraphics[width=0.49\textwidth]{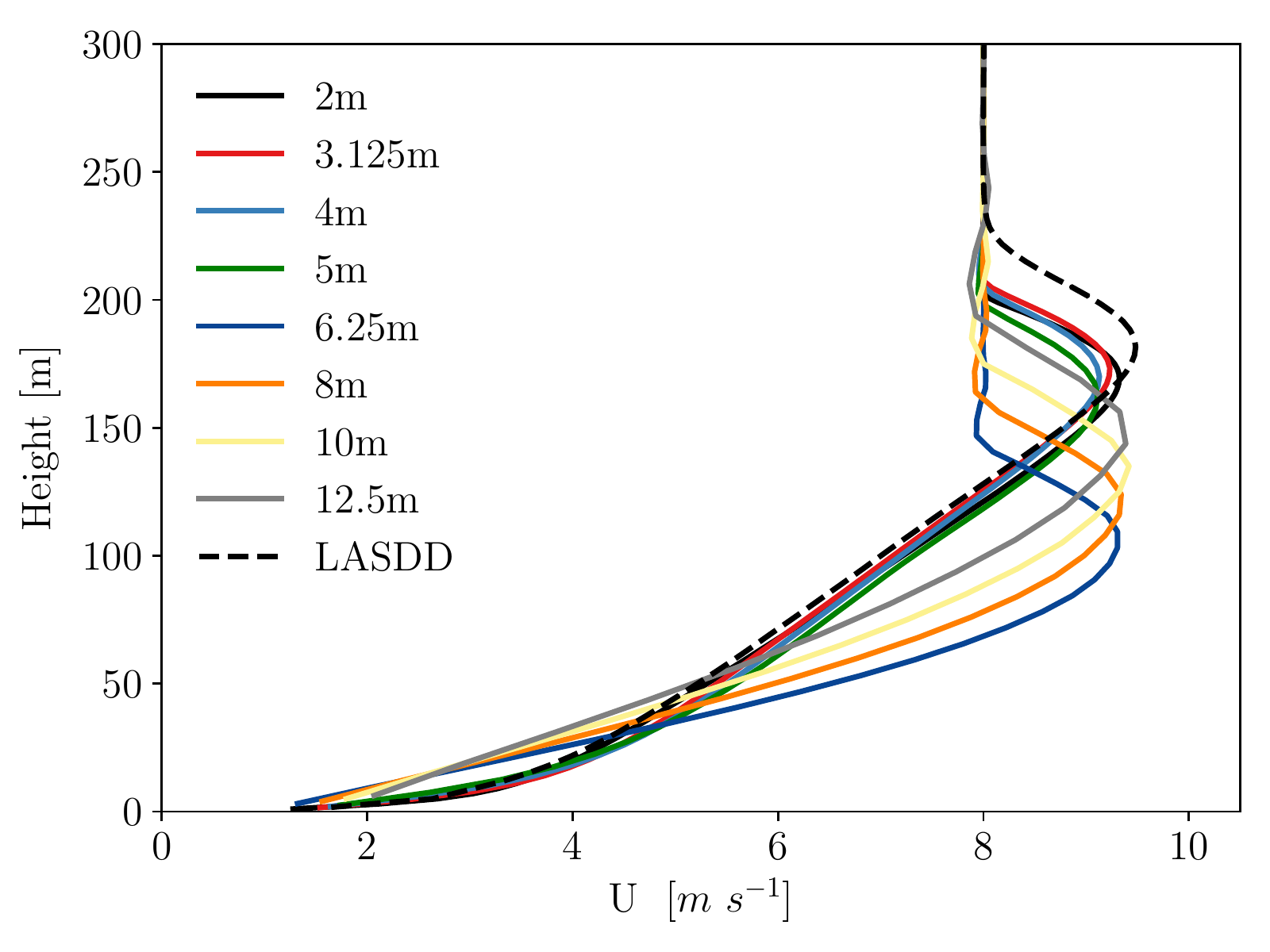}
  \includegraphics[width=0.49\textwidth]{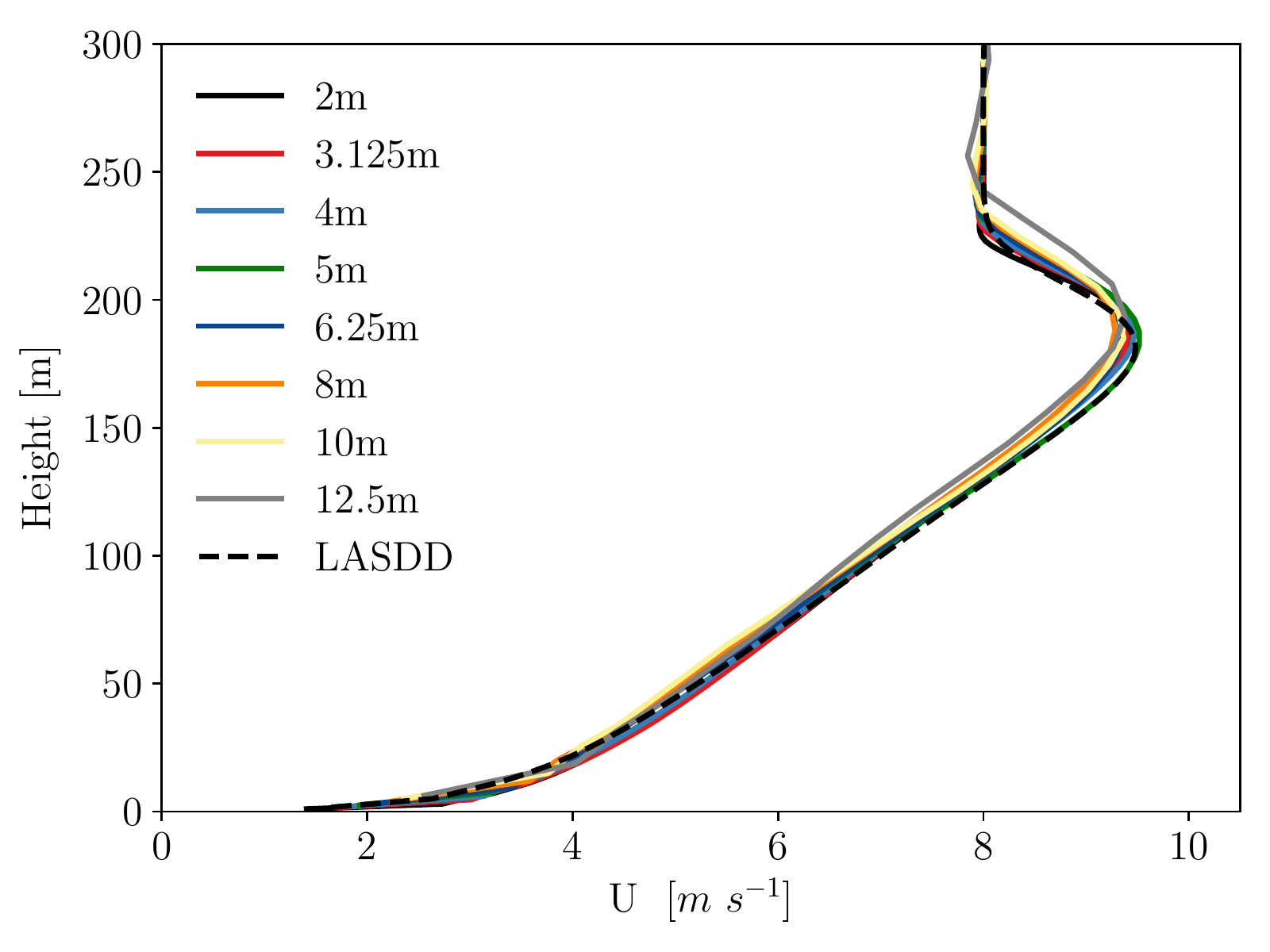}
  \includegraphics[width=0.49\textwidth]{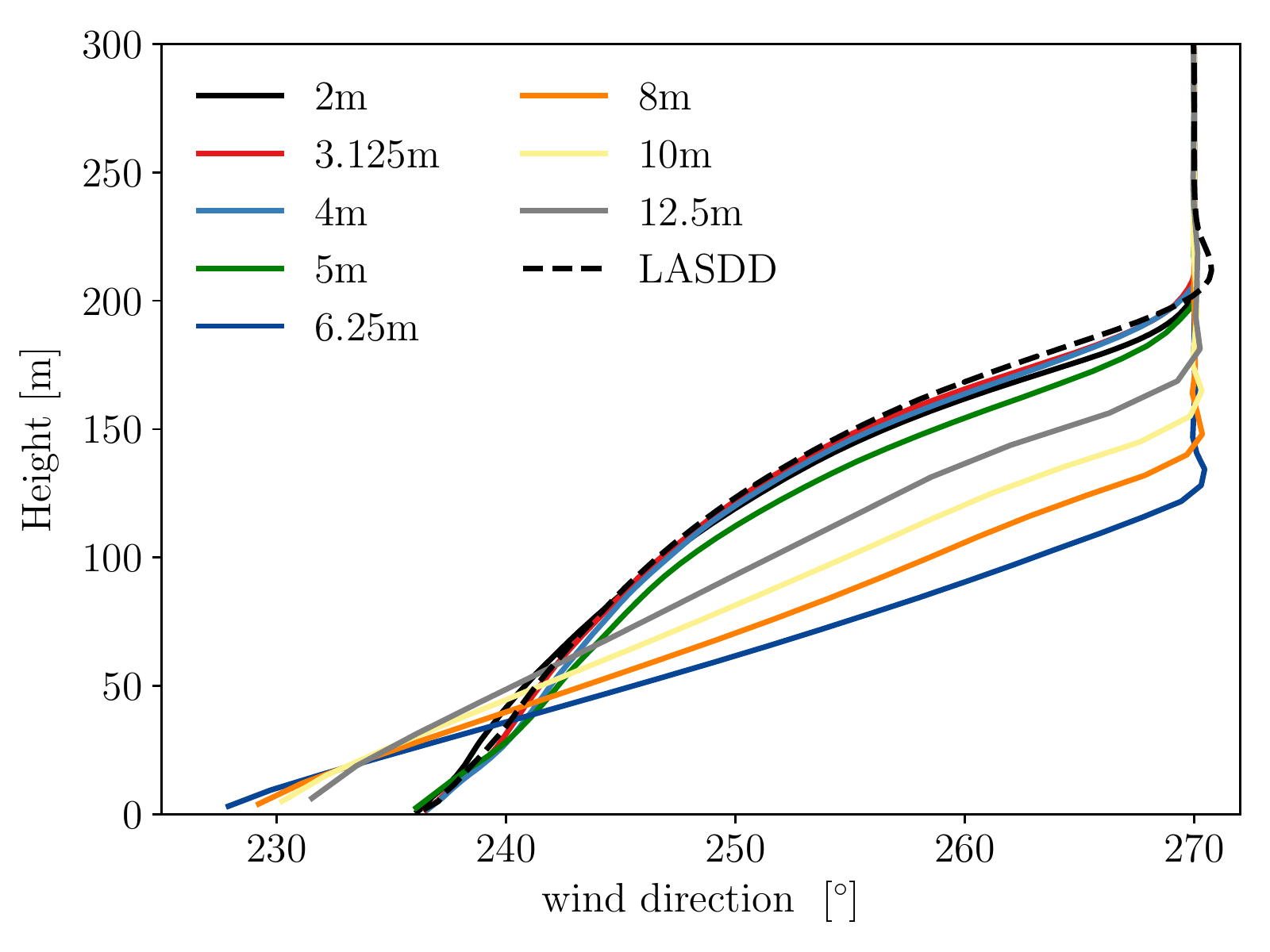}
  \includegraphics[width=0.49\textwidth]{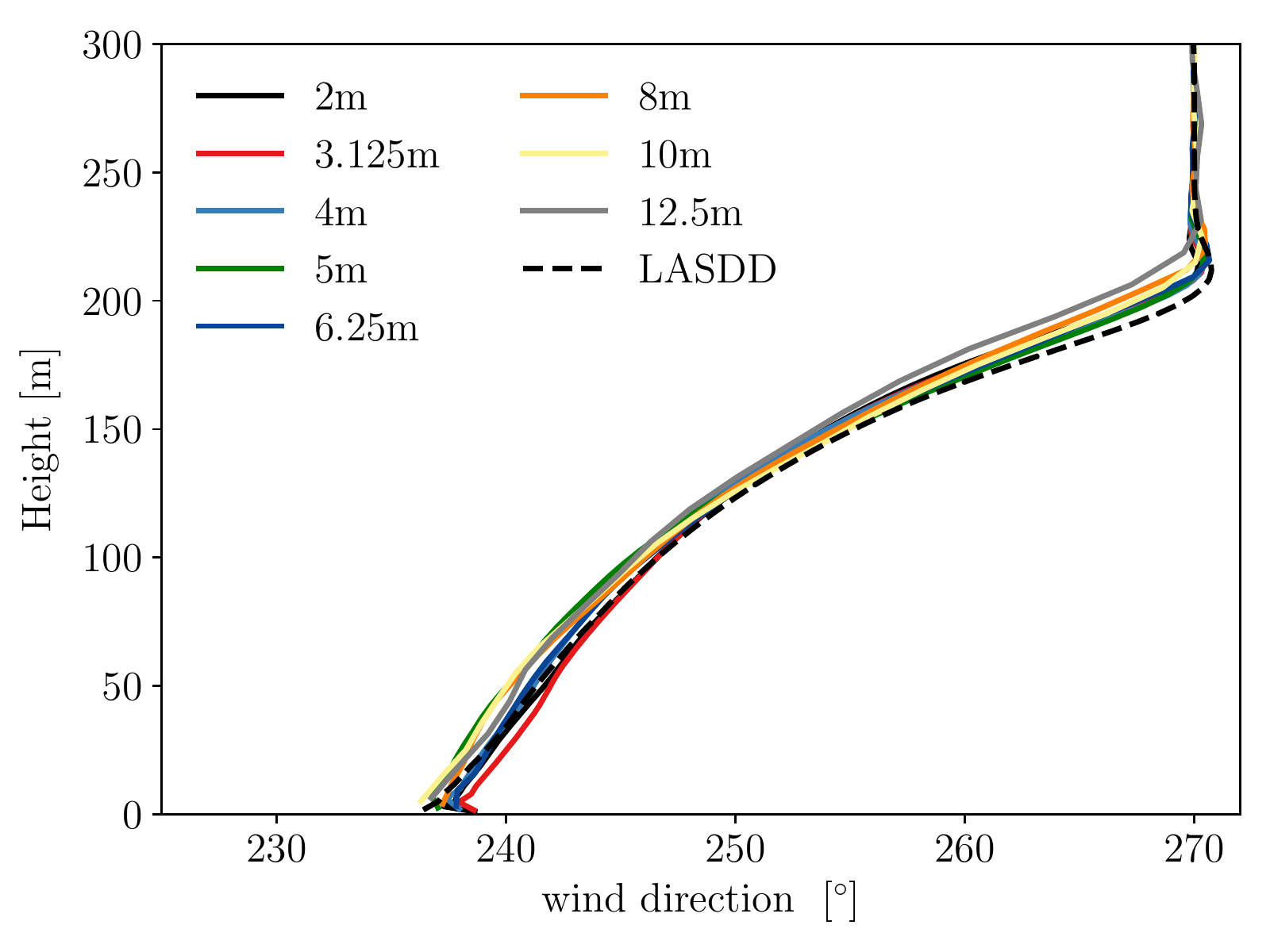}
  \includegraphics[width=0.49\textwidth]{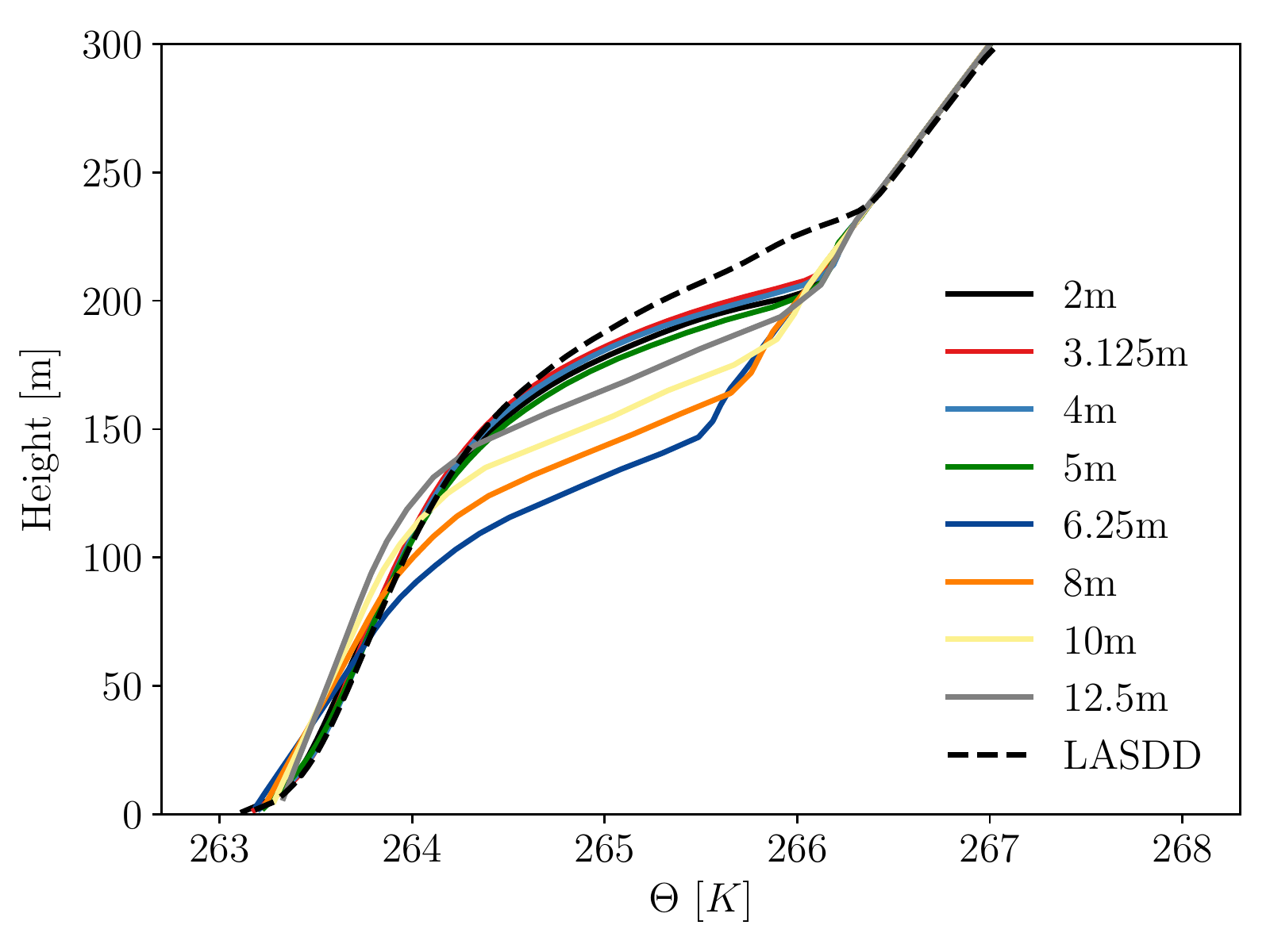}
  \includegraphics[width=0.49\textwidth]{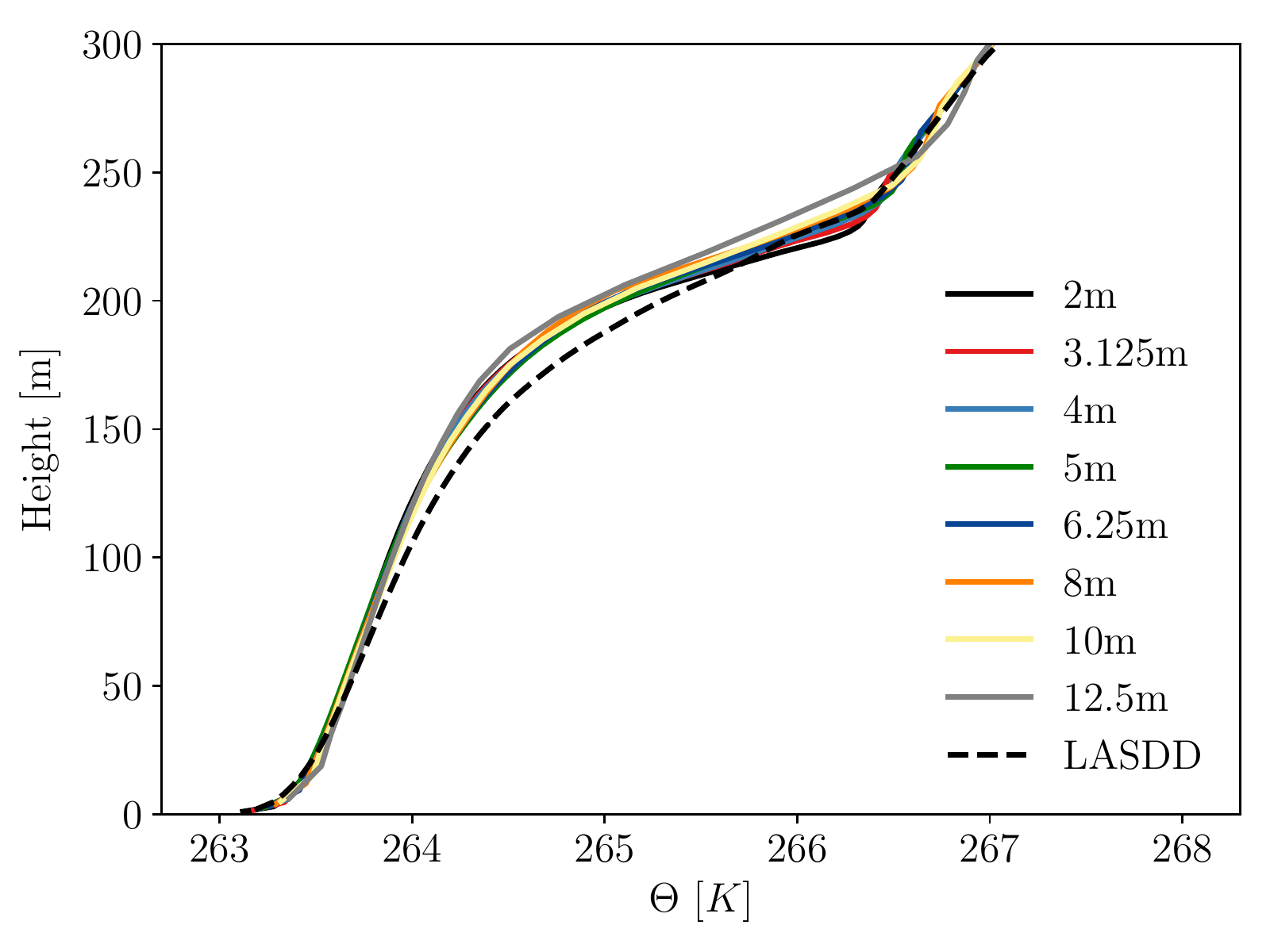}
\caption{Vertical profiles of mean wind speed (top panel), wind direction (middle panel), and potential temperature (bottom panel) from the D80 (left panel) and D80-R (right panel) based simulations using the DALES code. Different colored lines correspond to different grid sizes ($\Delta_g$). Results from the MATLES code are overlaid (dashed black lines) for comparison.}
\label{fig:fields}      
\end{figure*}

The vertical profiles of mean wind speeds are included in the top panel of Fig.~\ref{fig:fields}. The presence of the low-level jet (LLJ) is clearly visible in both the plots. However, the height of the LLJ peak shows strong sensitivity with respect to grid-size in the D80-based runs when $\Delta_g \ge $ 6.25 m. This unphysical behavior is solely due to the inherent limitations of the D80 SGS parameterization. Once $\Delta_g$ becomes smaller than 6.25 m, turbulence is sustained near the surface, and as a result of adequate diffusion, the simulated LLJ peak heights are elevated. With further increase in spatial resolution, the mean wind speed profiles reach convergence; albeit, the height of the DALES-based simulated LLJ peak remains lower and weaker than the one simulated by the MATLES code. The positive impacts of the revised mixing length parameterization in D80-R can be seen in the top-right panel of Fig.~\ref{fig:fields}. Almost all the DALES-based runs (with the exception of the one with $\Delta_g = $ 12.5~m) converge on a single line and also agree strongly with the MATLES-based results. 

Utilizing the results from the GABLS1 intercomparison study, Svensson and Holtslag \cite{svensson09} investigated the wind turning in SBLs in great detail. With certain assumptions, they analytically derived the relationship between the vertically integrated cross-isobaric flow and surface friction. Earlier, we have shown that the runs using the D80-R parameterization lead to $u_*$ series which are insensitive to grid sizes. Thus, it is not surprising that those simulations also lead to wind direction profiles which are in strong agreement with one another. From the middle panel of Fig.~\ref{fig:fields}, it is also clear that the original D80 parameterization performs quite poorly in capturing the turning of wind with height for coarse grid sizes.  

\begin{figure*}[ht!]
\centering
  \includegraphics[width=0.49\textwidth]{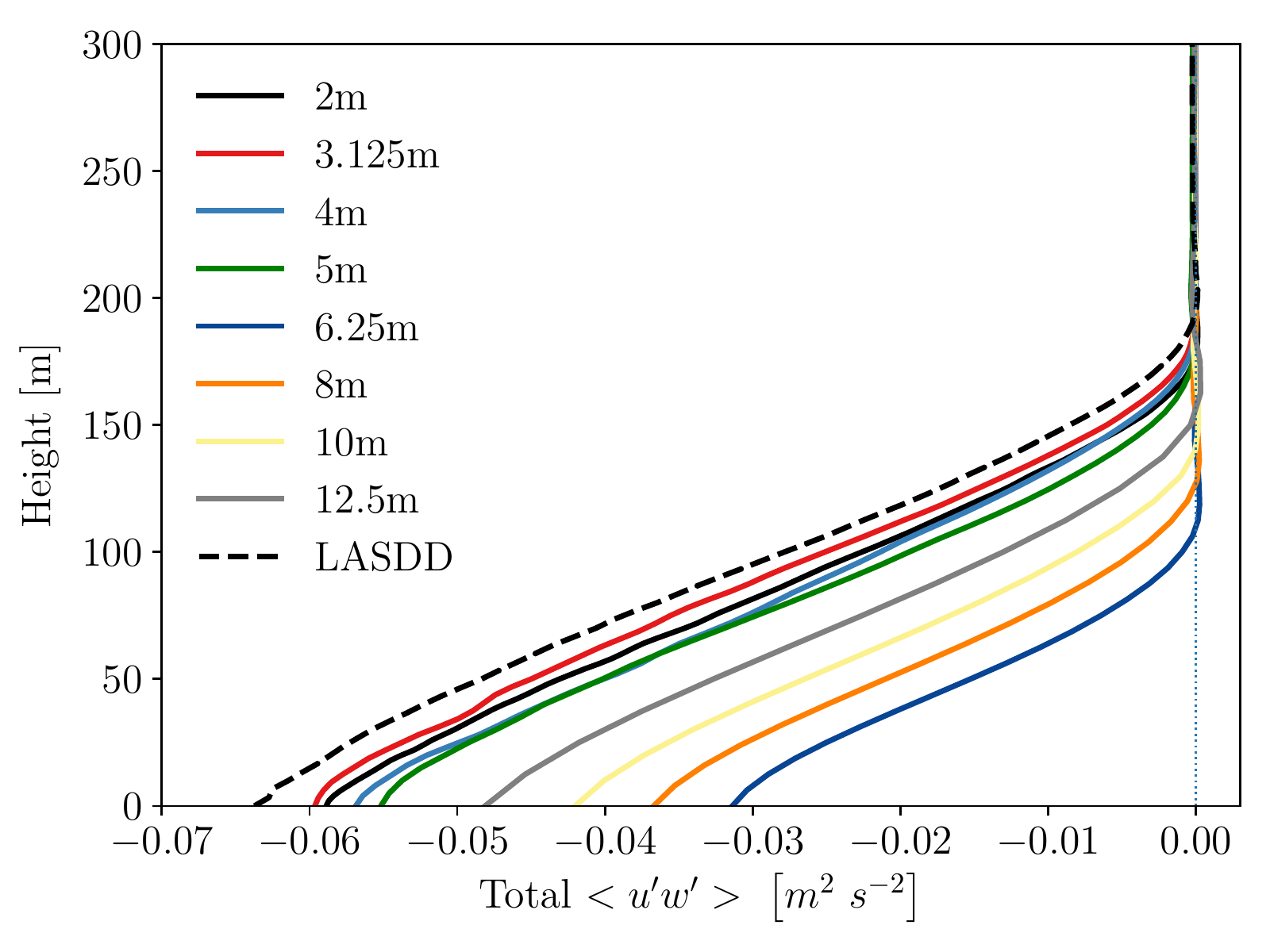}
  \includegraphics[width=0.49\textwidth]{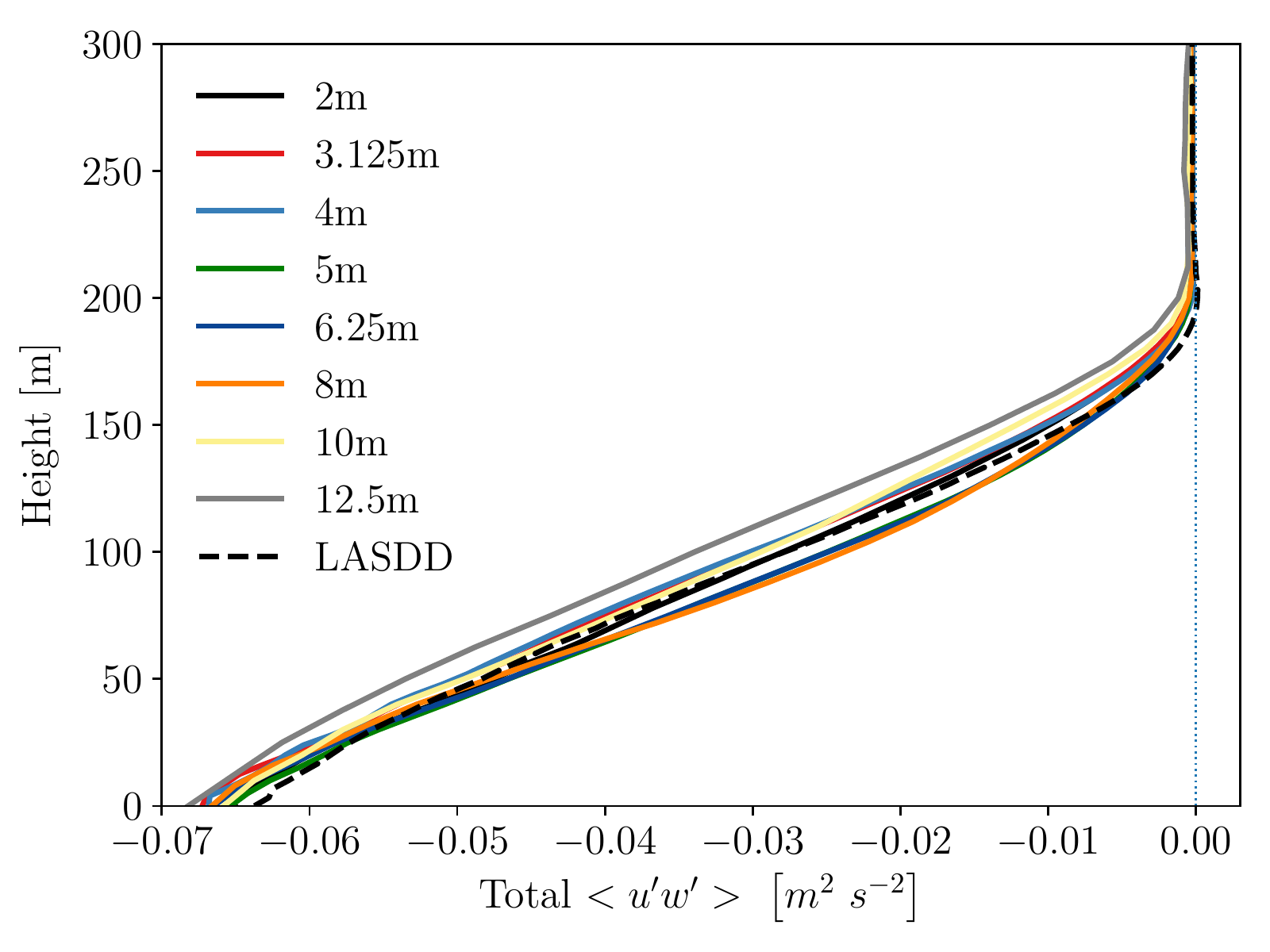}
  \includegraphics[width=0.49\textwidth]{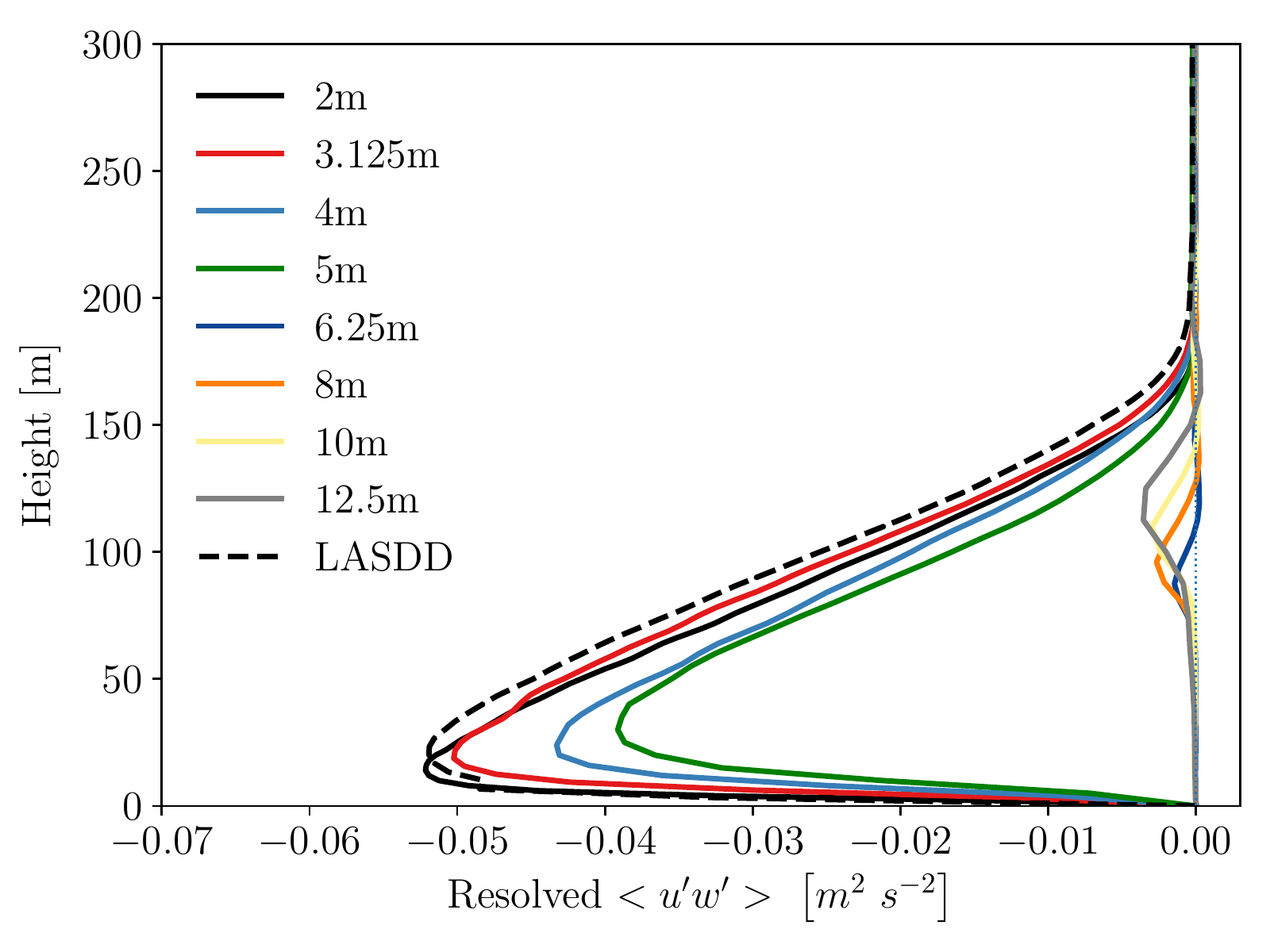}
  \includegraphics[width=0.49\textwidth]{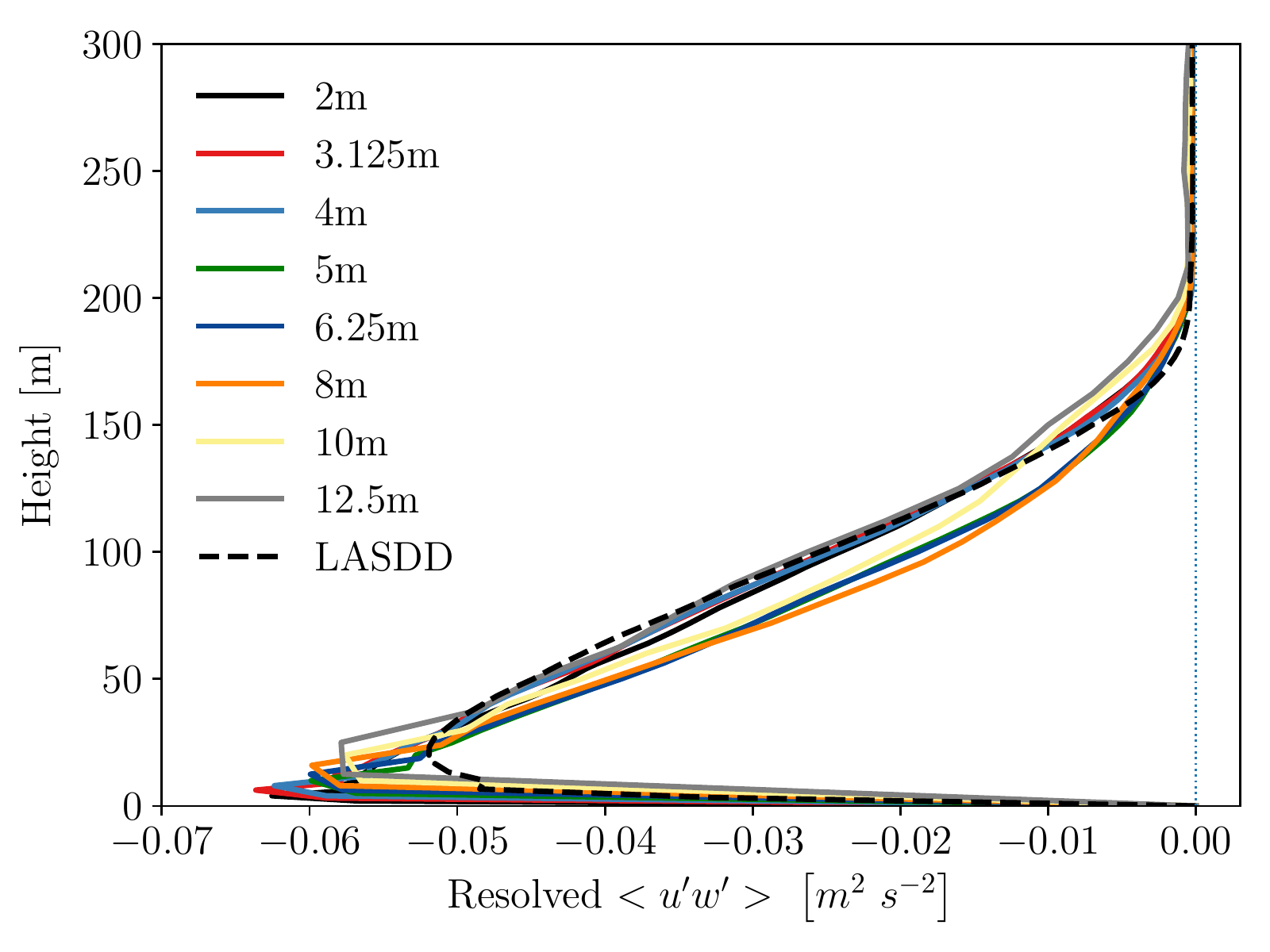}
\caption{Vertical profiles of total (top panel) and resolved (bottom panel) momentum flux ($u$-component) from the D80 (left panel) and D80-R (right panel) based simulations using the DALES code. Different colored lines correspond to different grid sizes ($\Delta_g$). Results from the MATLES code are overlaid (dashed black lines) for comparison.}
\label{fig:Flux1}      
\end{figure*}

The profiles of mean potential temperature are shown in the bottom panel of Fig.~\ref{fig:fields}. All the profiles do show similar convex curvature as reported by earlier studies on GABLS1-LES. However, as with the mean wind speed profiles, the D80-based runs exhibit strong dependence on grid-sizes. Once again, incorporation of the revised SGS mixing length parameterization leads to better convergence. However, in this case, lower values of $Pr_S$ in the middle part of the SBL (see bottom-right panel of Fig.~\ref{fig:lambda}) cause more heat diffusion; not surprisingly, the potential temperature profiles from the D80-R-based runs are more convex than the one simulated by the MATLES code. In contrast, by using a higher value of $Pr_S$, the PALM model generates potential temperature profiles which are indistinguishable from the MATLES-based ones (refer to Fig.~\ref{fig:PALMfields} in Appendix~1). 

In the D80-based runs with finer resolutions, the simulated mean potential temperature profiles also start to converge. However, for these cases, $Pr_S$ values are equal to 0.33 in the lower part of the SBL. The (negative) impact of such high eddy diffusivity values is hard to notice in the mean potential temperature profiles; however, the vertical profiles of variance of potential temperature ($\sigma_\theta^2$) do show the effect clearly (discussed later).  

\begin{figure*}[ht!]
\centering
  \includegraphics[width=0.49\textwidth]{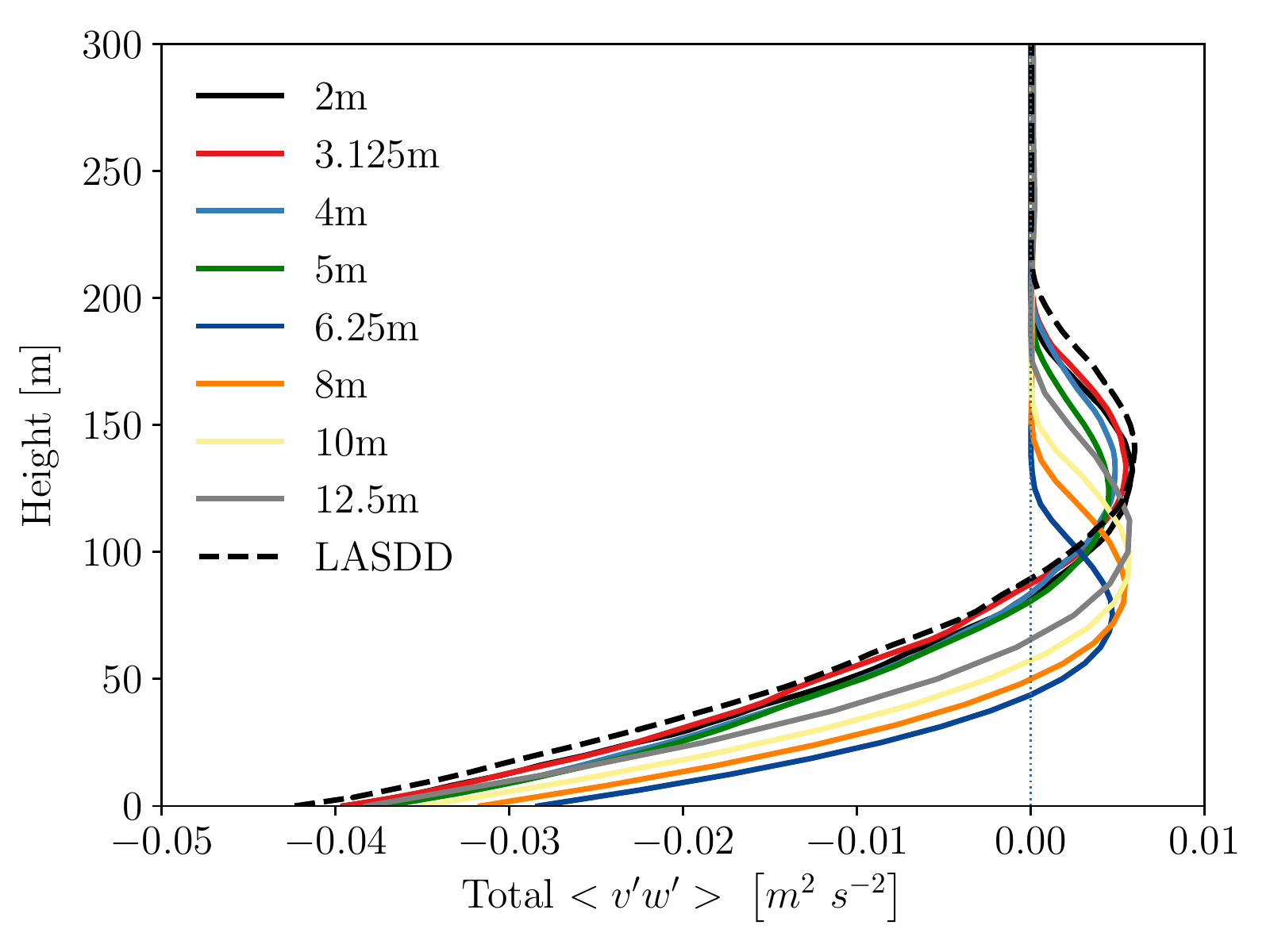}
  \includegraphics[width=0.49\textwidth]{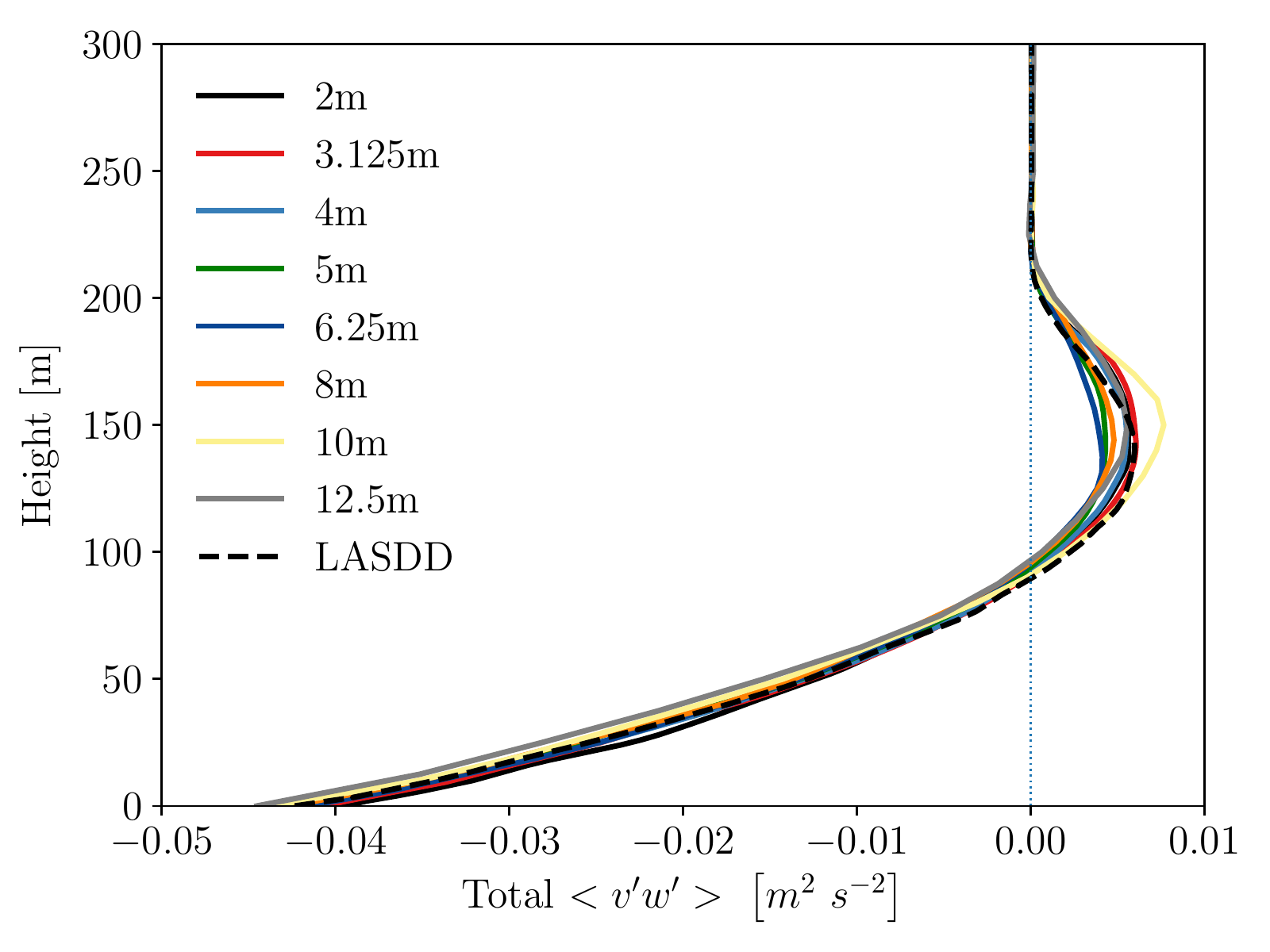}
  \includegraphics[width=0.49\textwidth]{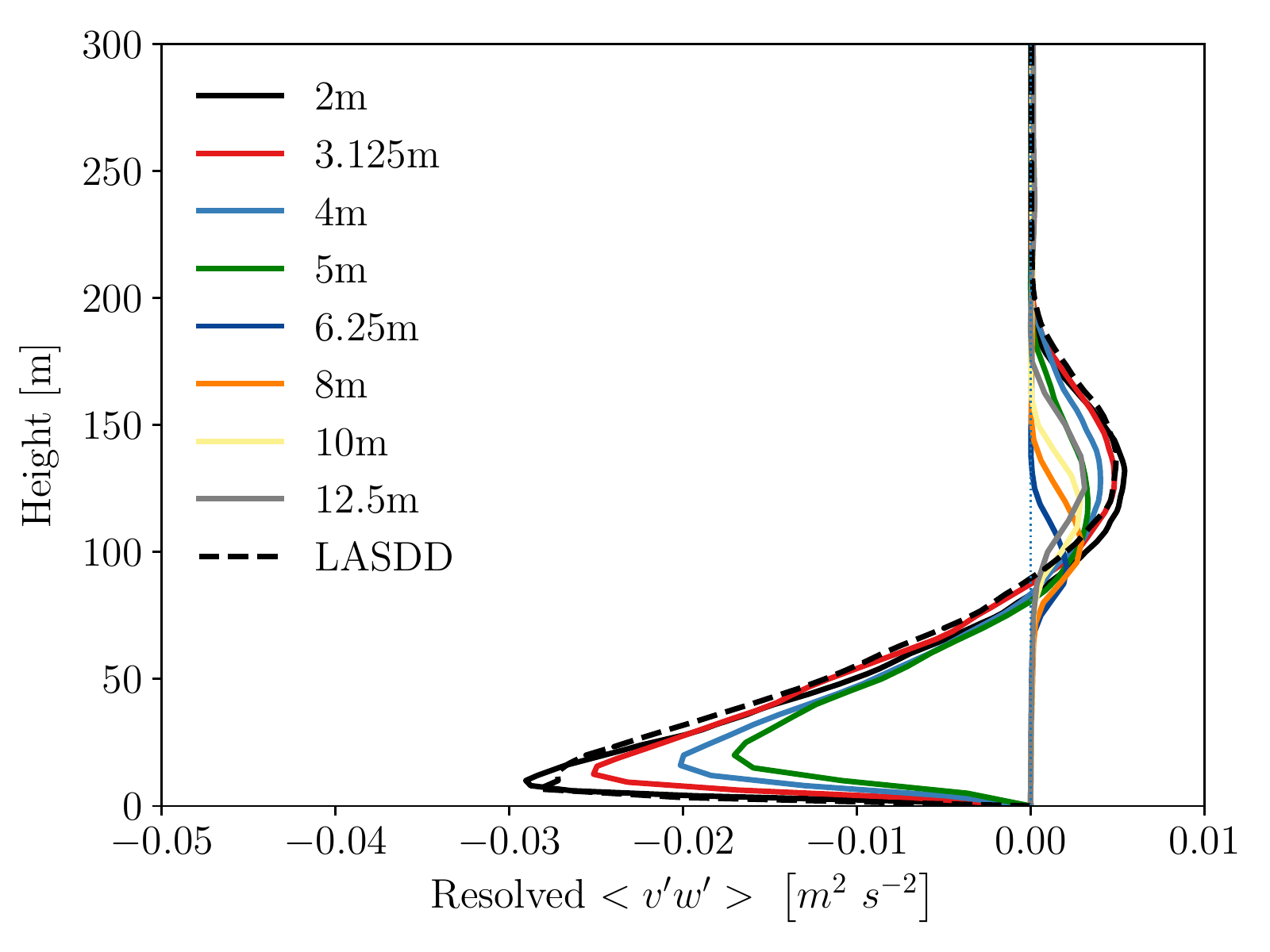}
  \includegraphics[width=0.49\textwidth]{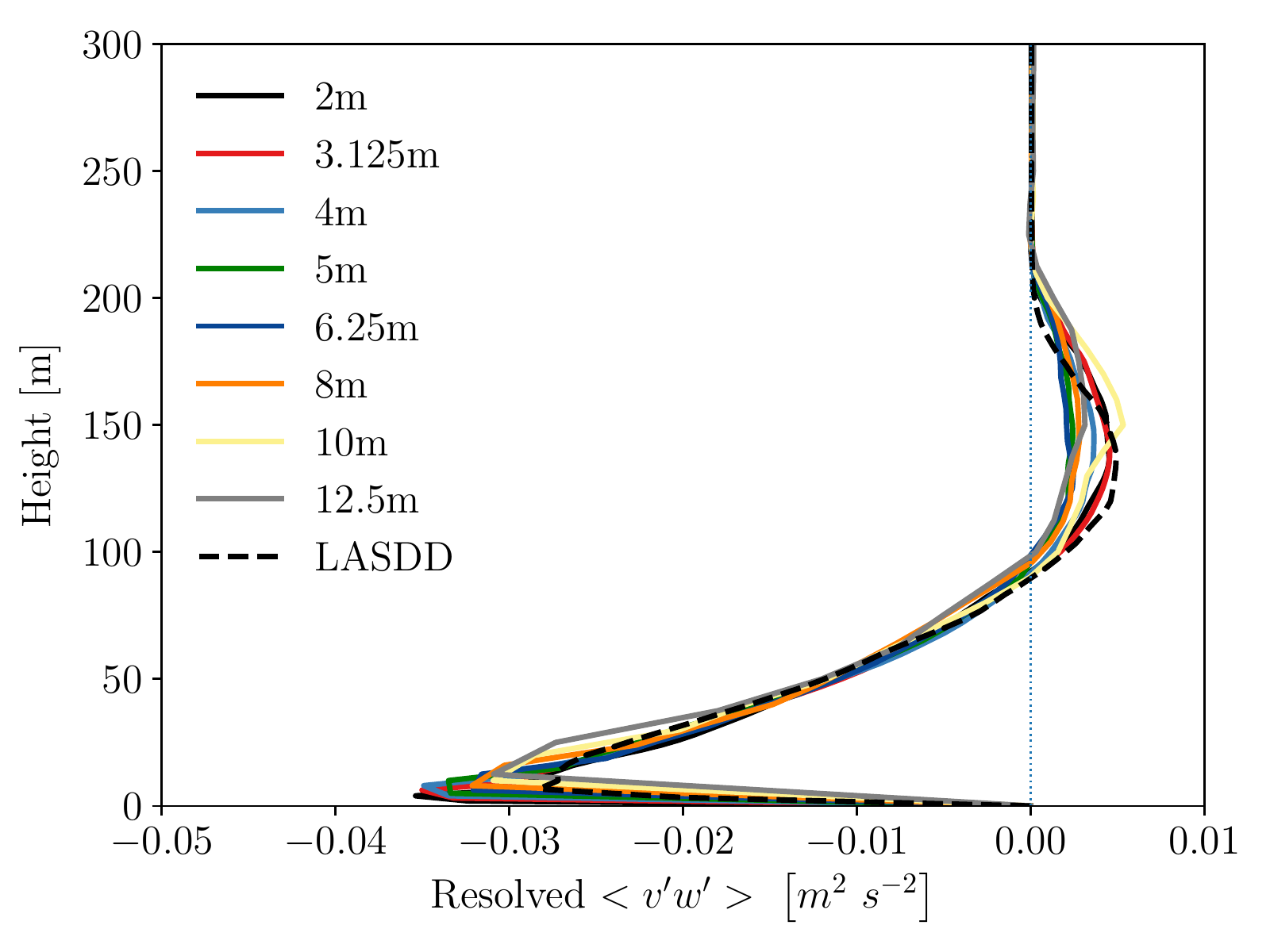}
\caption{Vertical profiles of total (top panel) and resolved (bottom panel) momentum flux ($v$-component) from the D80 (left panel) and D80-R (right panel) based simulations using the DALES code. Different colored lines correspond to different grid sizes ($\Delta_g$). Results from the MATLES code are overlaid (dashed black lines) for comparison.}
\label{fig:Flux2}      
\end{figure*}

The $u$- and $v$-components of momentum flux profiles are shown in Figs.~\ref{fig:Flux1} and ~\ref{fig:Flux2}, respectively. Several observations can be made from these plots. First of all, the resolved momentum fluxes are virtually non-existent for the D80-based runs with $\Delta_g \ge 6.25$ m. This result is not surprising given the other statistics shown in the previous plots. The total fluxes are strongly grid-size dependent for the D80-based runs; however, they are not in the D80-R-based ones. Also, the fluxes are well resolved in the revised case; near the surface, the resolved fluxes increase with increasing resolution as would be expected. The revised results are more-or-less in-line with the fluxes generated by the MATLES code. 

\begin{figure*}[ht!]
\centering
  \includegraphics[width=0.49\textwidth]{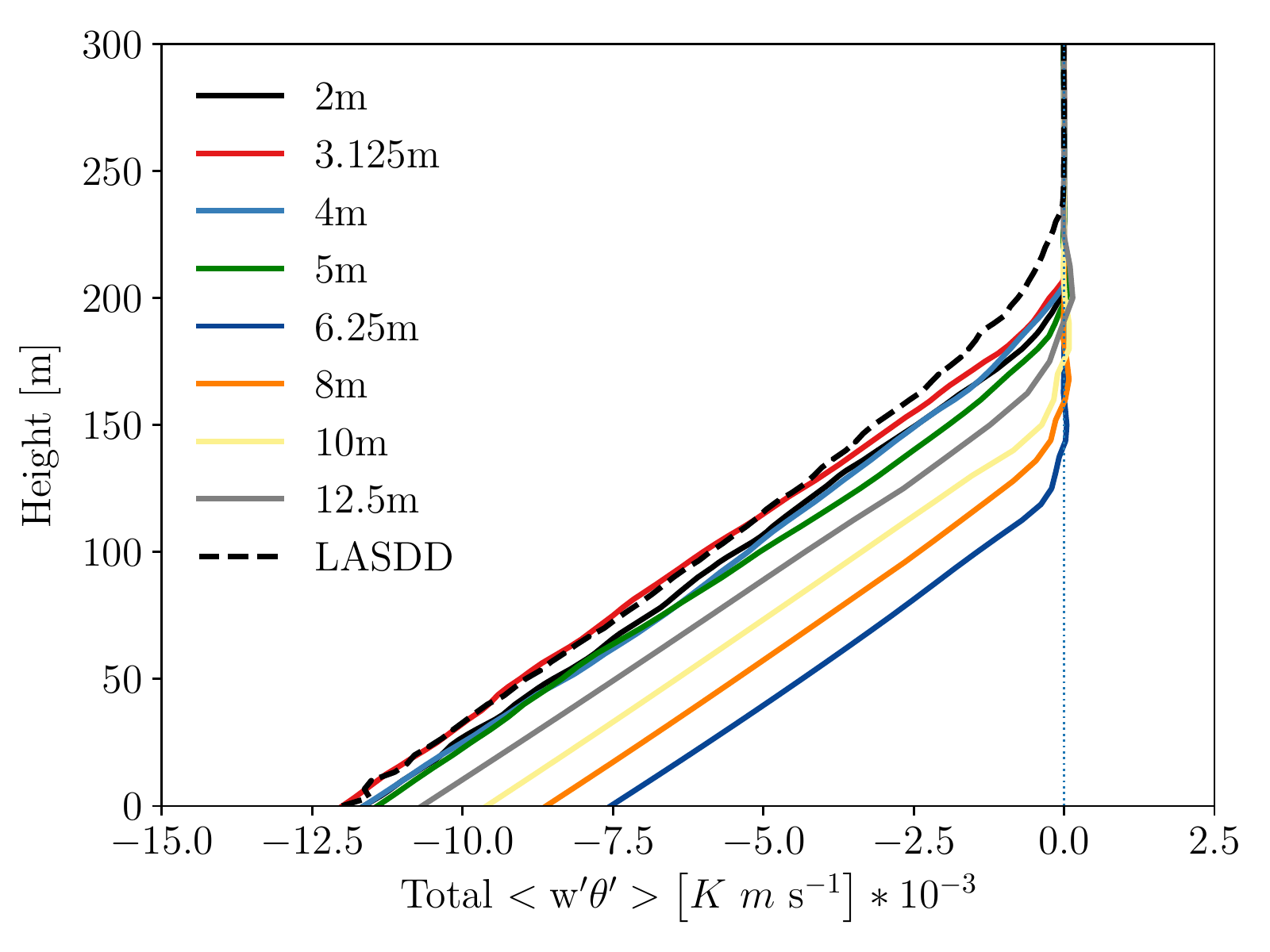}
  \includegraphics[width=0.49\textwidth]{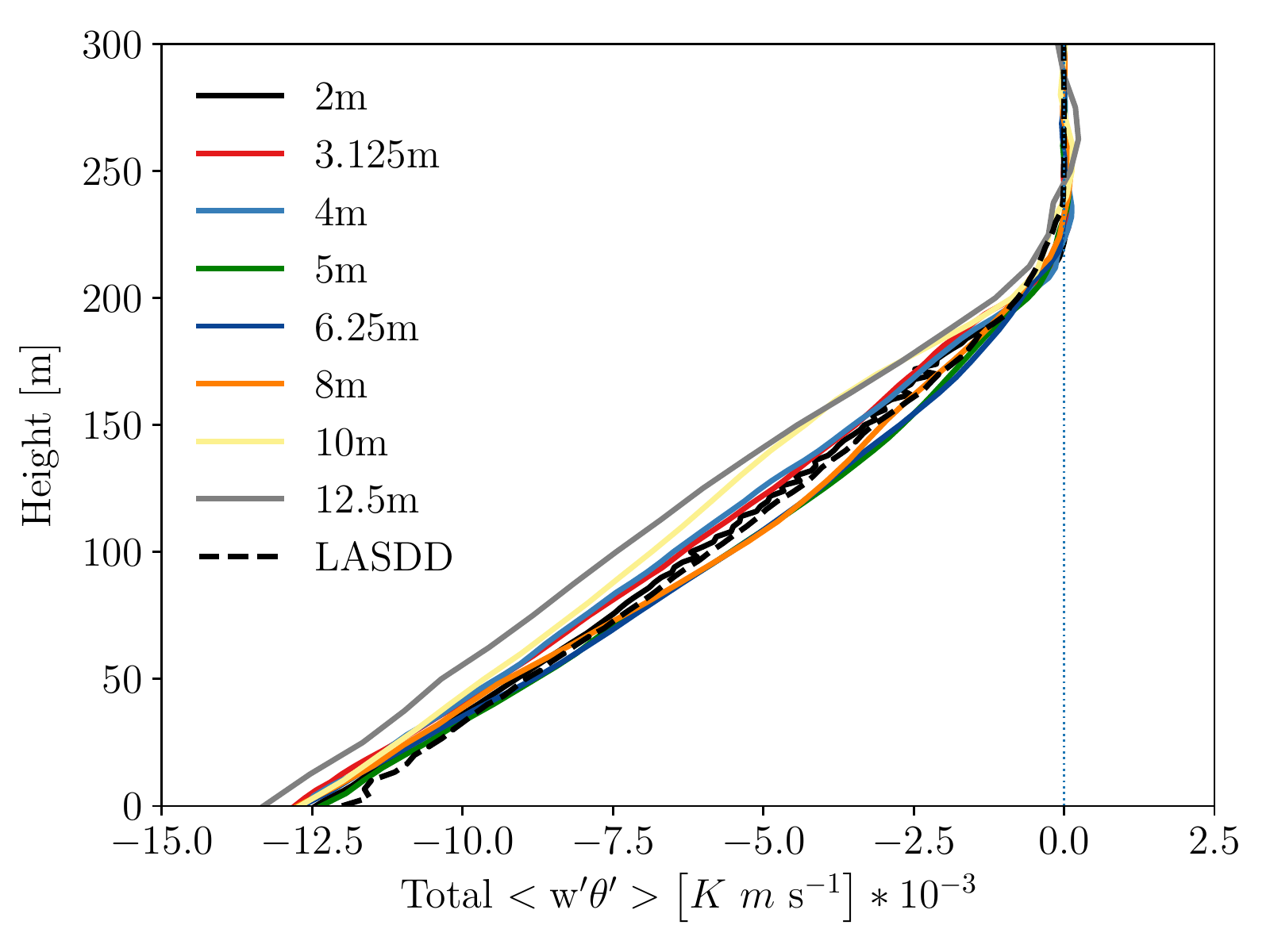}
  \includegraphics[width=0.49\textwidth]{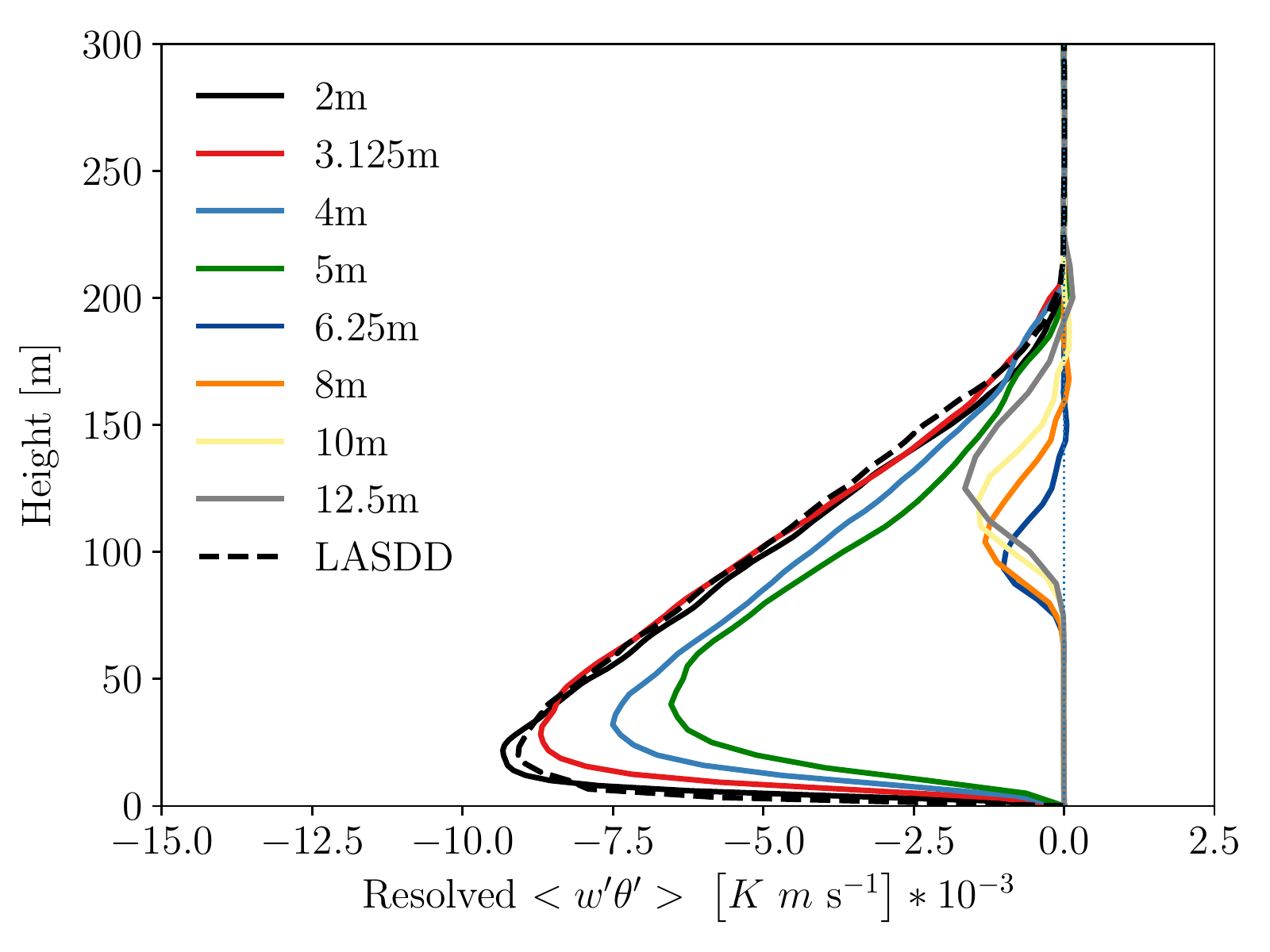}
  \includegraphics[width=0.49\textwidth]{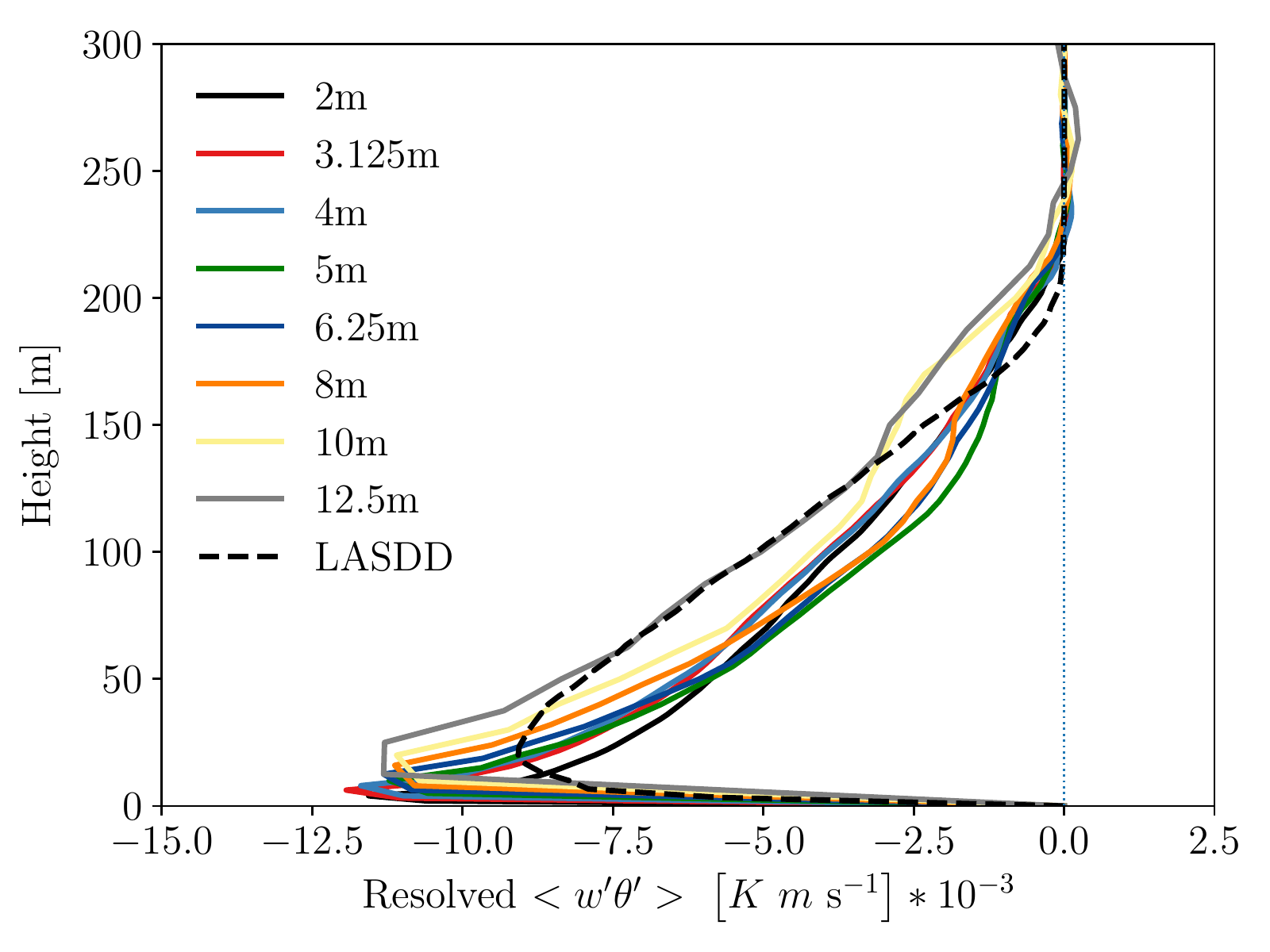}
\caption{Vertical profiles of total (top panel) and resolved (bottom panel) sensible heat flux from the D80 (left panel) and D80-R (right panel) based simulations using the DALES code. Different colored lines correspond to different grid sizes ($\Delta_g$). Results from the MATLES code are overlaid (dashed black lines) for comparison.}
\label{fig:HeatFlux}      
\end{figure*}

As depicted in Fig.~\ref{fig:HeatFlux}, the overall behavior of the total and resolved sensible heat flux profiles are qualitatively similar to those of the momentum flux profiles. In the case of the D80-R-based runs, the grid-convergence is slightly less than satisfactory for the total sensible heat flux profiles. Especially, the simulation with $\Delta_g = 12.5$ m consistently overestimates the magnitude of heat flux across the SBL.  

\begin{figure*}[ht!]
\centering
  \includegraphics[width=0.49\textwidth]{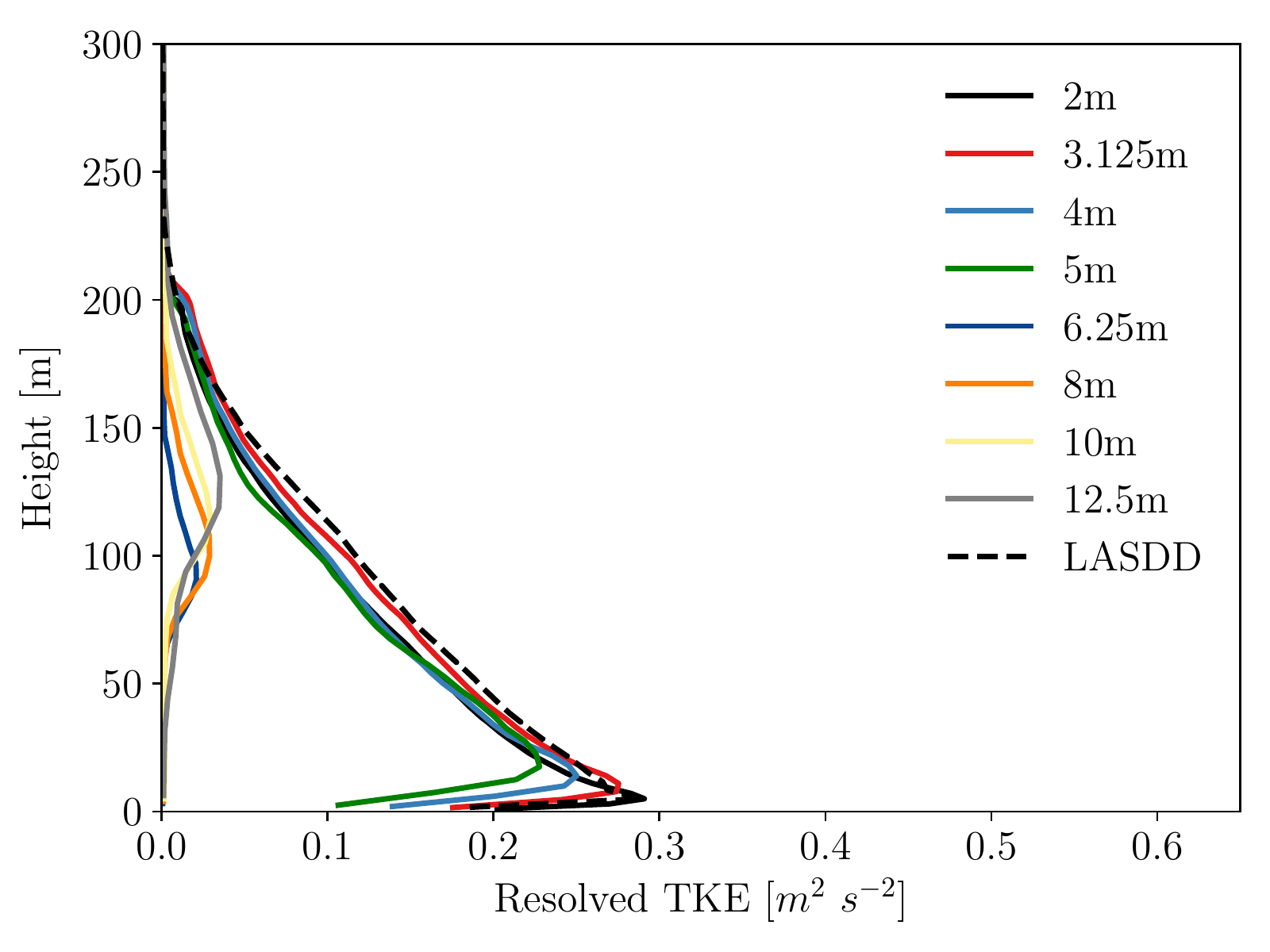}
  \includegraphics[width=0.49\textwidth]{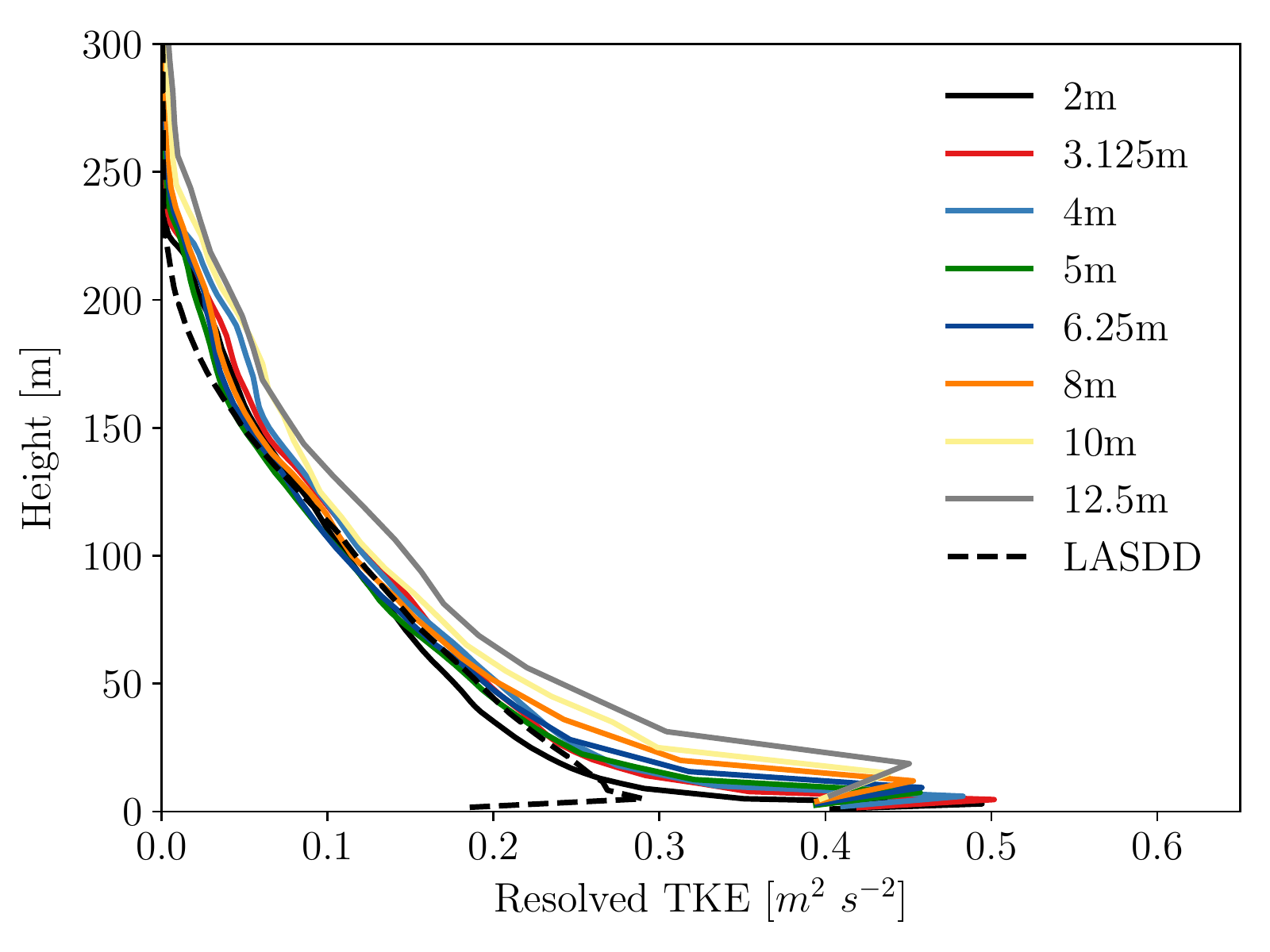}
  \includegraphics[width=0.49\textwidth]{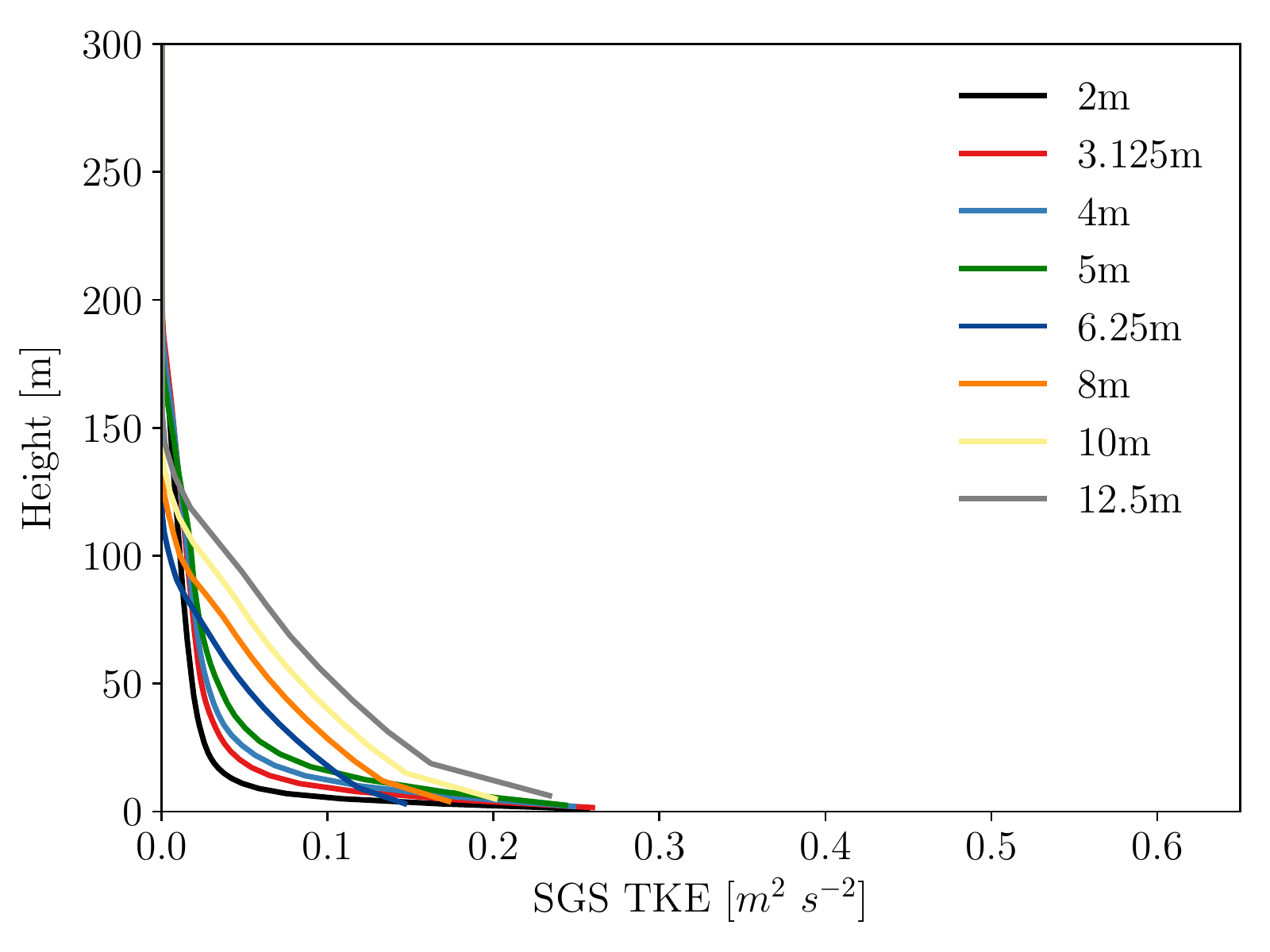}
  \includegraphics[width=0.49\textwidth]{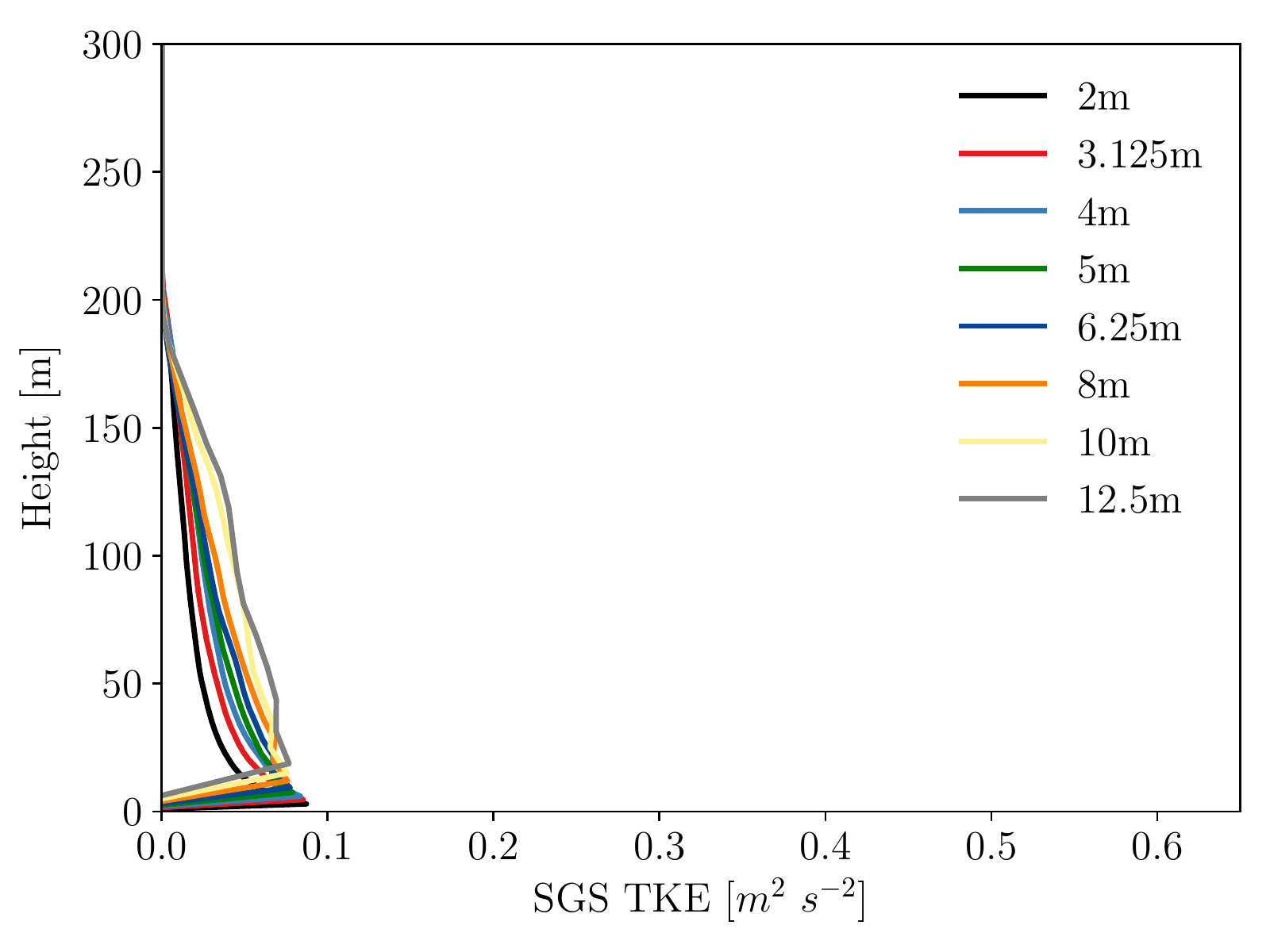}
\caption{Vertical profiles of resolved (top panel) and subgrid-scale (bottom panel) turbulent kinetic energy from the D80 (left panel) and D80-R (right panel) based simulations using the DALES code. Different colored lines correspond to different grid sizes ($\Delta_g$). Results from the MATLES code are overlaid (dashed black lines) for comparison in the resolved TKE plots.}
\label{fig:TKE}      
\end{figure*}

The resolved and SGS TKE plots are shown in Fig.~\ref{fig:TKE}. The MATLES code does not solve for the SGS TKE equation; thus, only the resolved TKE values are overlaid for comparison. As with the momentum and sensible heat flux profiles, the resolved TKE is non-existent in the bottom-half of the SBL for the D80-based simulations using $\Delta_g \ge$ 6.25 m. In the lower part of the SBL, the resolved TKE values are larger in the D80-R-based runs in comparison to the D80-based ones. In that region, with increasing resolution, resolved TKE values increase as would be physically expected. Most importantly, the SGS TKE values are much smaller in magnitude than their resolved counter-part (especially, in the lower part of the SBL). In other words, the flow is highly resolved for all the simulations involving the D80-R parameterization.  

\begin{figure*}[ht!]
\centering
  \includegraphics[width=0.49\textwidth]{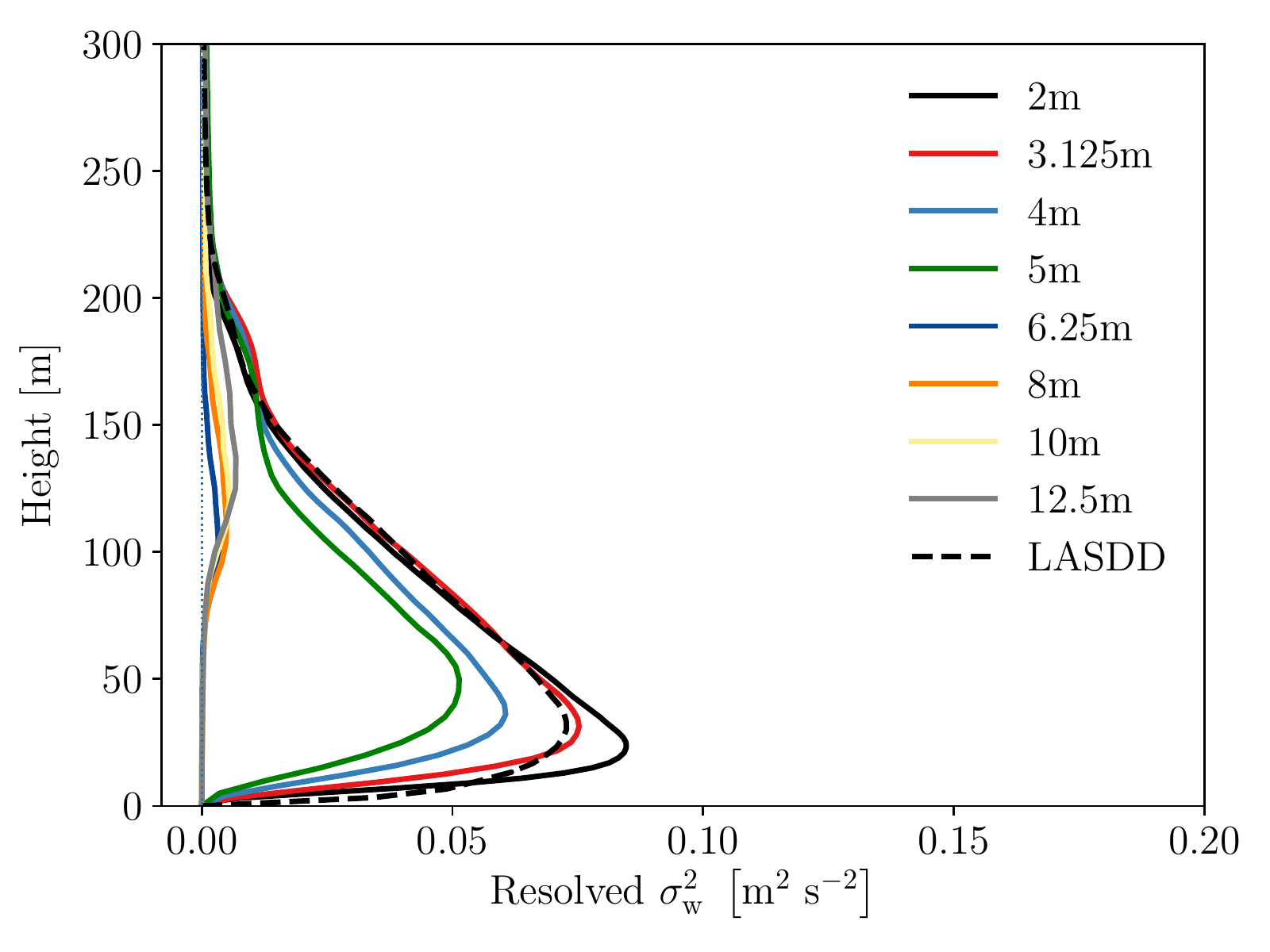}
  \includegraphics[width=0.49\textwidth]{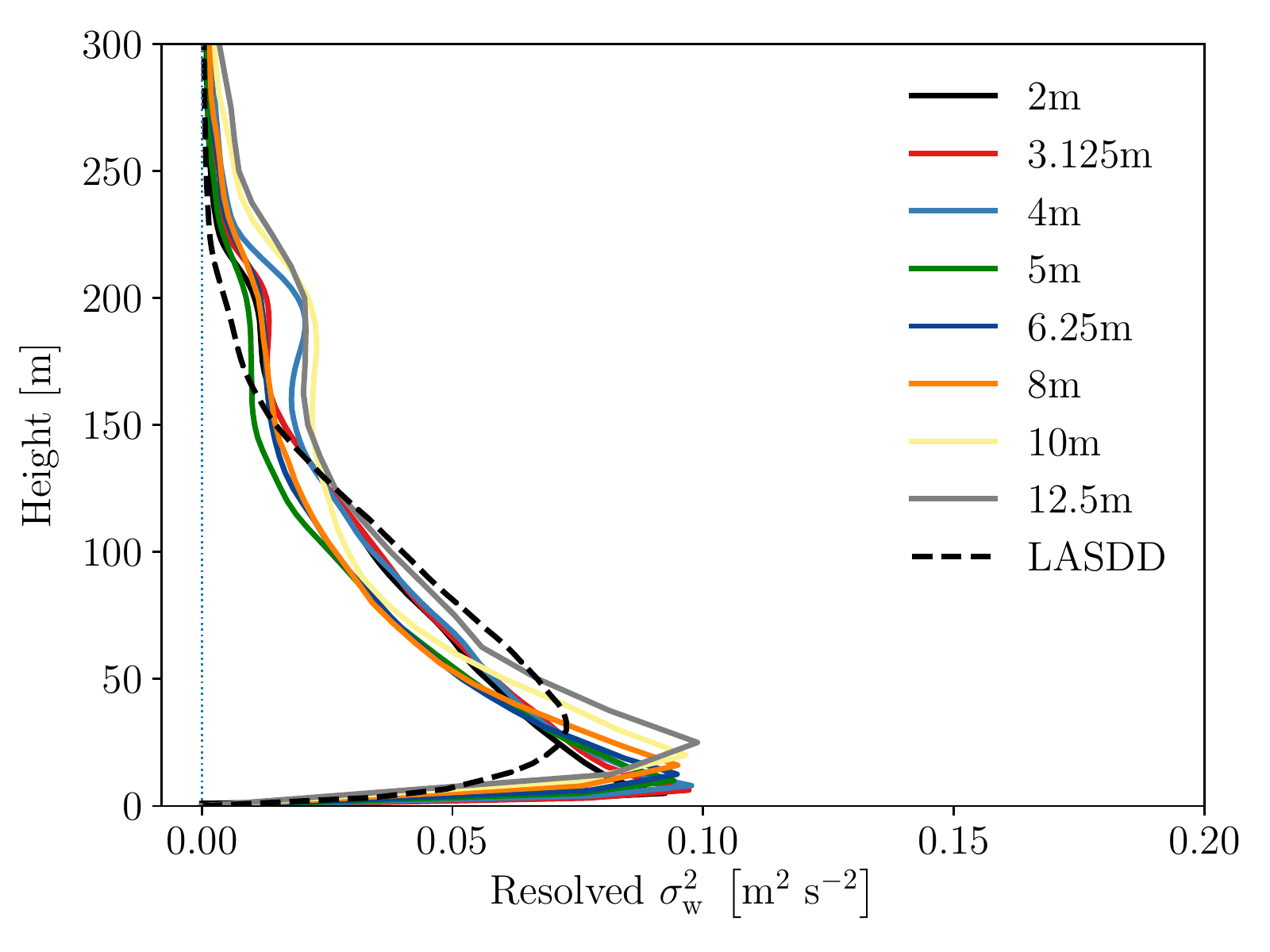}
  \includegraphics[width=0.49\textwidth]{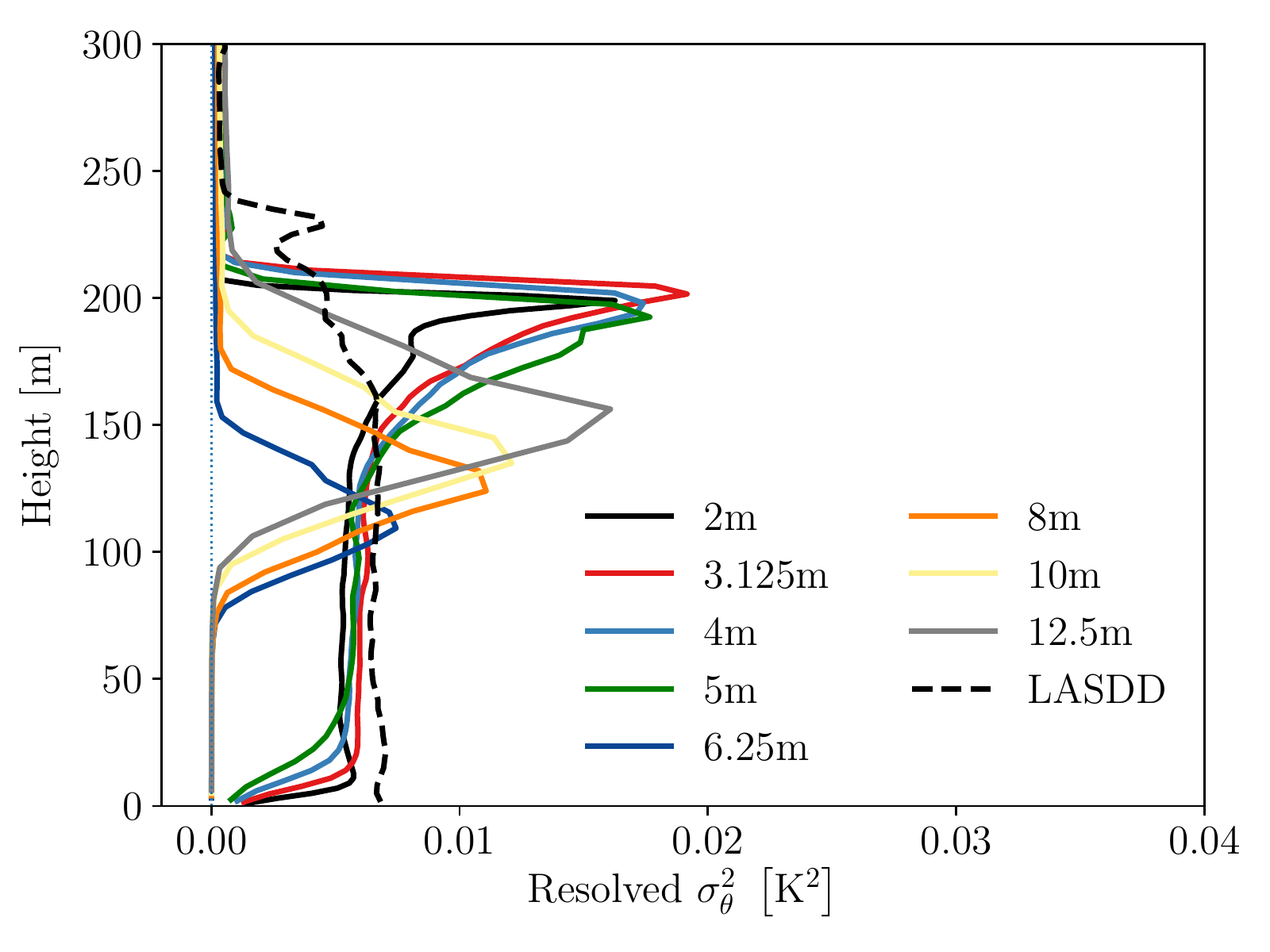}
  \includegraphics[width=0.49\textwidth]{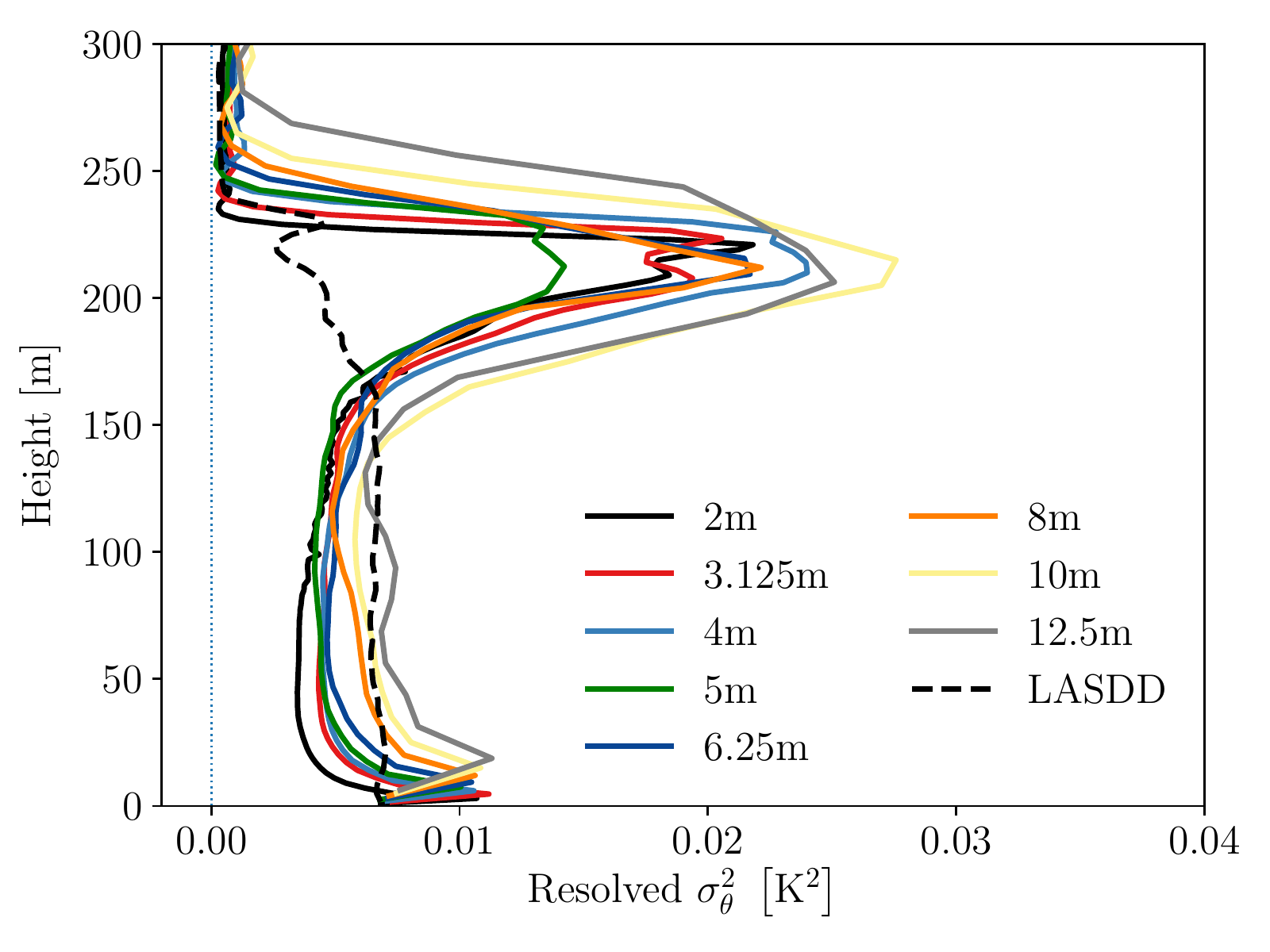}
\caption{Vertical profiles of resolved variances of vertical velocity (top panel) and potential temperature (bottom panel) from the D80 (left panel) and D80-R (right panel) based simulations using the DALES code. Different colored lines correspond to different grid sizes ($\Delta_g$). Results from the MATLES code are overlaid (dashed black lines) for comparison.}
\label{fig:Variance}      
\end{figure*}

Lastly, the resolved variances of vertical velocity and potential temperature are plotted in Fig.~\ref{fig:Variance}. The trends of the resolved $\sigma_w^2$ profiles are similar to the resolved TKE plots. The resolved $\sigma_\theta^2$ profiles from the D80-based runs are quite interesting. As discussed earlier, the coarse-resolution runs (i.e., $\Delta_g \ge$ 6.25 m) show the quasi-laminarization problem up to 75~m or so. Interestingly, for higher resolution runs with the original mixing length parameterization, the resolved $\sigma_\theta^2$ values decrease significantly near the surface. The opposite trend is seen in the D80-R-based simulations. We believe this decrease in the D80-based runs is due to the low value of $Pr_S$ (= 0.33) in the bottom part of the SBL. Similar decrease in $Pr_S$ in the core of the SBL also causes resolved $\sigma_\theta^2$ to significantly decrease in the D80-R-based cases.       

Even though most vertical profiles from the DALES code look physically realistic and agree quite well with the corresponding results from the MATLES code, some discrepancies are noticeable in the case of resolved variances and fluxes. It is possible that the LASDD SGS model is slightly over-dissipative near the surface; the alternative scenario is that the proposed D80-R SGS model is slightly under-dissipative near the surface. Typically, in pseudo-spectral codes, spectral analysis can shed light into such undesirable behaviors \citep[see][for some examples]{anderson07}. Spurious piling up of energy near the high-wavenumber part of the spectrum is a telltale sign of under-dissipation. Since DALES and PALM are finite-difference codes, detection of such a subtle feature in the energy spectrum is a challenging task due to strong numerical dissipation. As such, we will be implementing the D80-R SGS model in the MATLES code for in-depth spectral analysis. 

\section{Discussion}

In addition to Deardorff's SGS model, the Smagorinsky-Lilly SGS model \citep{smagorinsky63,Lilly66a,Lilly66b} is also quite popular in the boundary layer community. 
In this SGS model, the effective mixing length is $C_S \Delta_f$; where, $C_S$ is the so-called Smagorinsky coefficient. This coefficient is adjusted empirically \citep[e.g.,][]{mason90,brown94,kleissl04} or dynamically \citep[e.g.,][]{porte00,bouzeid05,basu06} to account for shear, stratification, and grid resolution. 

In contrast, the effective mixing length is $c_m \lambda$ in D80 SGS model. Most of the LES codes assume a constant value for $c_m$ and expects $\lambda$ to account for shear, stratification, and grid resolution. As elaborated in Sect.~2, the original mixing length formulation does not account for shear or stratification properly. The parameterization by Brost and Wyngaard \cite{brost78} provides a major improvement as both the shear and stratification effects are now explicitly included. The effects of grid size is somewhat indirect. The performance of the D80-R SGS model may be improved if $c_m$ is assumed to be $\Delta_g$-dependent and estimated via a dynamic approach \citep[e.g.,][]{ghosal95,krajnovic02}. Our future work will be in that direction. 

We will also need a better understanding of the SGS Prandtl number ($Pr_S$) and its relationship to the turbulent Prandtl number ($Pr_T$). Several field and laboratory experiments have reported that $Pr_T$ should be on the order of one for stably stratified conditions \citep[see][and the references therein]{sukoriansky06,li19}. However, within the SBL (excluding surface and inversion layers), the dynamic SGS models typically estimate $Pr_S$ and $Pr_T$ values to be around 0.5--0.6 and 0.7, respectively \citep{basu06,stoll08}. We agree with Li (2019) \cite{li19} that $Pr_S$ is less important than $Pr_T$ as the large-scale fluxes are resolved in LES. 
Nonetheless, we do believe the current $Pr_S$ parameterizations in the D80 and D80-R SGS models are not acceptable and definitely need amendments. The formulation proposed by Gibbs and Fedorovich \cite{gibbs16} is a good starting point and should be rigorously tested in future studies. 

\section{Concluding Remarks}

In this study, we have demonstrated that Deardorff's mixing length parameterization is not suitable for SBL simulations. Instead an older scheme, proposed by Brost and Wyngaard for RANS, gives promising results. Even though we have made some progress in the arena of LES of SBLs, many open questions still remain: 
\begin{itemize}
    \item Would this new mixing length parameterization work for very stable boundary layers? Would it allow us to simulate turbulent bursting events? 
    \item Is the proposed parameterization under-dissipative near the surface? How can we (dis)prove this behavior? 
    \item Is a buoyancy-based length scale really appropriate for weakly or moderately stable boundary layer? Or, should we opt for a shear-based length scale \citep{hunt88,hunt89,basu20}?
    \item How sensitive are the simulated results with respect to the chosen SGS coefficients (i.e., $c_n$, $c_m$, $c_\varepsilon$)? Should they be dynamically determined instead of being prescribed? 
    \item How should we parameterize the SGS Prandtl number? Should it be a function of point-wise gradient Richardson number? 
\end{itemize}

A few years ago, Basu and Lacser\cite{basu17} cautioned against violating MOST in LES runs with very high resolutions. To overcome this issue, Maronga et al \cite{maronga19} proposed certain innovative strategies; however, the results were somewhat inconclusive (possibly) due to the usage of the D80 mixing length parameterization in all their simulations. In light of the findings from the present work, we will revisit the MOST issue in very high resolution LES in conjunction with the D80-R SGS model. In addition, we will investigate the interaction of the D80-R parameterization with a coupled land-surface model.  

We further recommend the SBL-LES community to revisit some of the past LES intercomparison studies organized under GABLS with revised or newly proposed SGS models. We speculate that some of the conclusions from the previous studies will no longer be valid. 

\section*{Data and Code Availability}
The DALES code is available from: \url{https://github.com/dalesteam/dales}. The PALM model system is available from: \url{https://palm.muk.uni-hannover.de/trac}. The MATLES code is available from S.~Basu upon request. Upon acceptance of the manuscript, all the analysis codes and processed data will be made publicly available via \url{zenodo.org}.

\begin{acknowledgements}
The authors are grateful to Patrick Hawbecker, Bert Holtslag, Branko Kosovi\'{c}, Arnold Moene, Siegfried Raasch, and Chiel van Heerwaarden for useful discussions. 
\end{acknowledgements}

\section*{Appendix 1: SGS Models with Implicit $\Delta_f$-dependence}

Traditional eddy-viscosity models, commonly used by the boundary-layer meteorology community, explicitly involves the filter-size ($\Delta_f$) as the mixing length scale. For example, in the case of Smagorinsky-Lilly SGS model, one uses:
\begin{subequations}
\begin{equation}
    K_m = \left(C_S \Delta_f\right)^2 |\tilde{S}|,
\end{equation}
\begin{equation}
    \lambda = \left(C_S \Delta_f\right),
\end{equation}
\end{subequations}
where, $|\tilde{S}|$ is the magnitude of the resolved strain rate tensor. $C_S$ is known as the Smagorinsky coefficient. 

In contrast to these popular SGS models, the proposed D80-R SGS model depends on $\Delta_f$ in an implicit manner. Similar implicit $\Delta_f$-dependence can be found in the engineering LES literature. Recently, \cite{piomelli11} proposed an eddy-viscosity model: 
\begin{subequations}
\begin{equation}
    K_m \propto \frac{\overline{e}_r}{\omega_r^2}|\tilde{S}|,
\end{equation}
\begin{equation}
    \lambda \propto \frac{\overline{e}_r^{1/2}}{\omega_r},
    \label{Omega}
\end{equation}
\end{subequations}
where, $\overline{e}_r$ and $\omega_r$ denote resolved TKE and resolved magnitude of vorticity, respectively. In a more recent work, \cite{piomelli15} formulated the integral length scale approximation (ILSA) model: 
\begin{subequations}
\begin{equation}
    K_m \propto \frac{\overline{e}_r^3}{\overline{\varepsilon}_T^2}|\tilde{S}|,
\end{equation}
\begin{equation}
    \lambda \propto \frac{\overline{e}_r^{3/2}}{\overline{\varepsilon}_T},
    \label{ILSA}
\end{equation}
\end{subequations}
where, $\overline{\varepsilon}_T$ represent the total energy dissipation rate. A dynamic version of the ILSA model was developed by \cite{rouhi2016} and utilized for simulations in complex geometries by \cite{lehmkuhl2019}. The advantages of these flow-physics-dependent length scales (i.e., Eq.~\ref{Omega} and Eq.~\ref{ILSA}) over $\Delta_f$ as a mixing length scale have been extensively discussed by \cite{piomelli14} and \cite{geurts2019} and will not be repeated here for brevity. 

In addition to the aforementioned eddy-viscosity type SGS models, certain non-eddy-viscosity type SGS models also do not include explicit dependency on $\Delta_f$. A case in point is the similarity model and its variants \citep{bardina80,liu94}.

\section*{Appendix 2: Results from the PALM Model System}

The simulated results from the PALM model are documented in this appendix. The trends are very similar to those reported in Sect.~4. Thus, we do not discuss most of these figures for brevity. However, we would like to point out a noticeable feature in the D80-based runs. Even though most of the simulated profiles converge for 2 m~$\le \Delta \le$~5 m, the results from $\Delta = 1$~m run stands out. We believe that $\lambda$ is quite low for this particular case; even though it is small enough to sustain turbulence near the surface, it is not large enough to promote diffusion. The D80-R run with $\Delta = 1$~m does not portray such an unusual behavior.   

\begin{figure*}[ht]
\centering
  \includegraphics[width=0.49\textwidth]{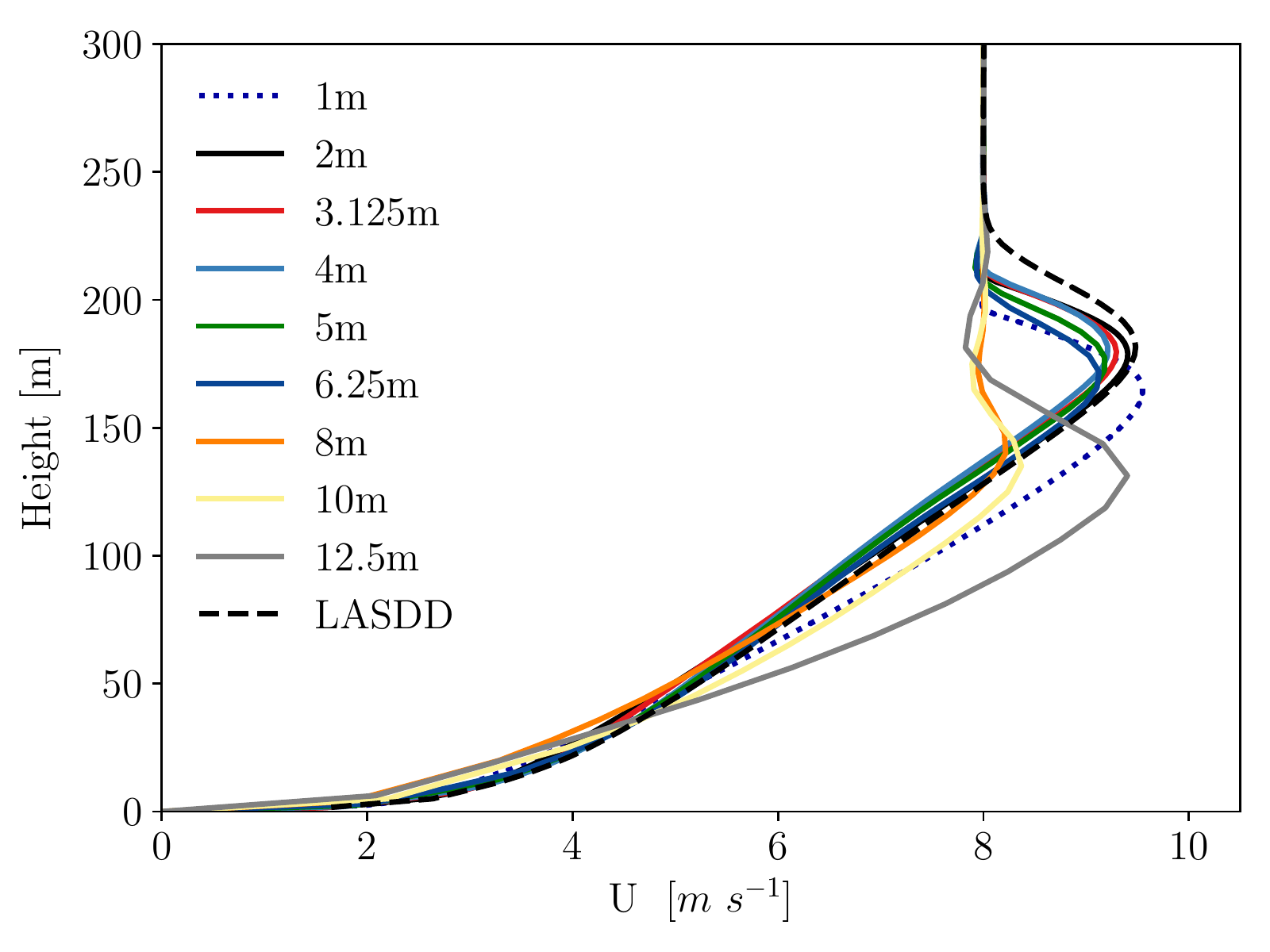}
  \includegraphics[width=0.49\textwidth]{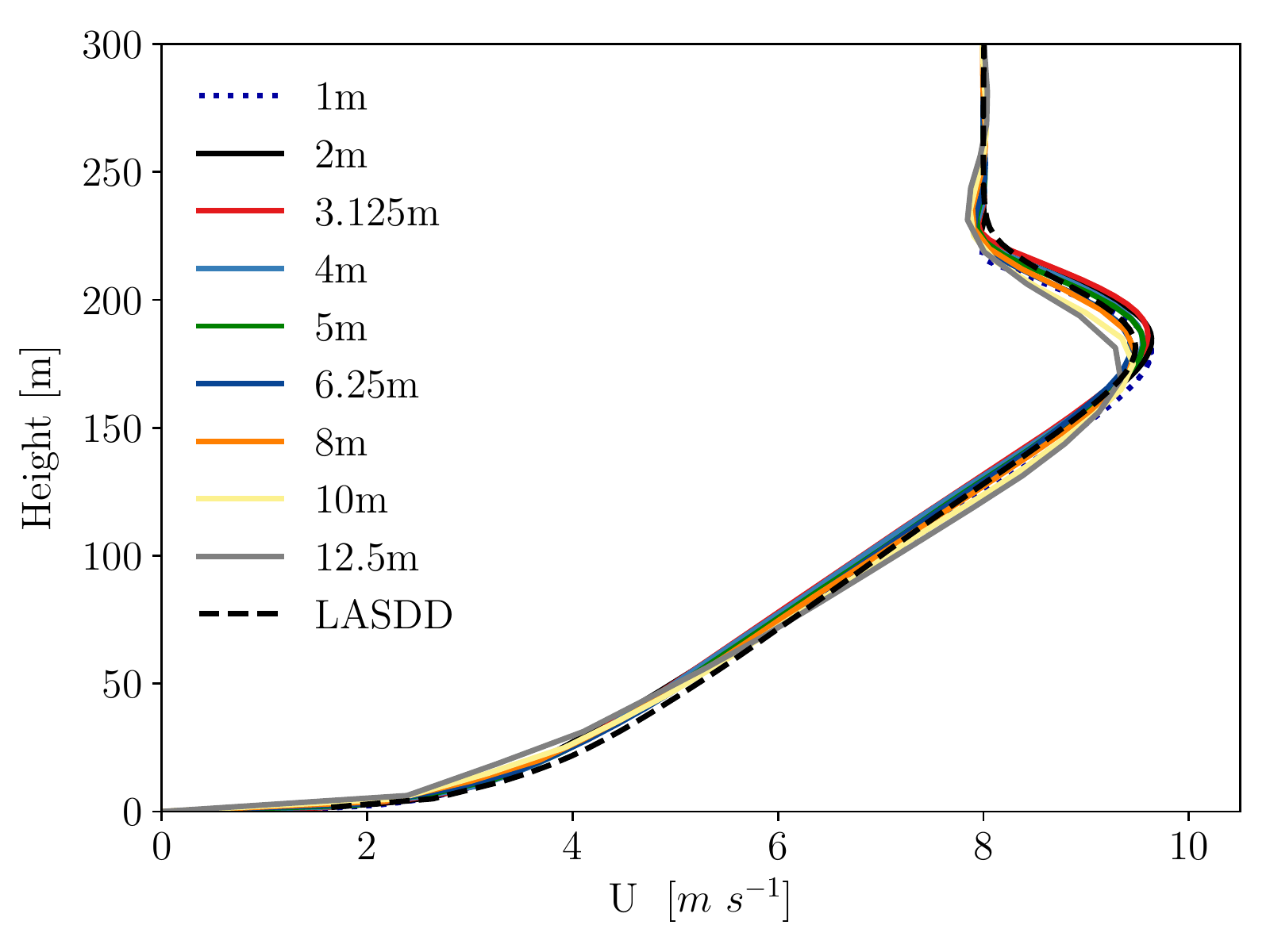}\\
  \includegraphics[width=0.49\textwidth]{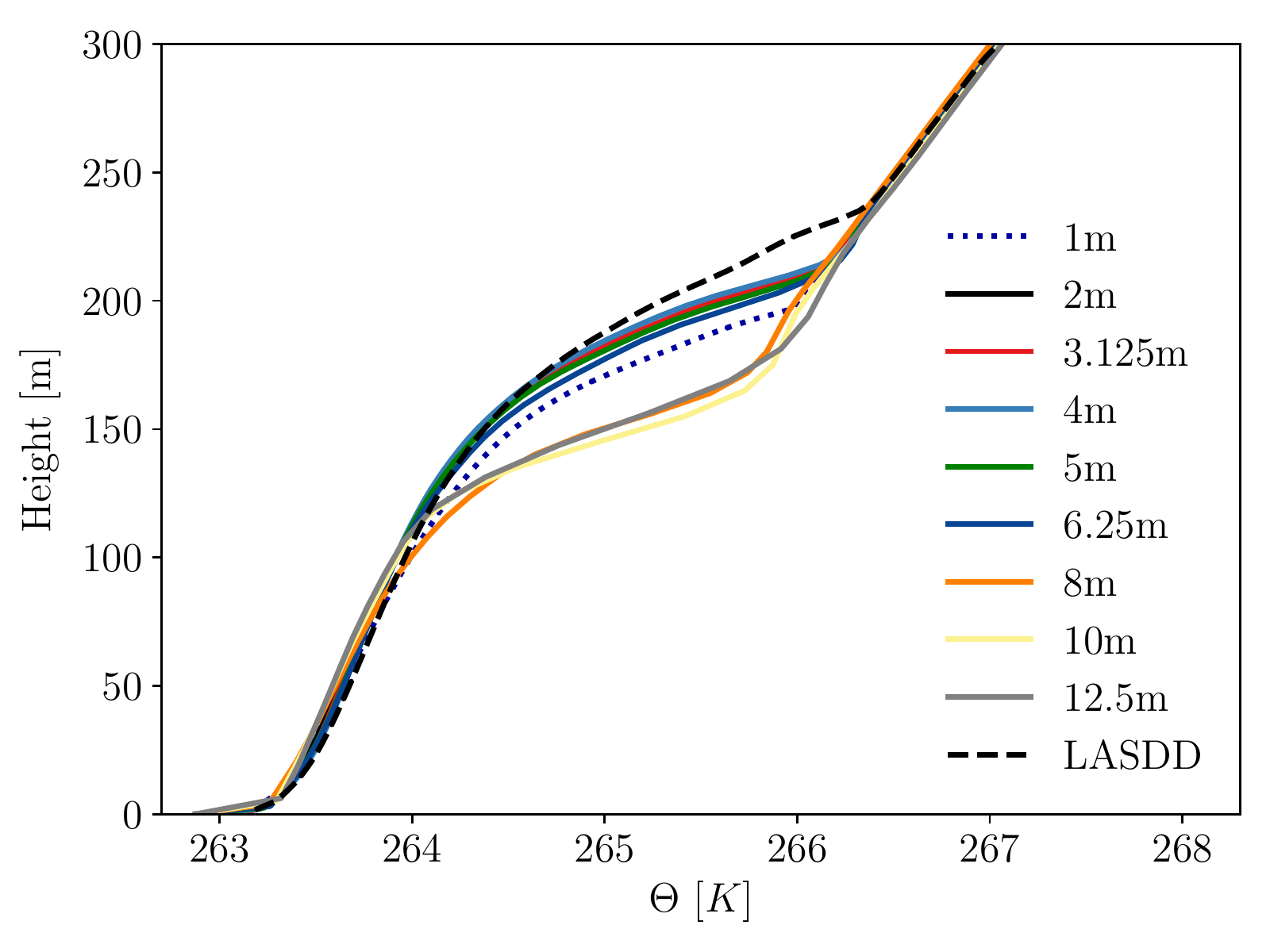}
  \includegraphics[width=0.49\textwidth]{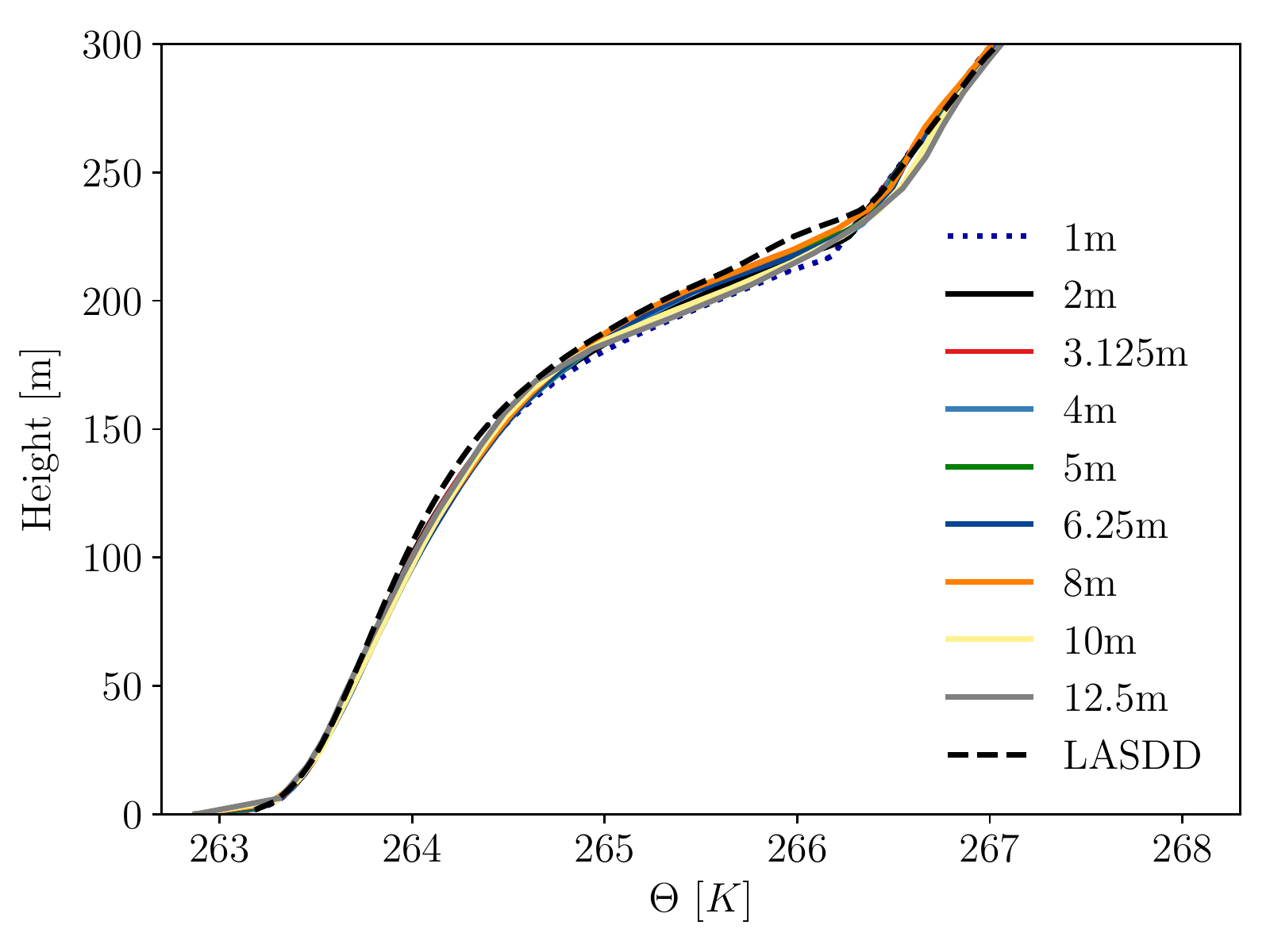}
\caption{Vertical profiles of mean wind speed (top panel), wind direction (middle panel), and potential temperature (bottom panel) from the D80 (left panel) and D80-R (right panel) based simulations using the PALM model system. Different colored lines correspond to different grid sizes ($\Delta$). Results from the MATLES code are overlaid (dashed black lines) for comparison.}
\label{fig:PALMfields}      
\end{figure*}

The (positive) impacts of higher SGS Prandtl number ($Pr_S$) utilized in the D80-R-based runs can be seen in some of the figures. First of all, Fig.~\ref{fig:PALMfields} (bottom-right panel) shows that the vertical profiles of potential temperature are less convex than the corresponding profiles from the DALES code (i.e., bottom-right panel of Fig.~\ref{fig:fields}). This is due to less heat diffusion. For the very same reason, the PALM model-generated resolved $\sigma_\theta^2$ values are much larger than their DALES counterpart.  

\begin{figure*}[ht!]
\centering
  \includegraphics[width=0.49\textwidth]{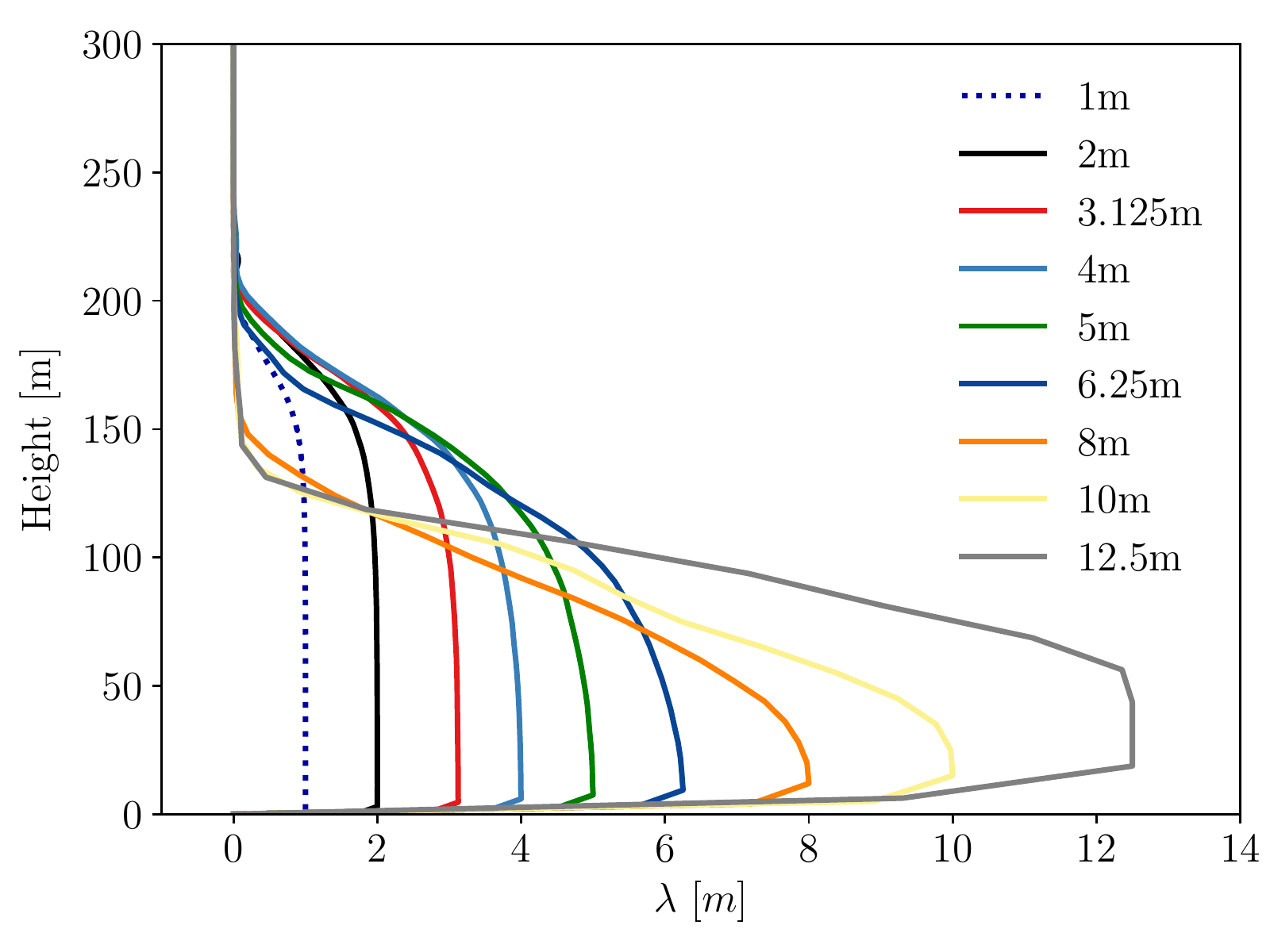}
  \includegraphics[width=0.49\textwidth]{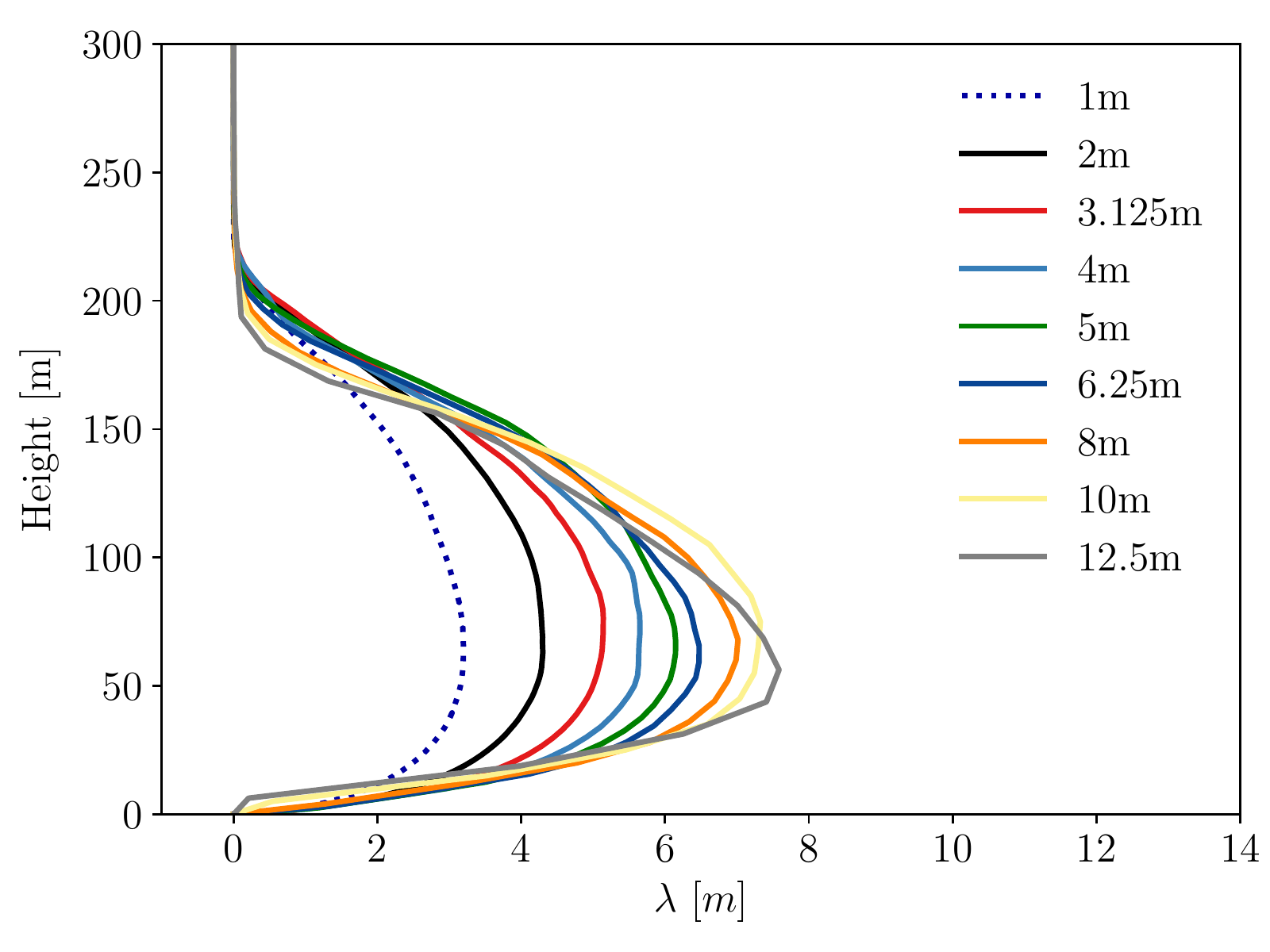}
\caption{Vertical profiles of mixing length ($\lambda$) from the D80 (left panel) and D80-R (right panel) based simulations using the PALM model system. Different colored lines correspond to different grid sizes ($\Delta$). Results from the MATLES code are overlaid (dashed black lines) for comparison.}
\label{fig:PALMlambda}      
\end{figure*}

\begin{figure*}[ht!]
\centering
  \includegraphics[width=0.49\textwidth]{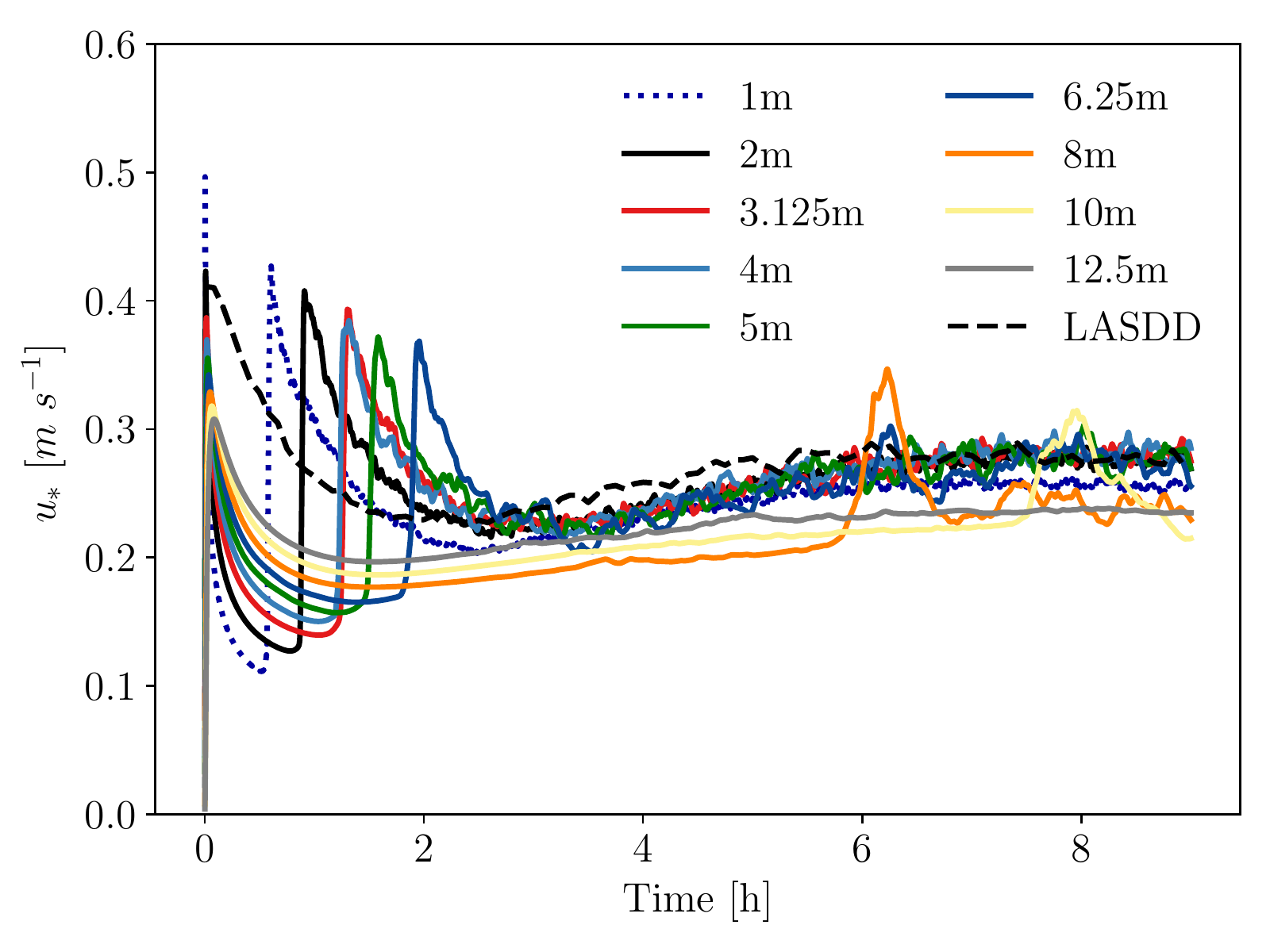}
  \includegraphics[width=0.49\textwidth]{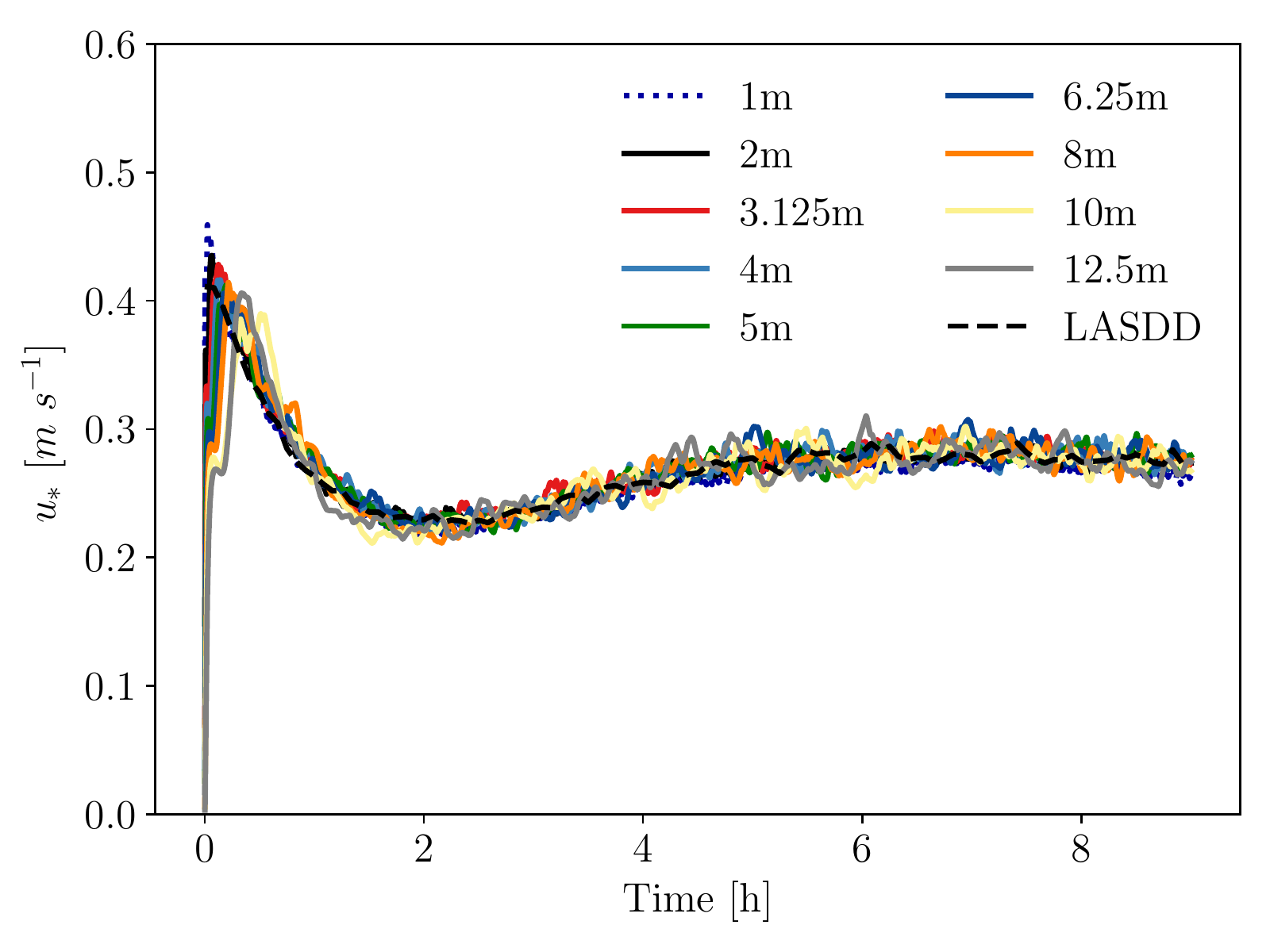}\\
  \includegraphics[width=0.49\textwidth]{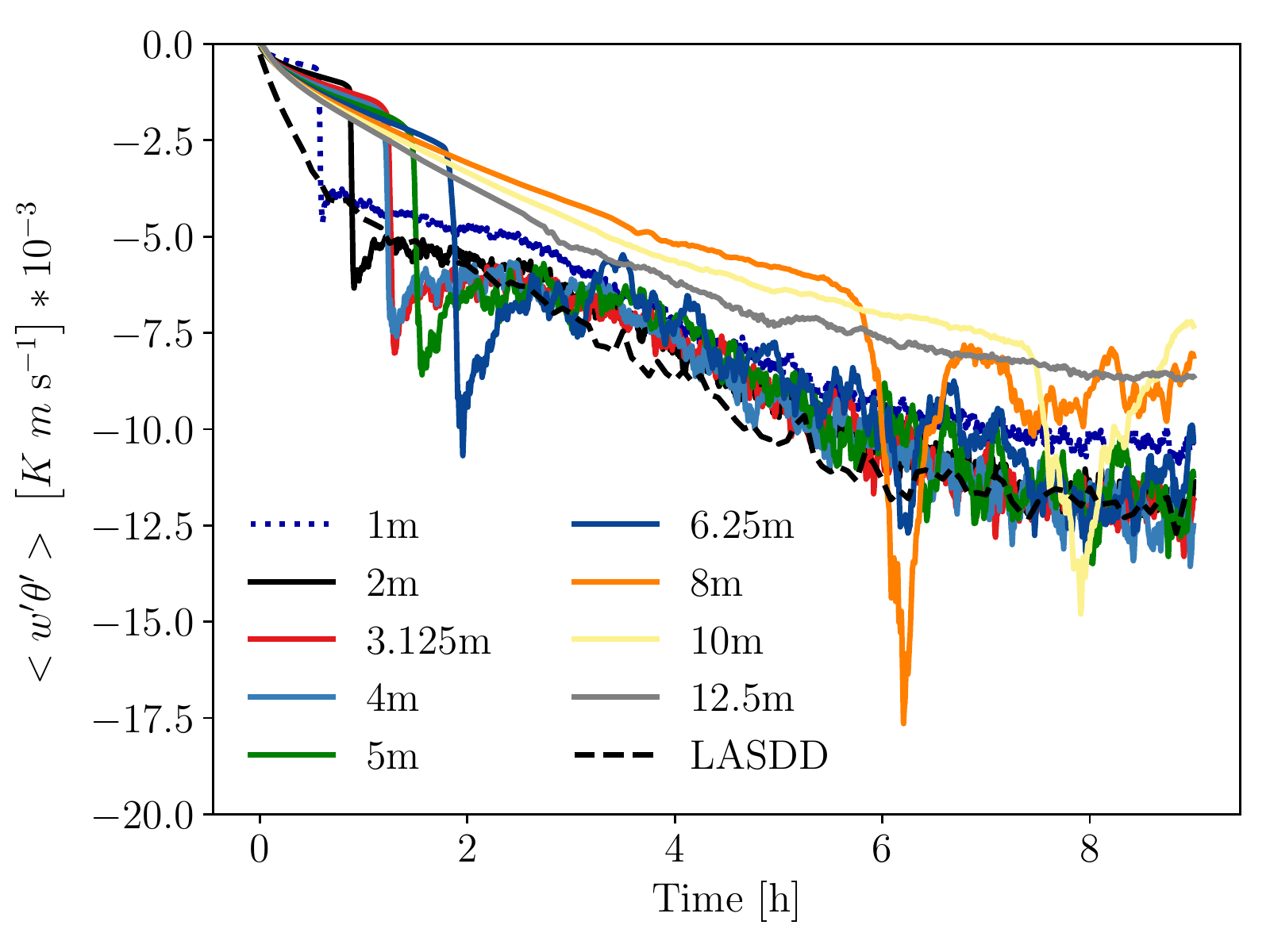}
  \includegraphics[width=0.49\textwidth]{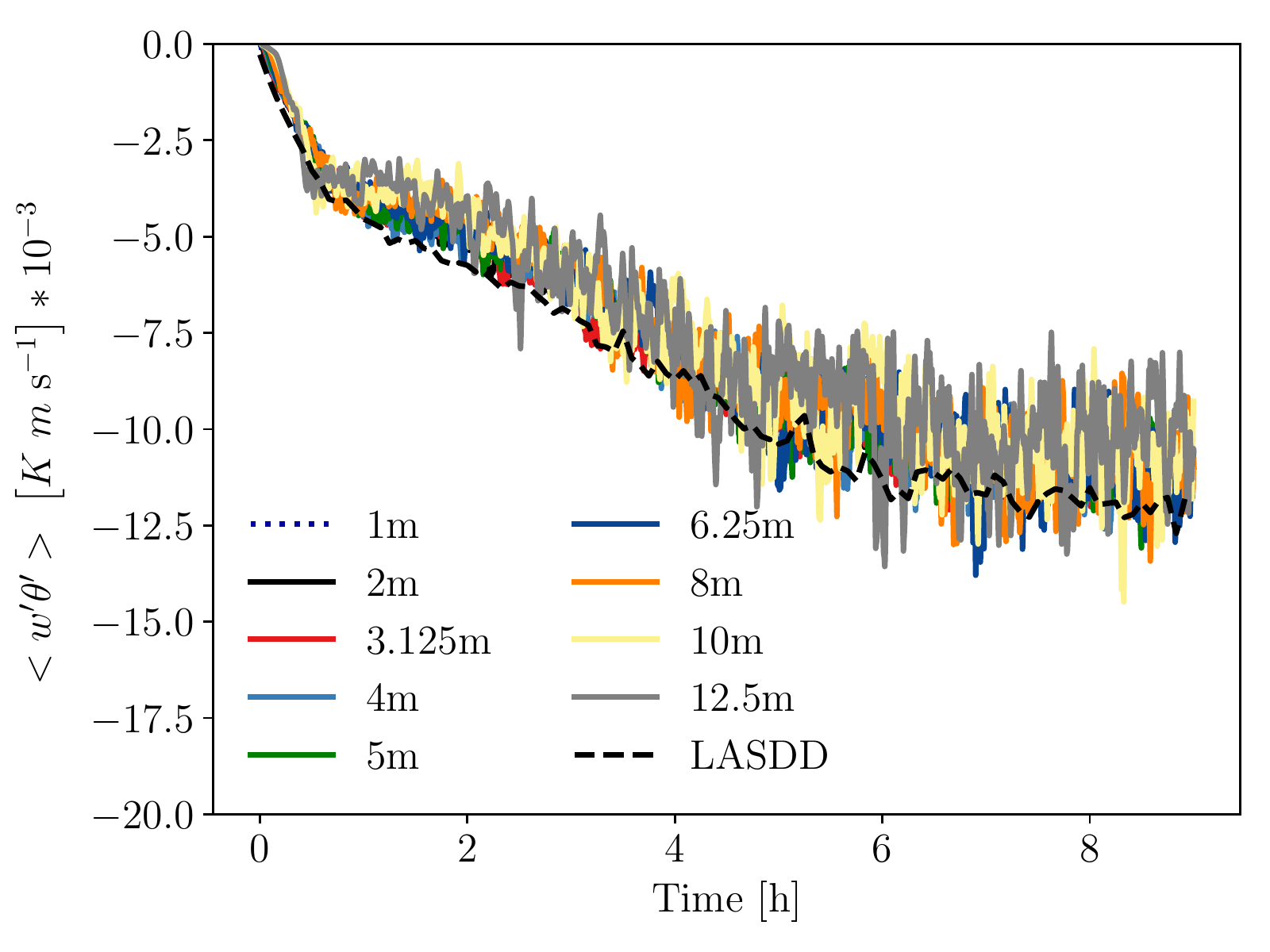}
\caption{Time series of surface friction velocity (top panel) and sensible heat flux (bottom panel) from the D80 (left panel) and D80-R (right panel) based simulations using the PALM model system. Different colored lines correspond to different grid sizes ($\Delta$). Results from the MATLES code are overlaid (dashed black lines) for comparison.}
\label{fig:PALMsurface}      
\end{figure*}

\begin{figure*}[ht]
\centering
  \includegraphics[width=0.49\textwidth]{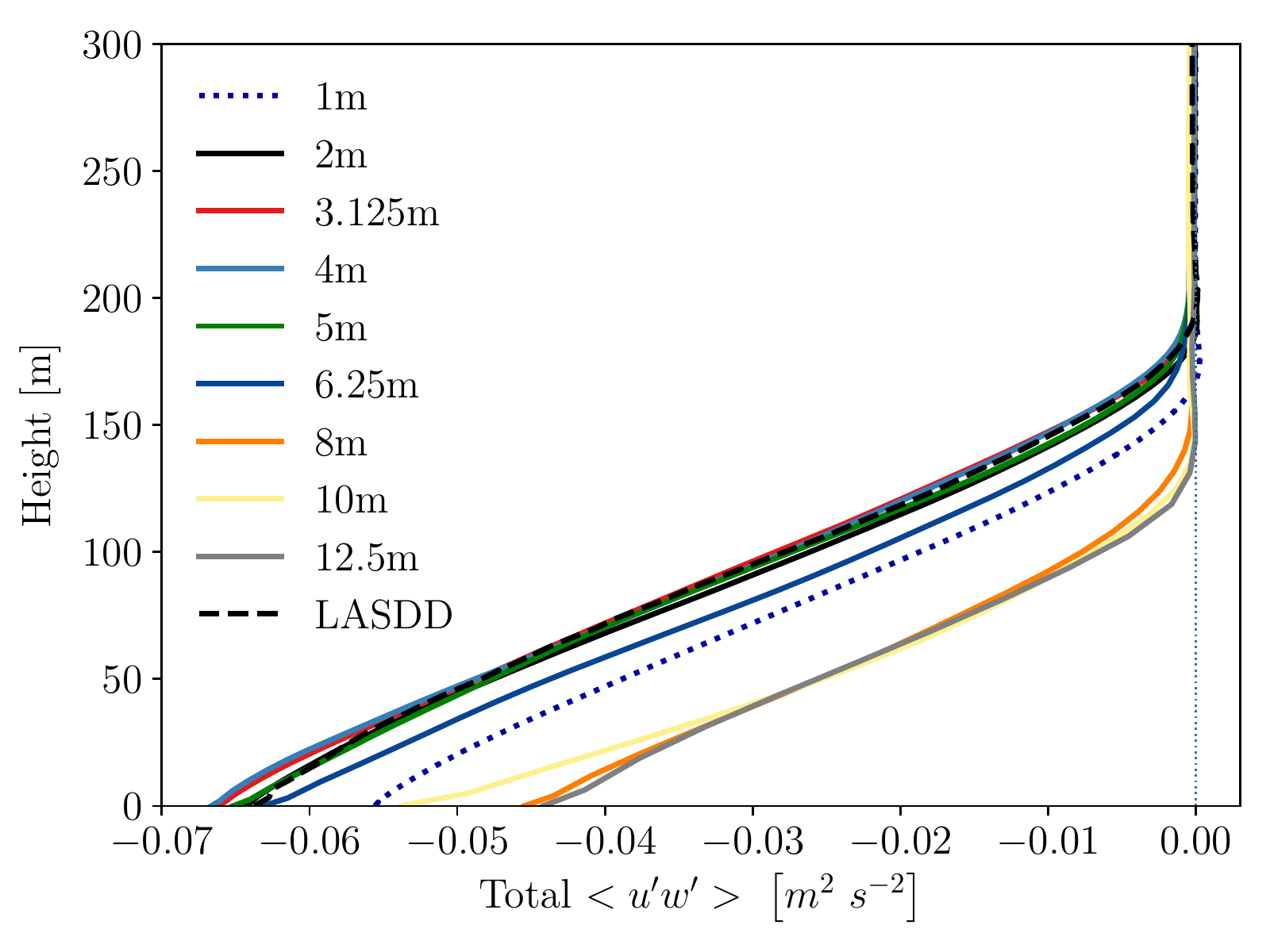}
  \includegraphics[width=0.49\textwidth]{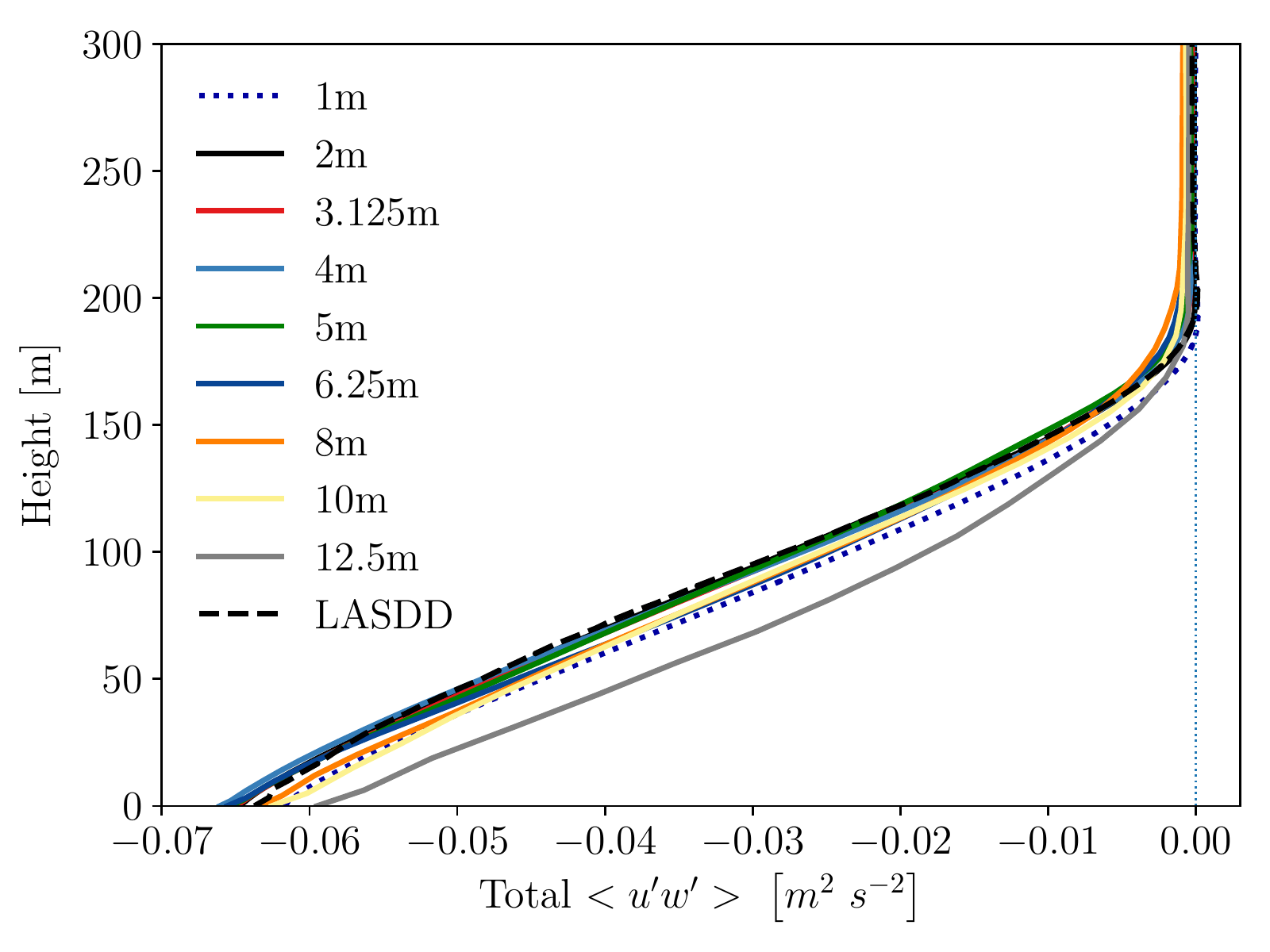}\\
  \includegraphics[width=0.49\textwidth]{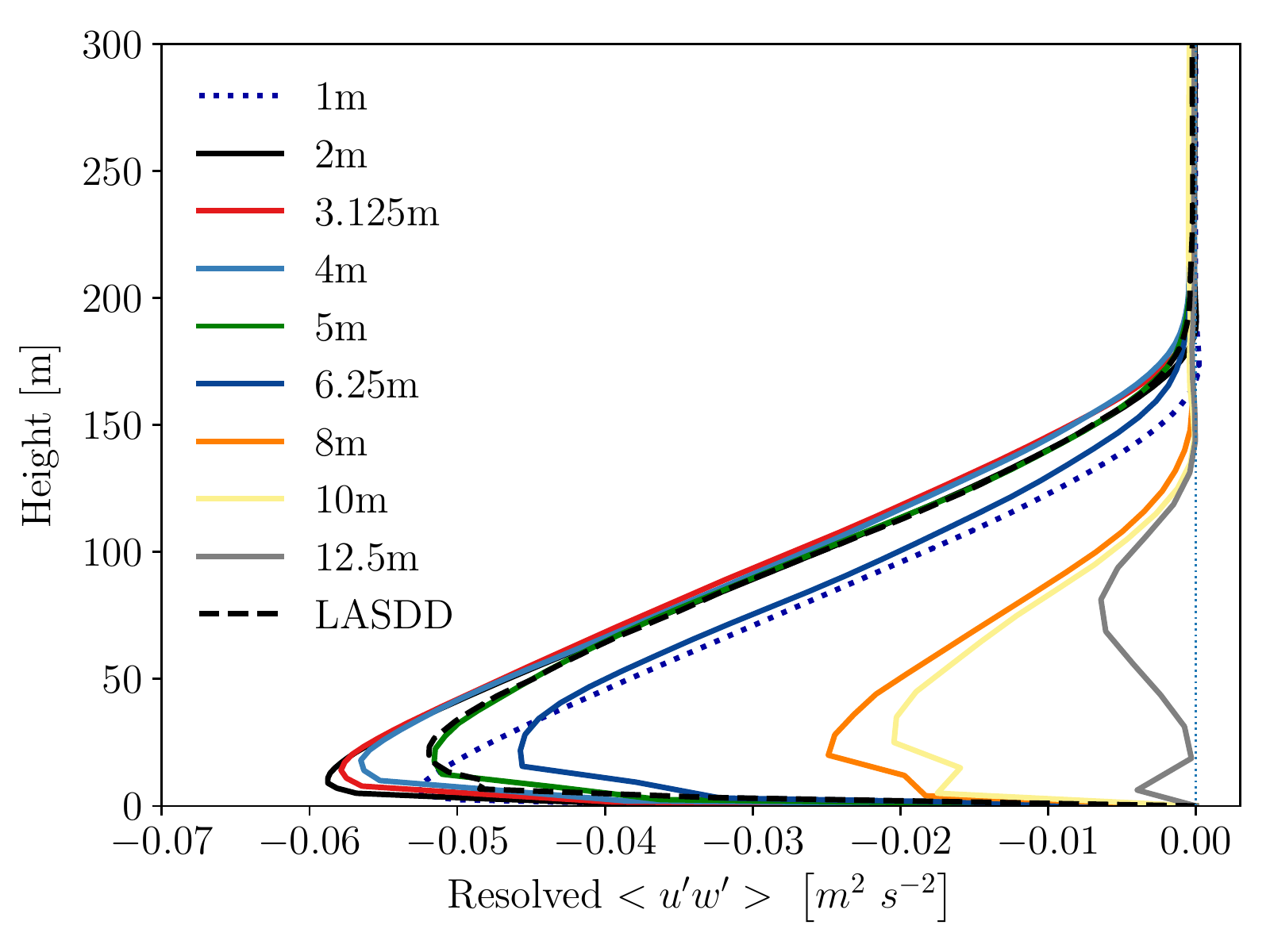}
  \includegraphics[width=0.49\textwidth]{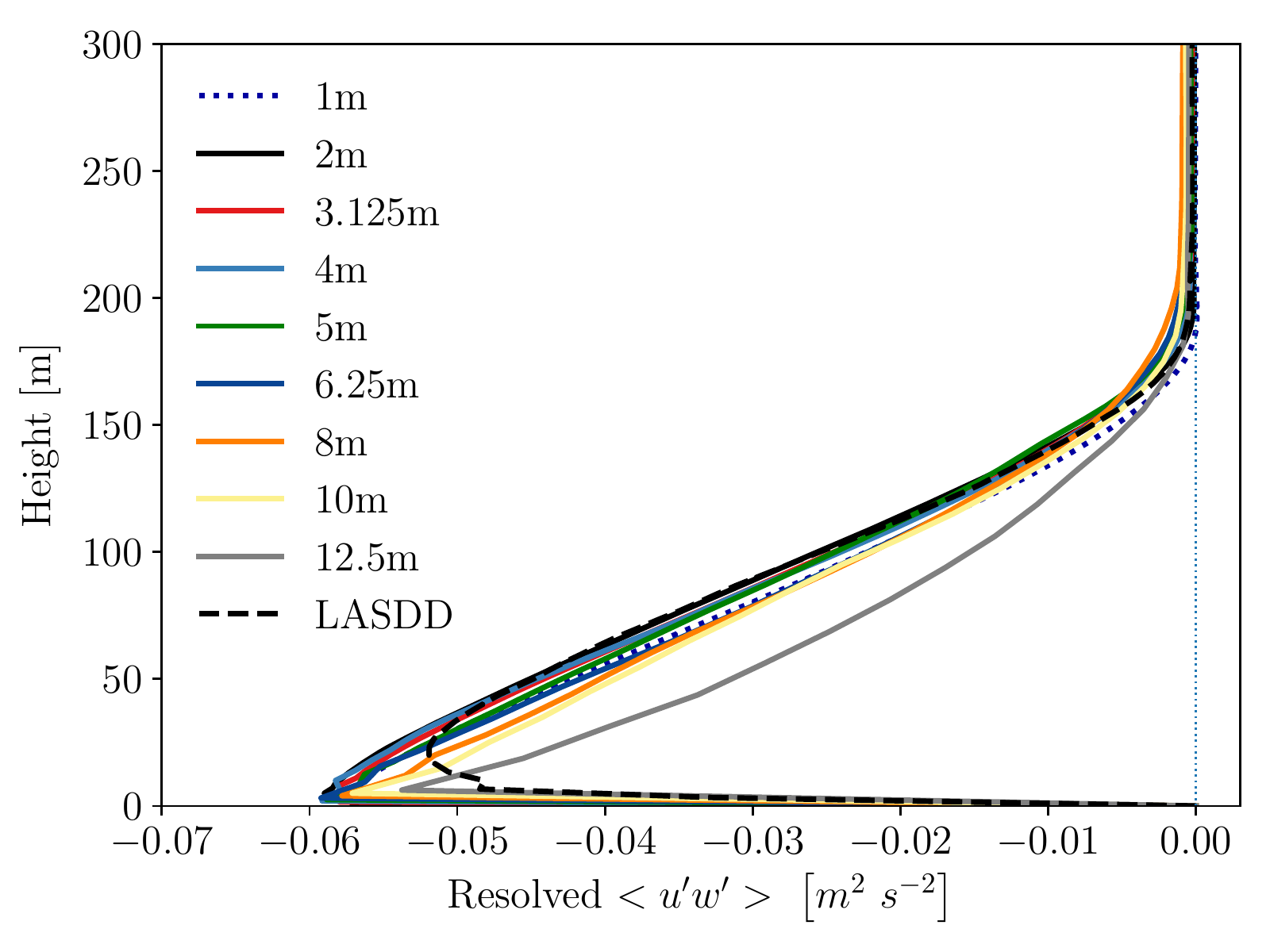}
\caption{Vertical profiles of total (top panel) and resolved (bottom panel) momentum flux ($u$-component) from the D80 (left panel) and D80-R (right panel) based simulations using the PALM model system. Different colored lines correspond to different grid sizes ($\Delta$). Results from the MATLES code are overlaid (dashed black lines) for comparison.}
\label{fig:PALMFlux1}      
\end{figure*}

\begin{figure*}[ht]
\centering
  \includegraphics[width=0.49\textwidth]{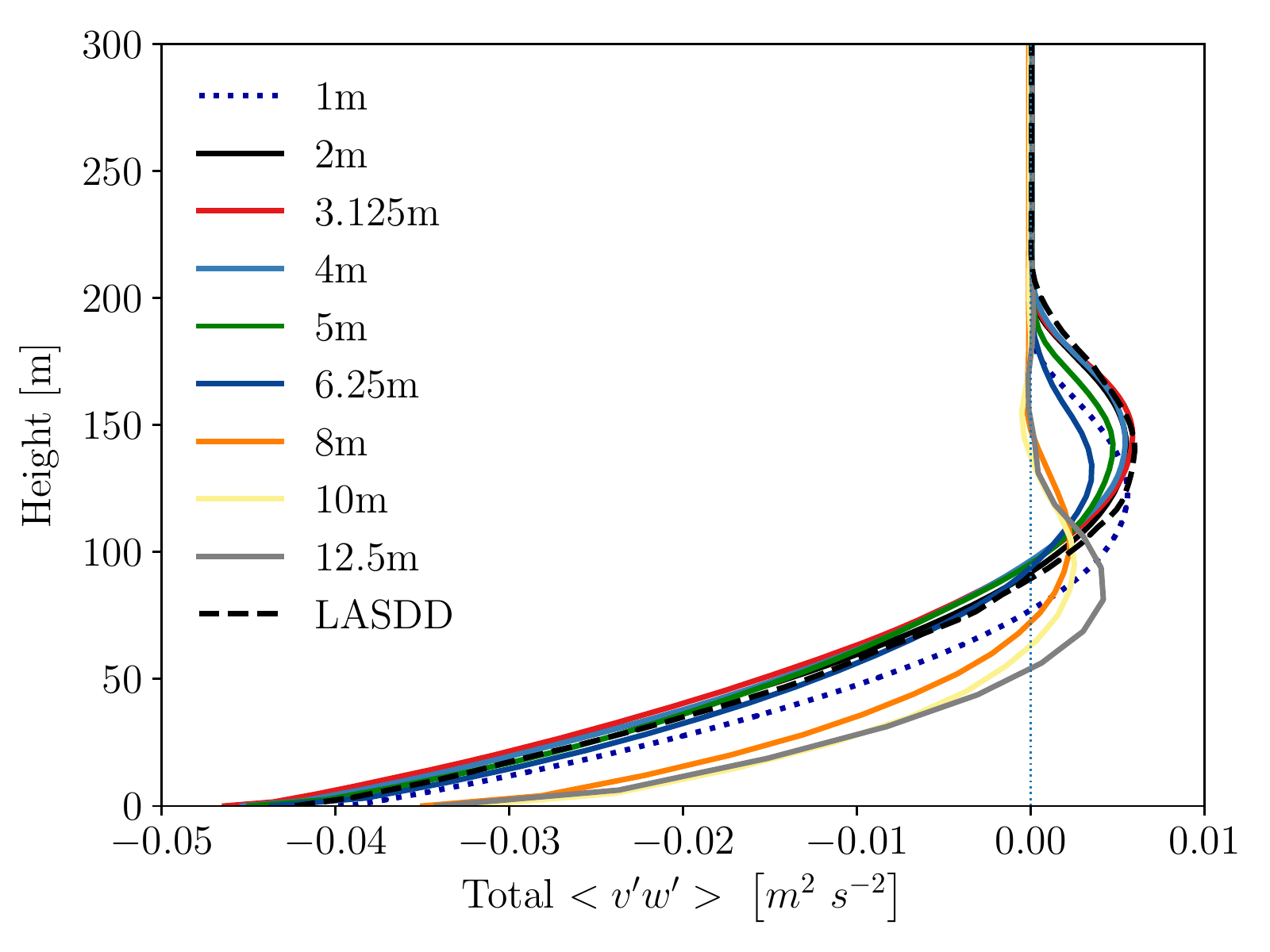}
  \includegraphics[width=0.49\textwidth]{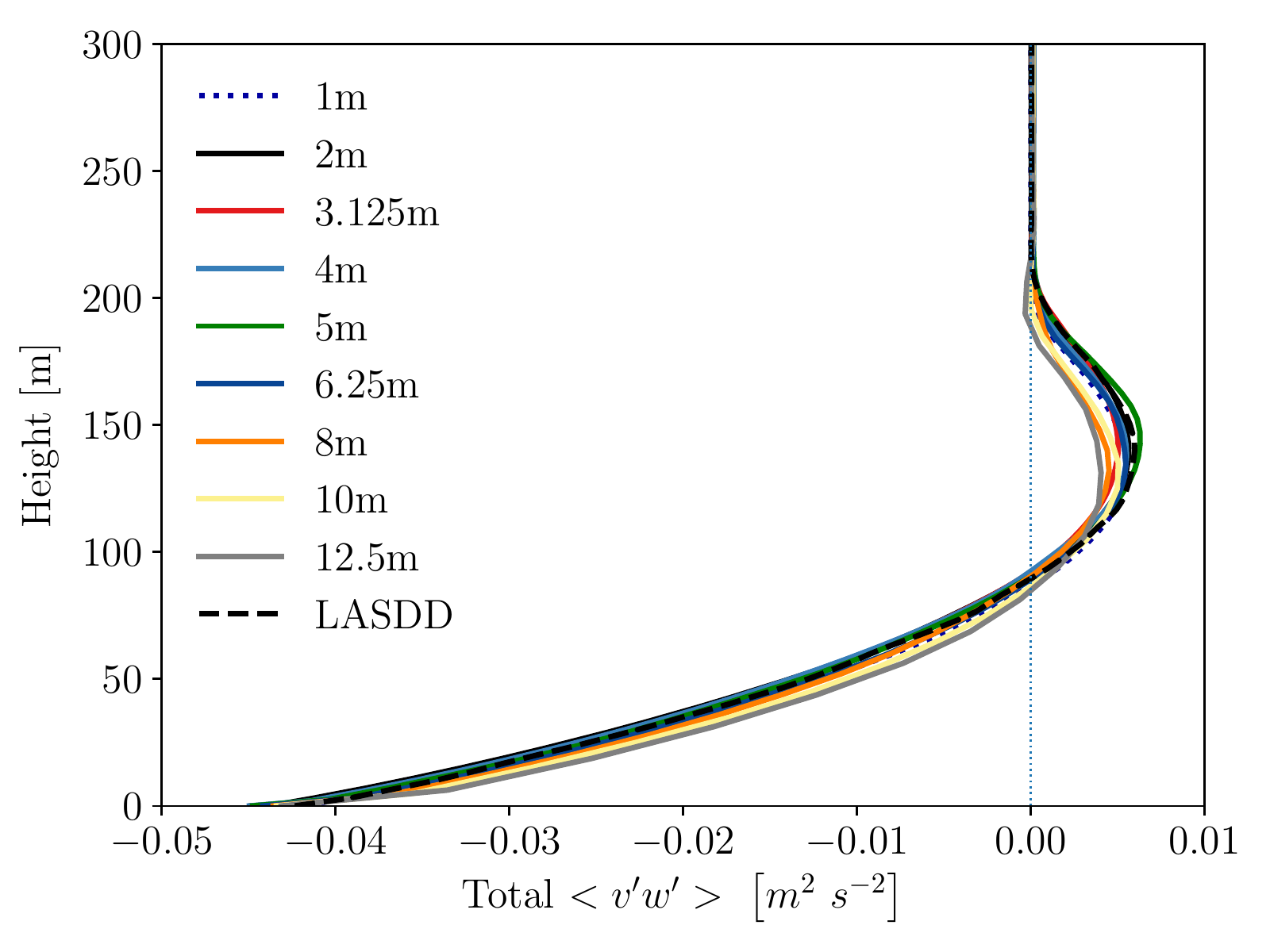}\\
  \includegraphics[width=0.49\textwidth]{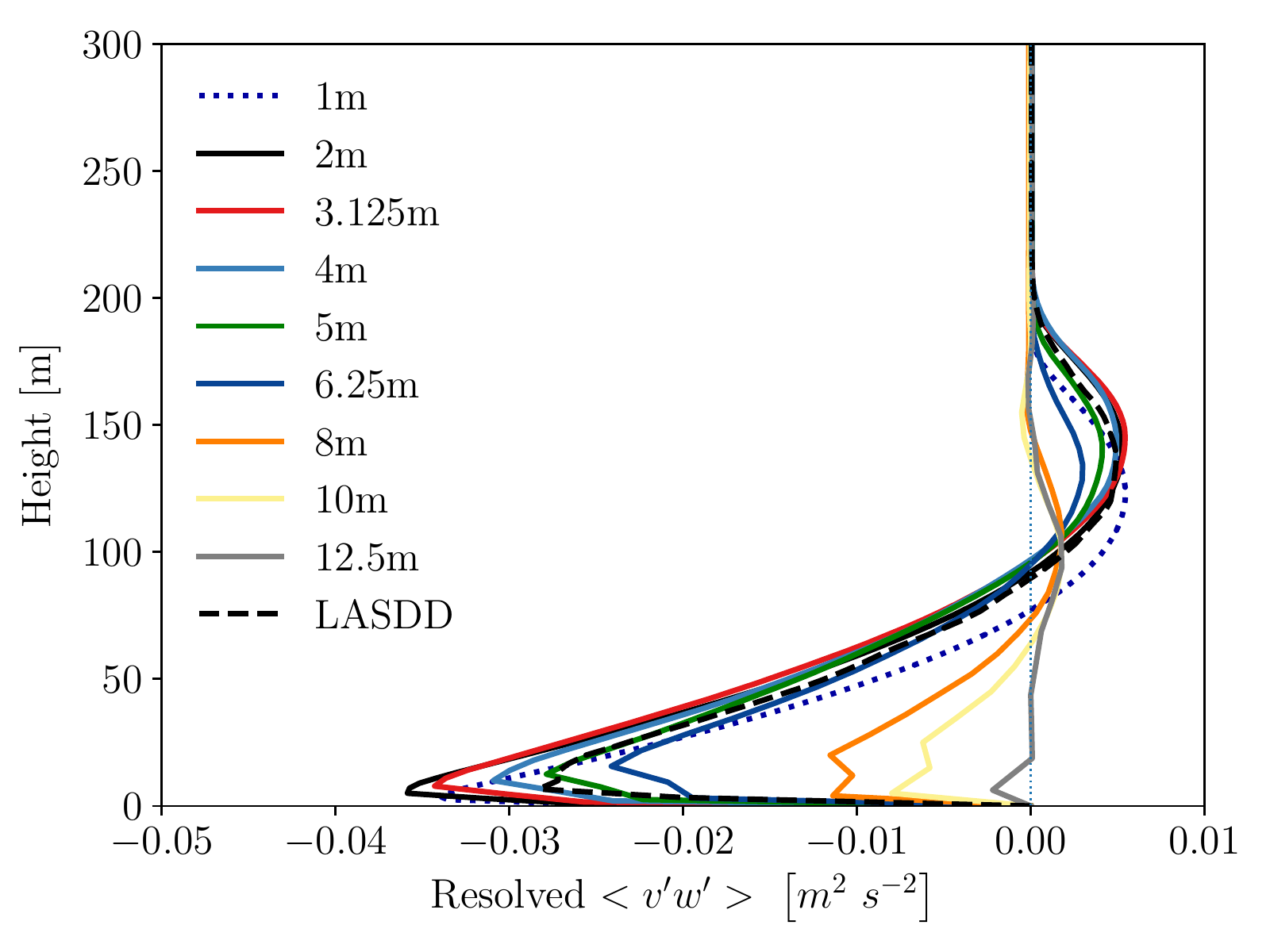}
  \includegraphics[width=0.49\textwidth]{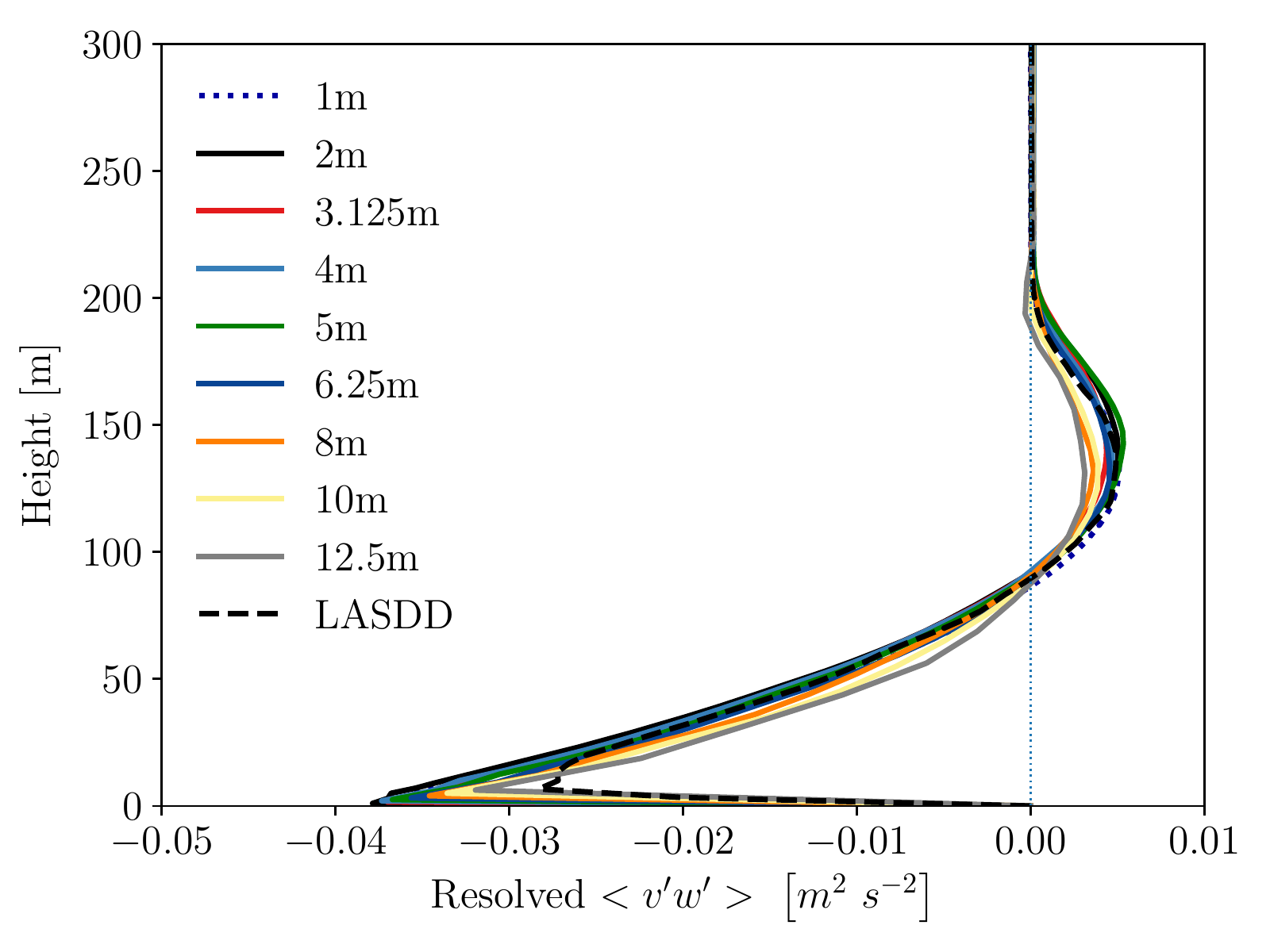}
\caption{Vertical profiles of total (top panel) and resolved (bottom panel) momentum flux ($v$-component) from the D80 (left panel) and D80-R (right panel) based simulations using the PALM model system. Different colored lines correspond to different grid sizes ($\Delta$). Results from the MATLES code are overlaid (dashed black lines) for comparison.}
\label{fig:PALMFlux2}      
\end{figure*}

\begin{figure*}[ht]
\centering
  \includegraphics[width=0.49\textwidth]{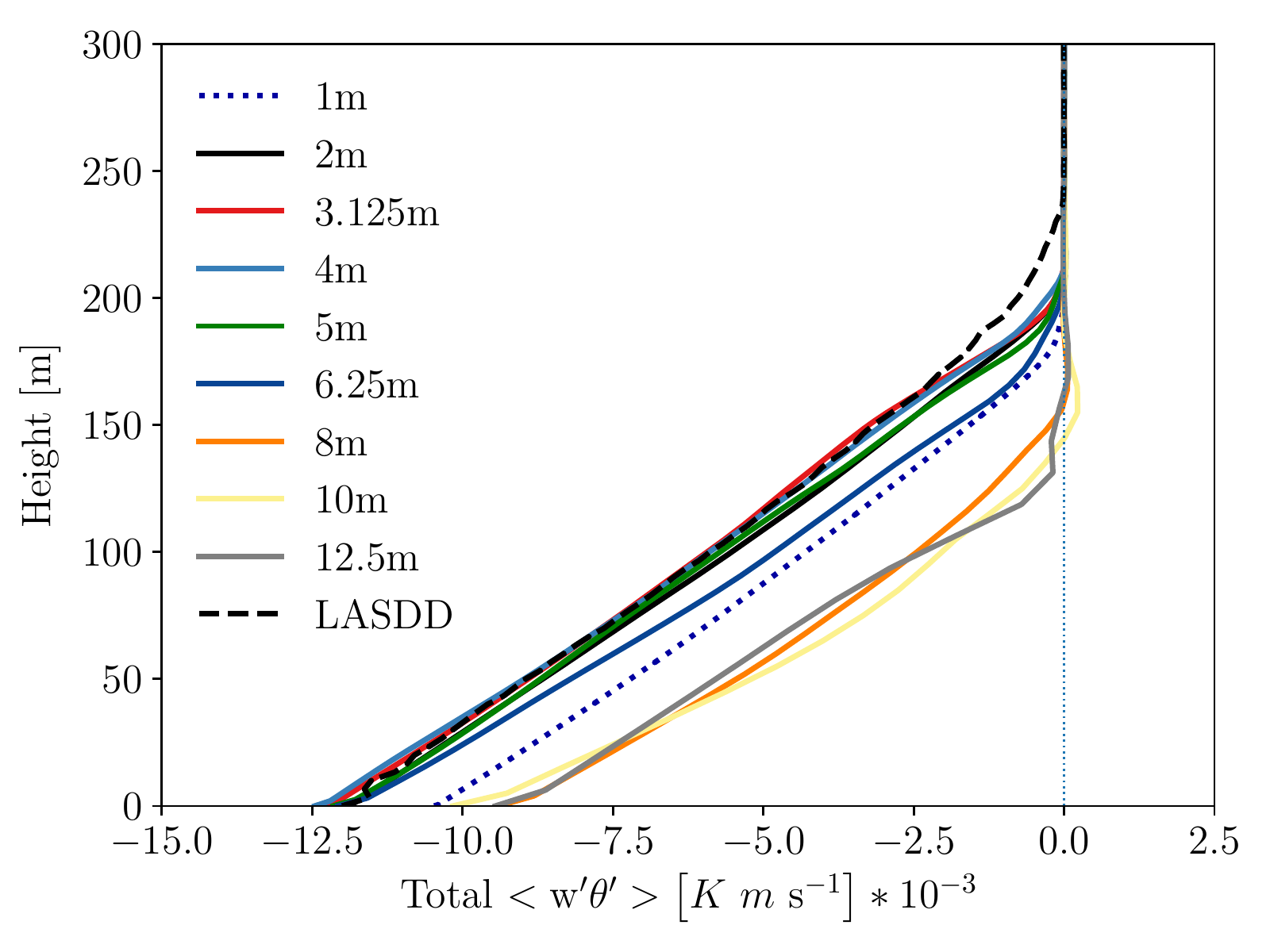}
  \includegraphics[width=0.49\textwidth]{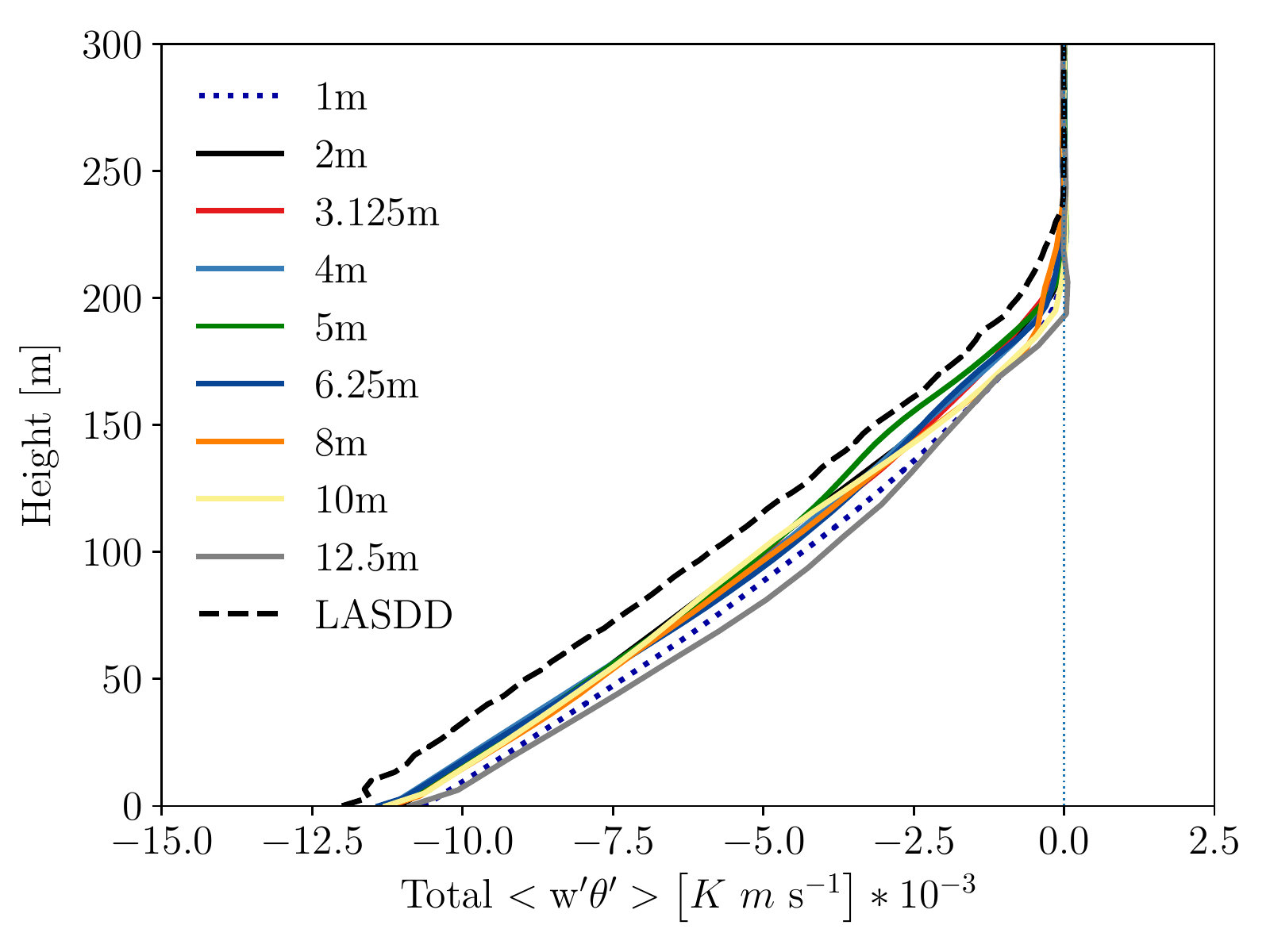}\\
  \includegraphics[width=0.49\textwidth]{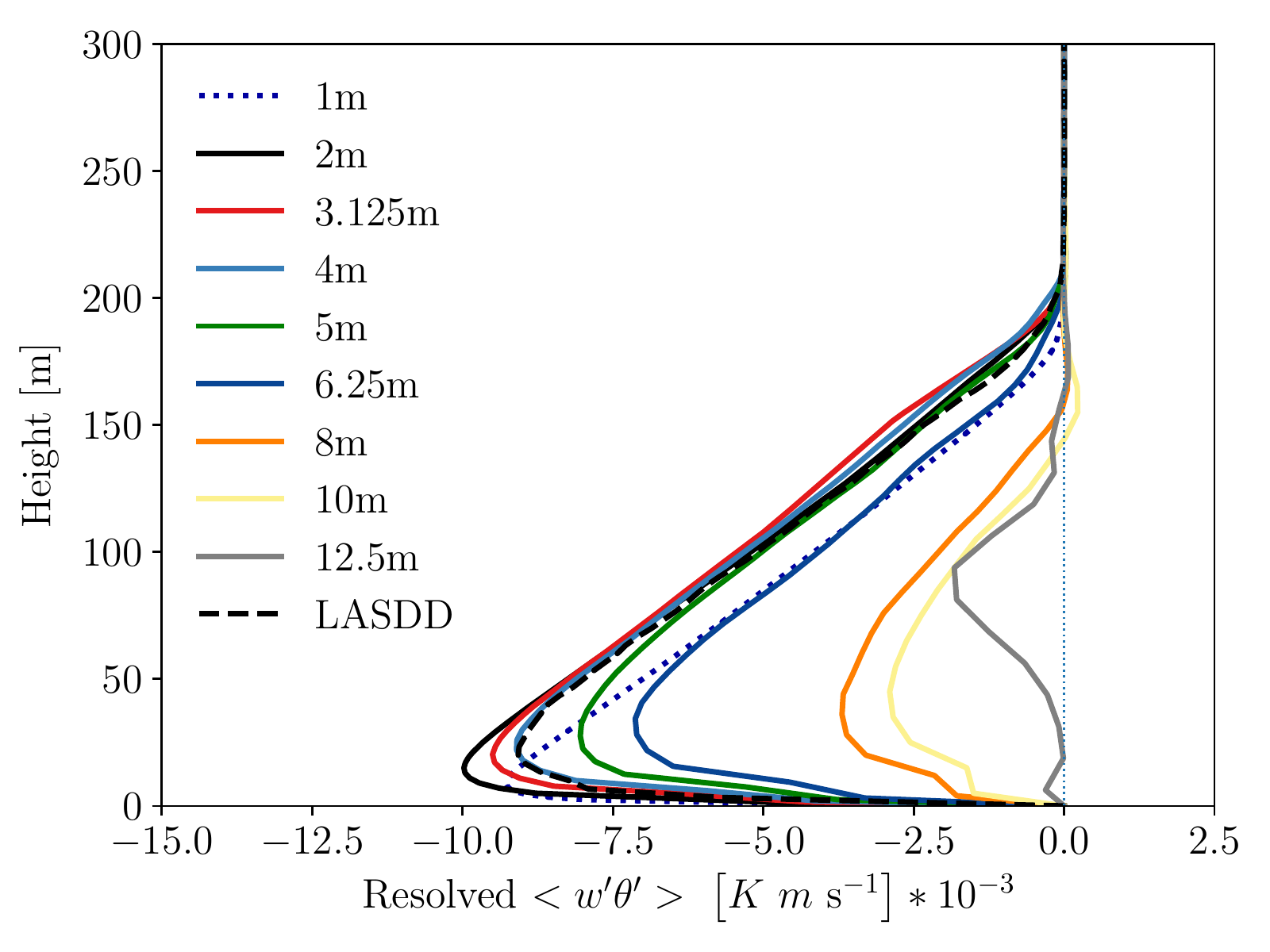}
  \includegraphics[width=0.49\textwidth]{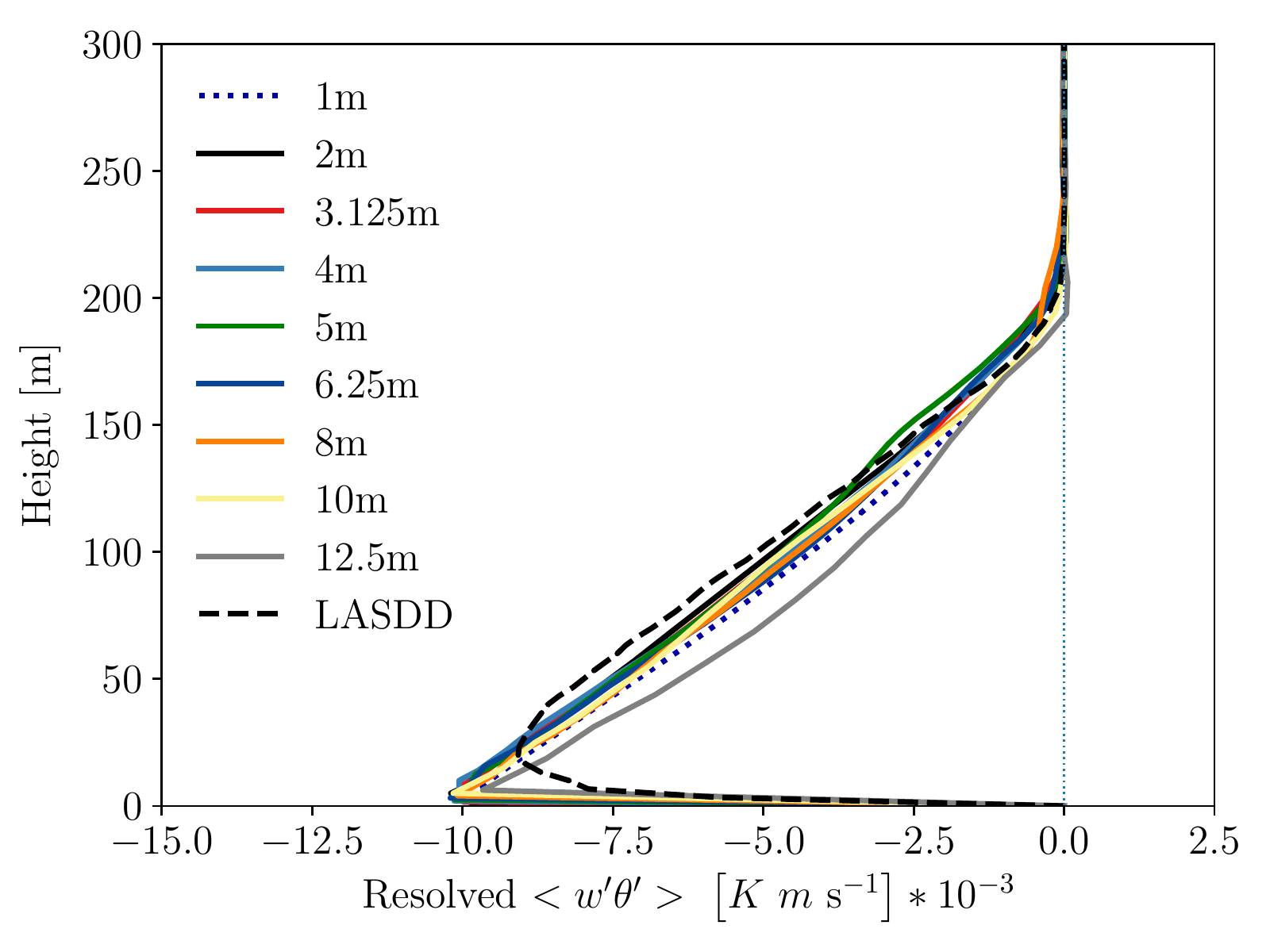}
\caption{Vertical profiles of total (top panel) and resolved (bottom panel) sensible heat flux from the D80 (left panel) and D80-R (right panel) based simulations using the PALM model system. Different colored lines correspond to different grid sizes ($\Delta$). Results from the MATLES code are overlaid (dashed black lines) for comparison.}
\label{fig:PALMFlux3}      
\end{figure*}

\begin{figure*}[ht]
\centering
  \includegraphics[width=0.49\textwidth]{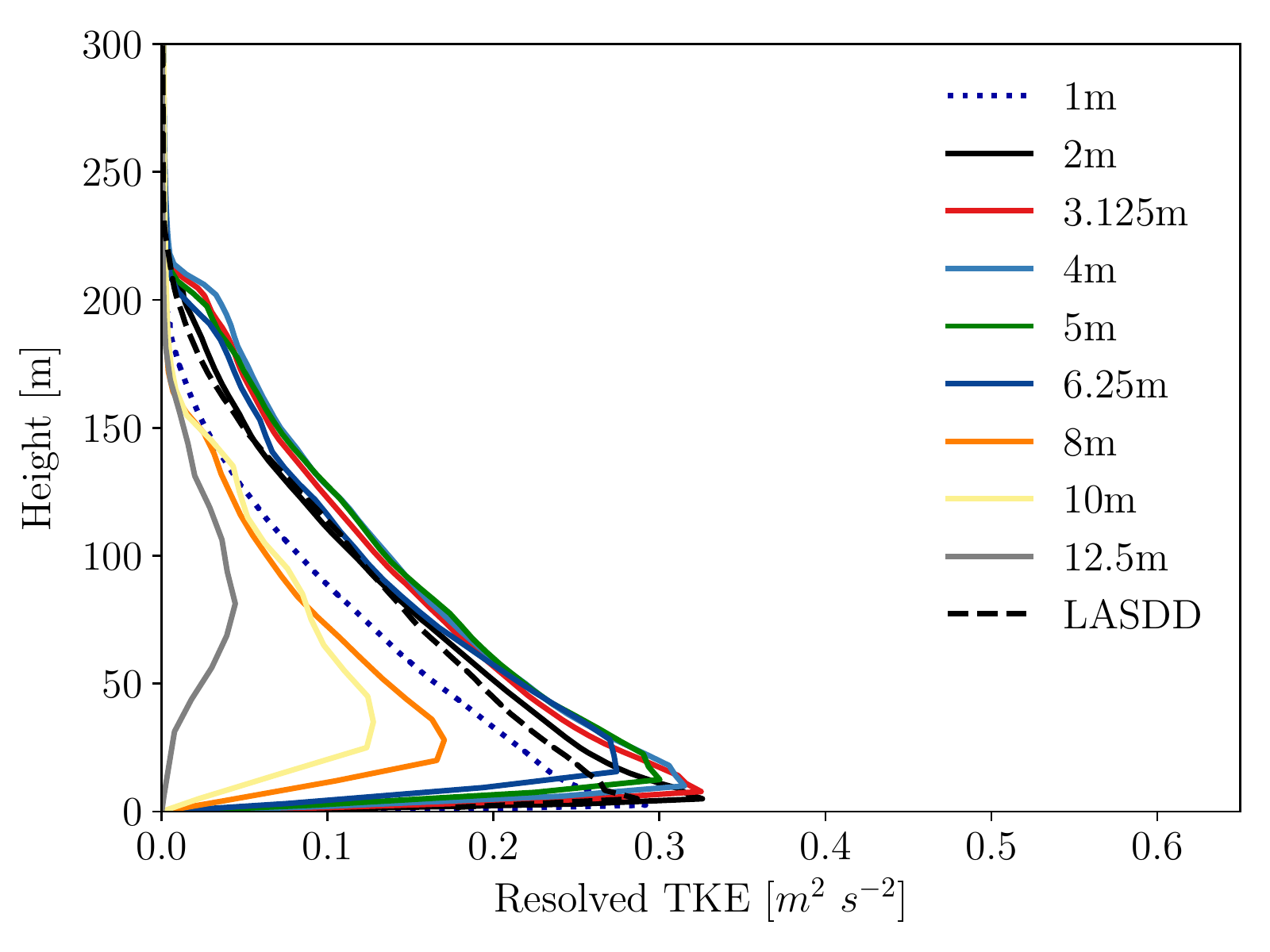}
  \includegraphics[width=0.49\textwidth]{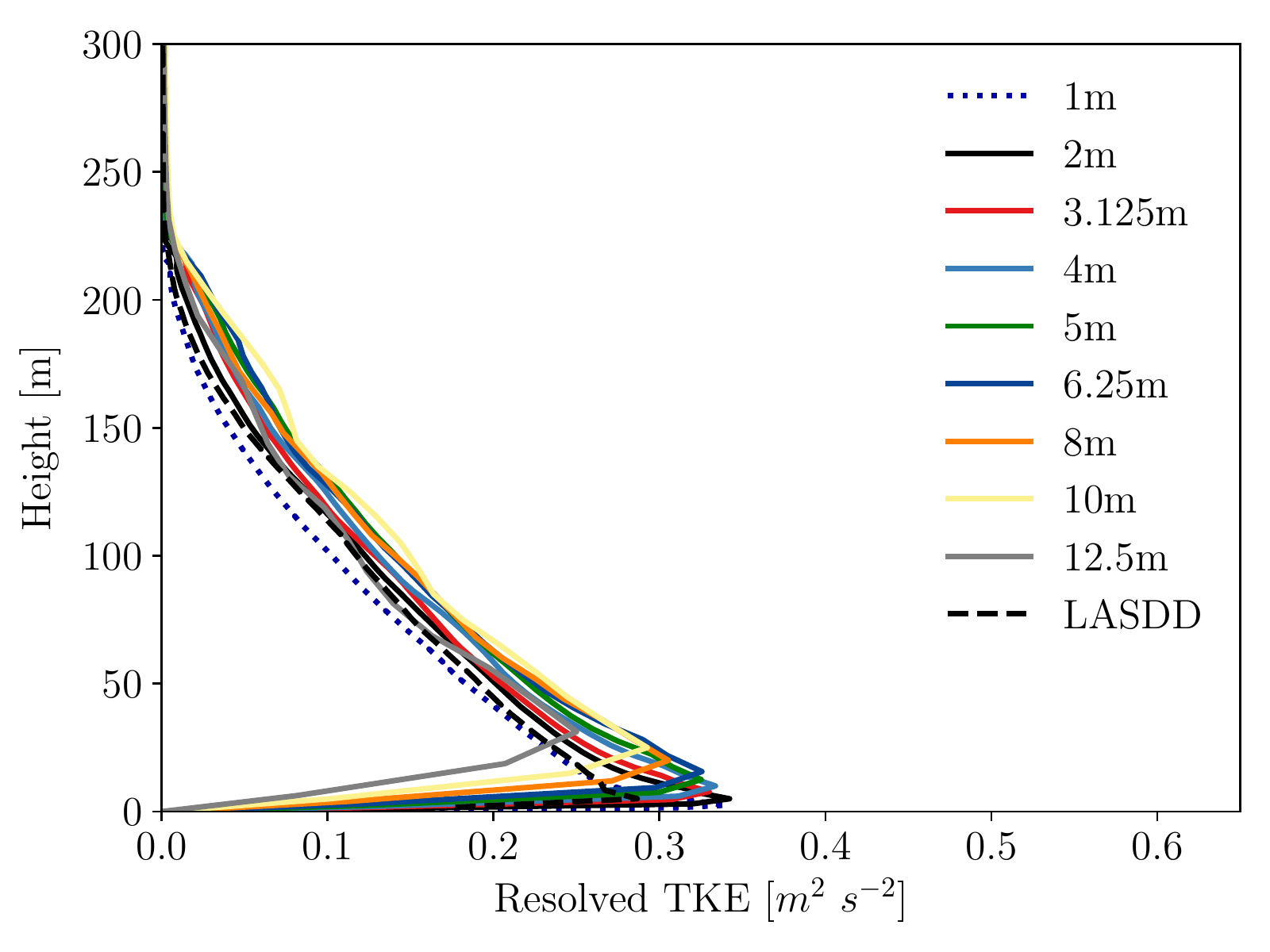}\\
  \includegraphics[width=0.49\textwidth]{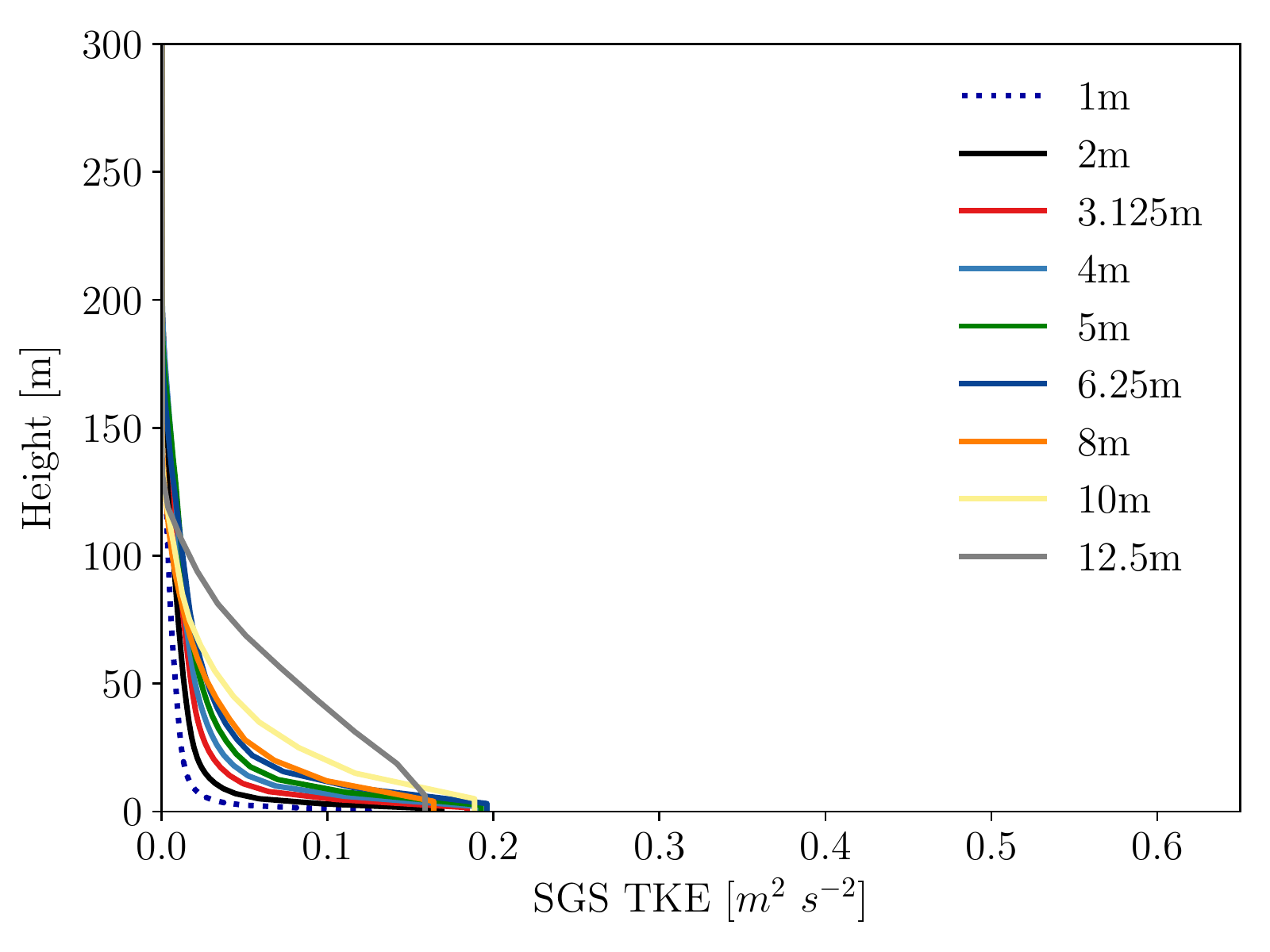}
  \includegraphics[width=0.49\textwidth]{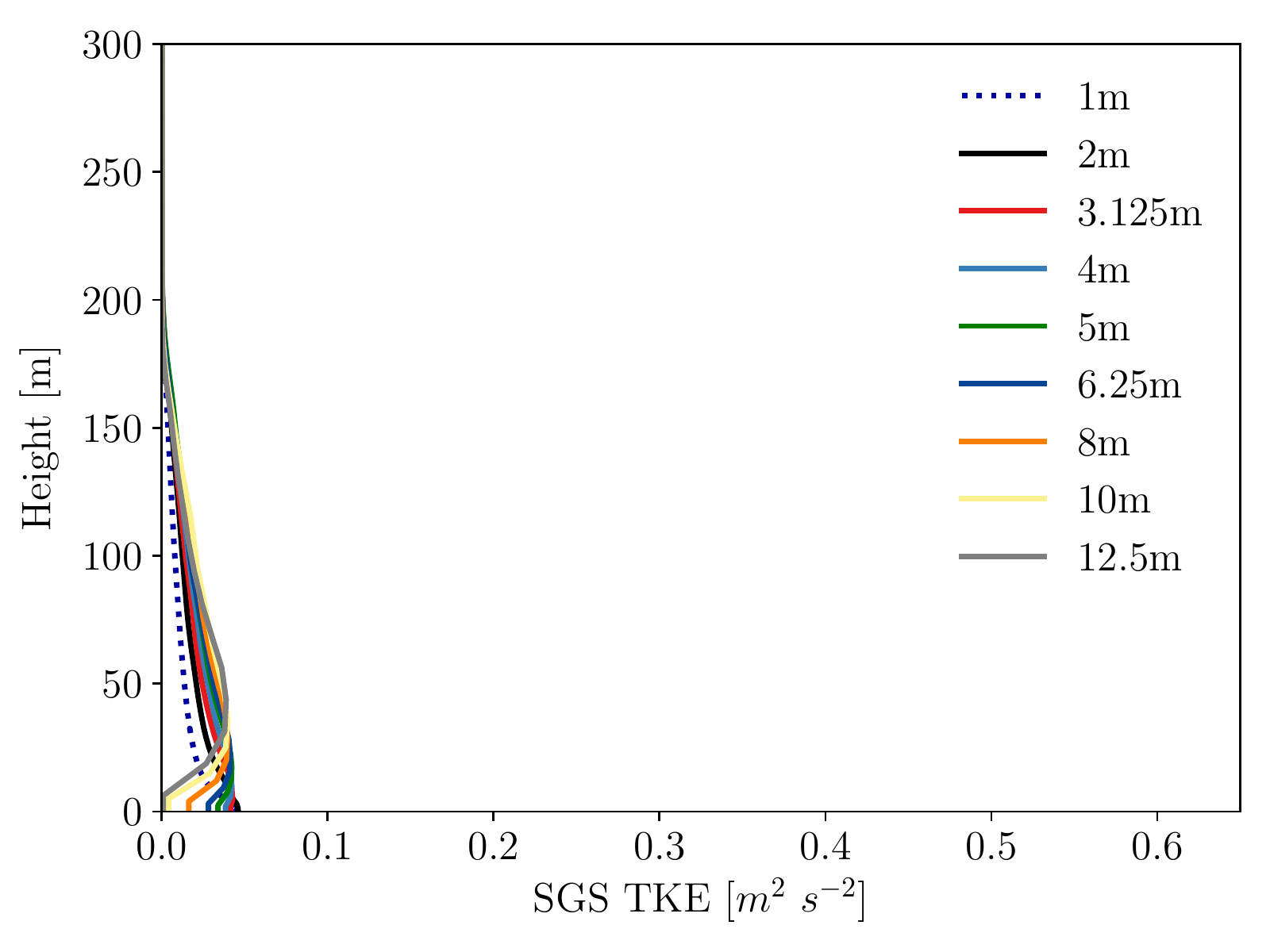}
\caption{Vertical profiles of resolved (top panel) and subgrid-scale (bottom panel) turbulent kinetic energy from the D80 (left panel) and D80-R (right panel) based simulations using the PALM model system. Different colored lines correspond to different grid sizes ($\Delta$). Results from the MATLES code are overlaid (dashed black lines) for comparison in the resolved TKE plots.}
\label{fig:PALMTKE}      
\end{figure*}

\begin{figure*}[ht]
\centering
  \includegraphics[width=0.49\textwidth]{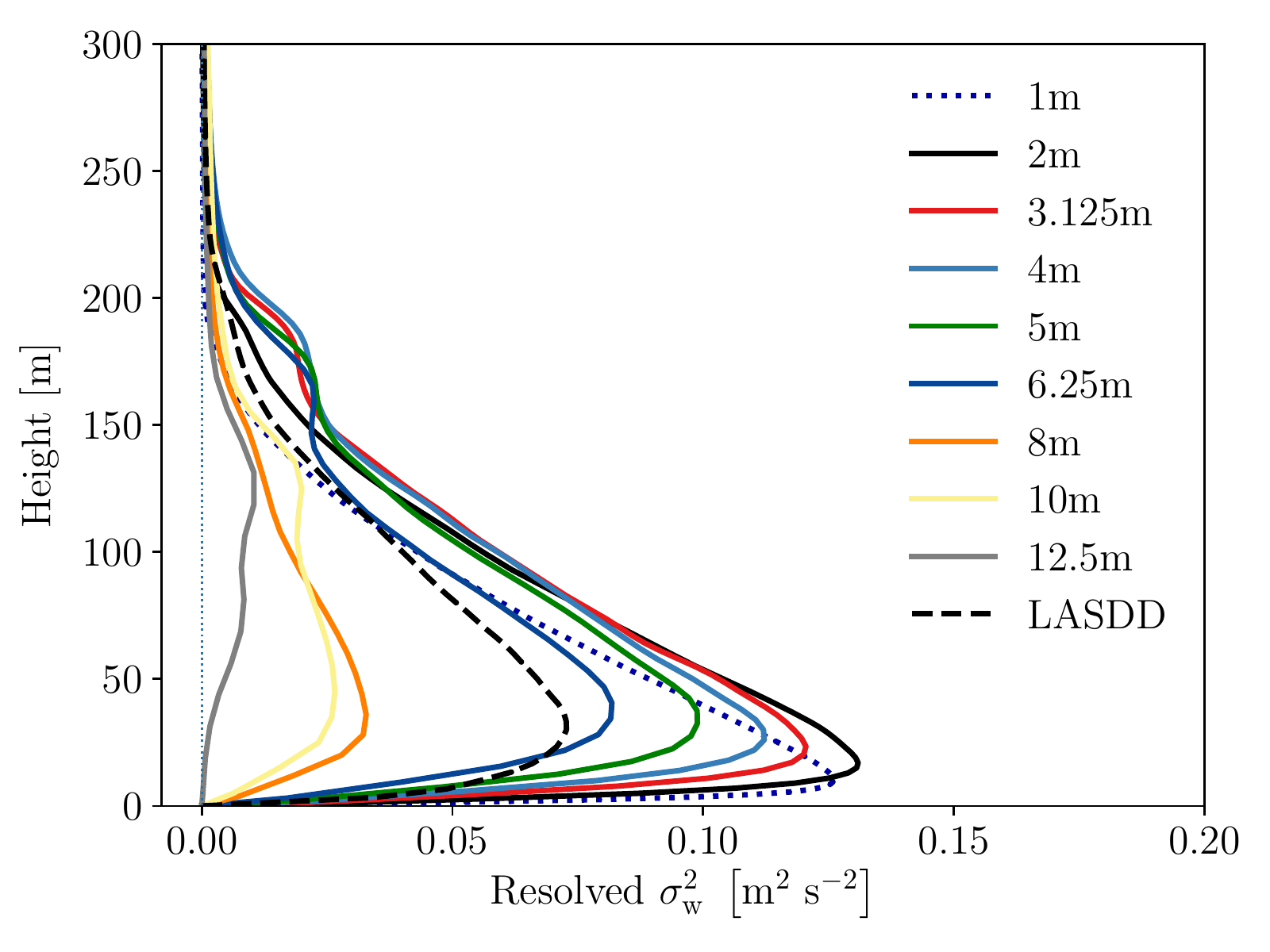}
  \includegraphics[width=0.49\textwidth]{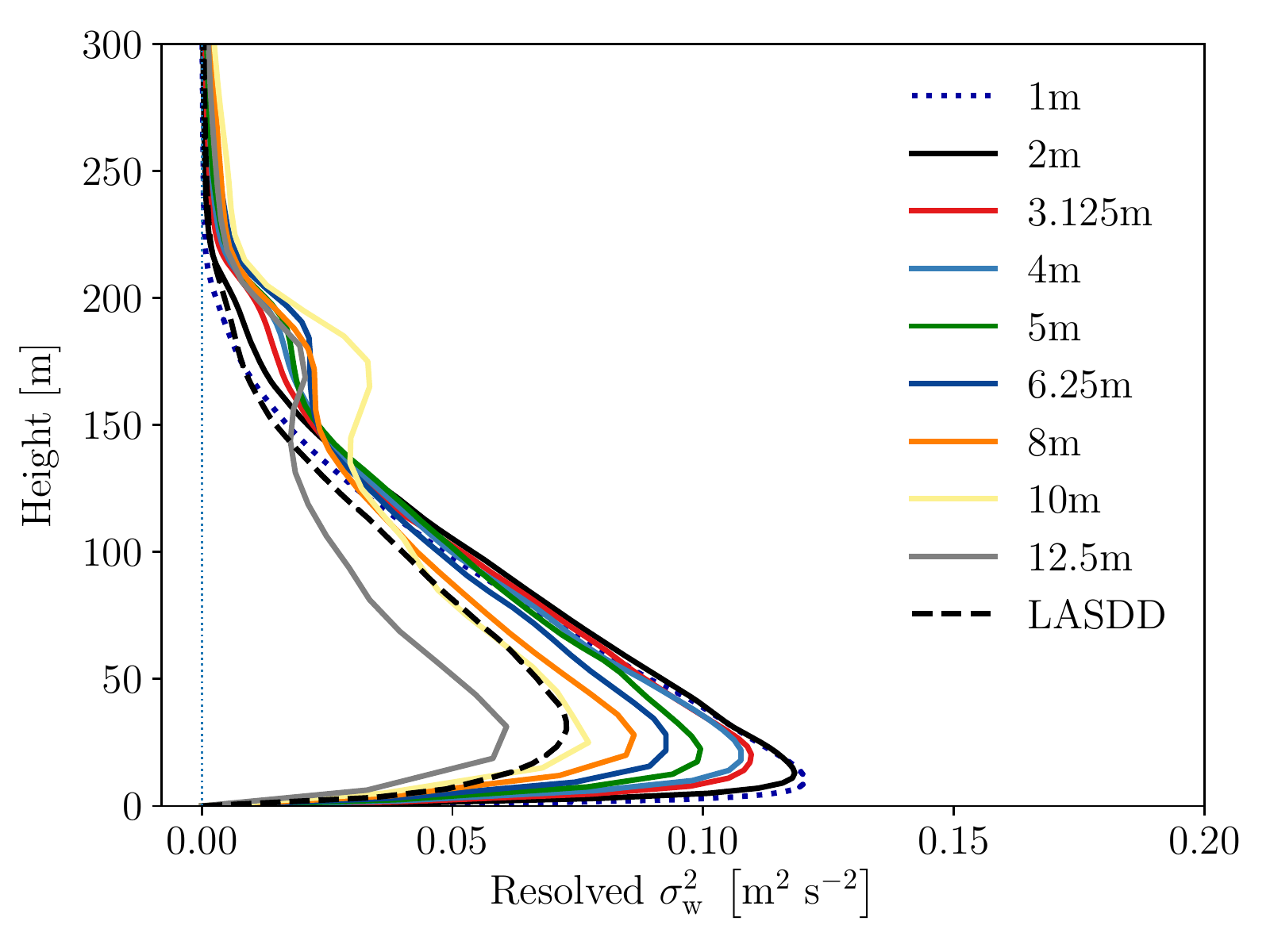}\\
  \includegraphics[width=0.49\textwidth]{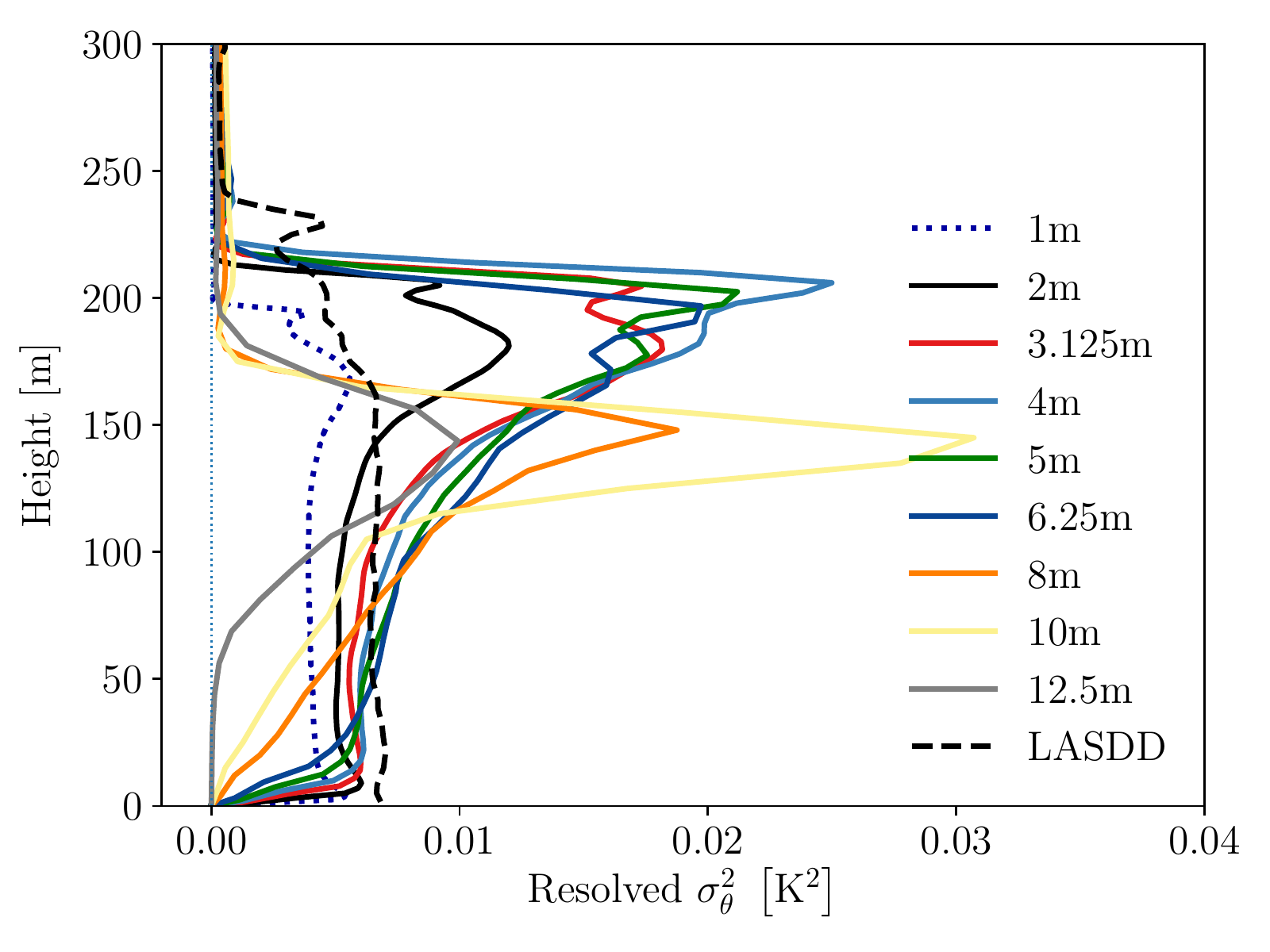}
  \includegraphics[width=0.49\textwidth]{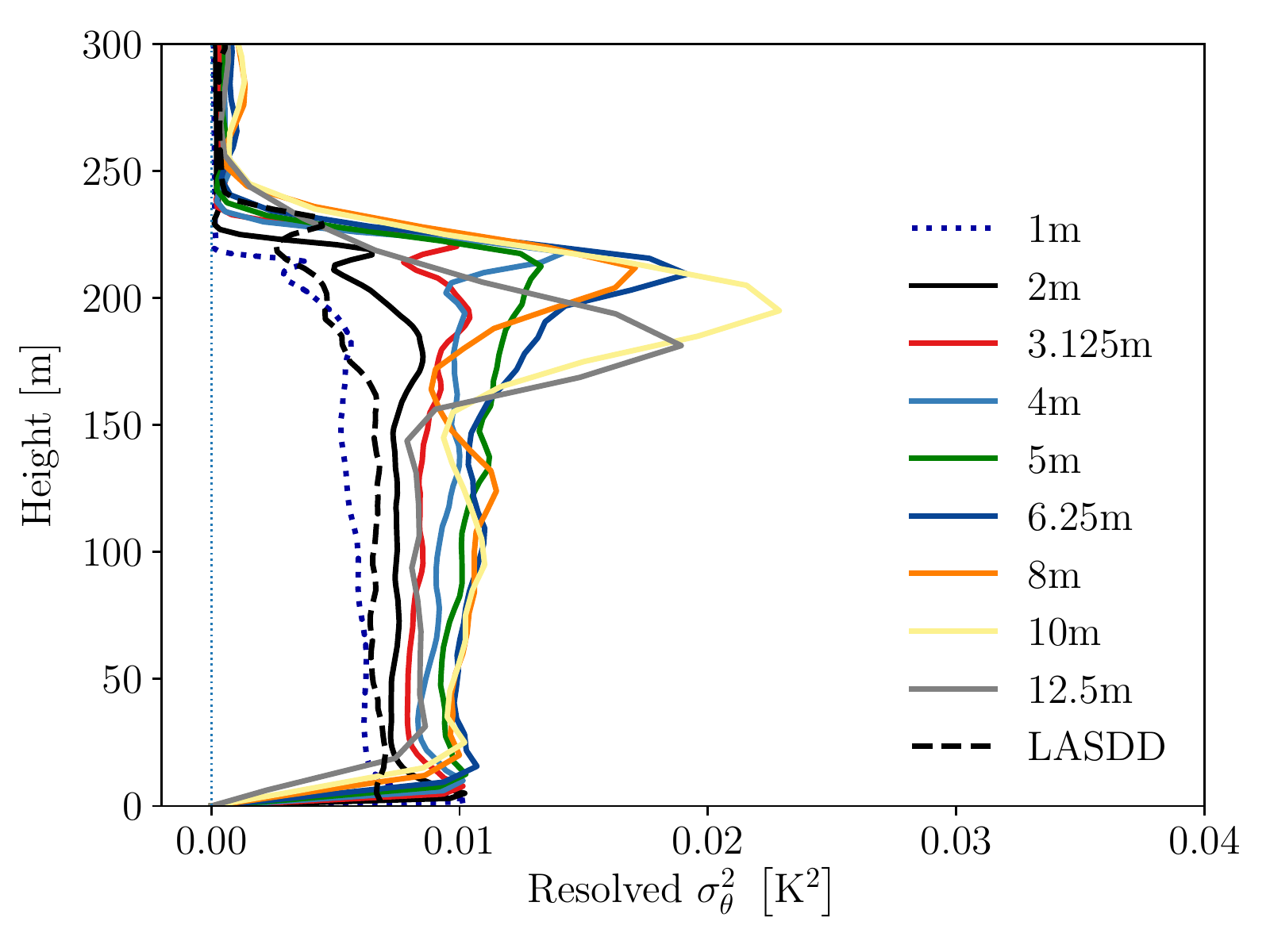}
\caption{Vertical profiles of resolved variances of vertical velocity (top panel) and potential temperature (bottom panel) from the D80 (left panel) and D80-R (right panel) based simulations using the PALM model system. Different colored lines correspond to different grid sizes ($\Delta$). Results from the MATLES code are overlaid (dashed black lines) for comparison.}
\label{fig:PALMVariance}      
\end{figure*}

\bibliography{GridSize}   

\end{document}